\documentclass[12pt]{article}
\usepackage{amsfonts}
\usepackage{mathrsfs}
\usepackage{amssymb}
\usepackage{amsmath}
\usepackage{amsthm}
\usepackage{graphicx}

\usepackage{subfigure}
\input epsf.sty
\usepackage{color}
\def\disp{\displaystyle}

\def\dref#1{(\ref{#1})}

\def\dfrac{\displaystyle\frac}

\numberwithin{equation}{section}
\DeclareMathOperator*{\res}{Res}
\begin{document}

\title{{\bf From $r$-Spin Intersection Numbers to Hodge Integrals}}
\author{\; Xiang-Mao Ding$^1$\footnote{\small
Email: xmding@amss.ac.cn},\;  Yuping Li$^1$\footnote{\small
Email: liyuping@amss.ac.cn},\; Lingxian Meng$^{1,2}$\footnote{\small Email: menglingxian@amss.ac.cn}
\\{\it ~$^1$ Institute of Applied Mathematics, Academy of
}\\ {\it  Mathematics and Systems Science; Chinese Academy of Sciences,}\\{\it Beijing 100190, People's Republic of China}\\{\it ~$^2$College of Mathematics and Information Science,}\\{\it Zhengzhou University of Light Industry}\date{}}

%\date{}
%February 2, 2012}

 \maketitle

\begin{abstract}
Generalized Kontsevich Matrix Model (GKMM) with a certain given potential is the partition function of $r$-spin intersection numbers. We represent this GKMM in terms of fermions and expand it in terms of the Schur polynomials by boson-fermion correspondence, and link it with a Hurwitz partition function and a Hodge partition by operators in a $\widehat{GL}(\infty)$ group. Then, from a $W_{1+\infty}$ constraint of the partition function of $r$-spin intersection numbers, we get a $W_{1+\infty}$ constraint for the Hodge partition function. The $W_{1+\infty}$ constraint completely determines the Schur polynomials expansion of the Hodge partition function.
\vspace{0.3cm}


\vspace{0.3cm}

%\noindent {\bf AMS subject classifications:}~35J10, 93C20, 93C25.
\end{abstract}
\section{Introduction}
It is commonly assumed that the generating functions in enumerative geometry constitute a particular subclass of the string theory partition functions. This subclass possesses nice integrable properties and matrix model representations, and from it one can find some universal properties of string theory partition functions. We call the generating function of certain type as the partition function for the case. There are kinds of partition functions for different purpose. The most known ones are partition function of $r$-spin intersection numbers, Hurwitz partition function and Hodge partition function.

The well-known Witten Conjecture \cite{E.Witten} was proved by M.Kontsevich \cite{M.Kontsevich}, it stated the identitical between the generating function of intersection numbers on moduli space of stable curves and the $\tau$-function of KdV hierarchies. Then and there, E.Witten introduced $r$-spin curves and their moduli spaces, and he conjectured that the generating function of the $r$-spin intersection numbers is a solution to $r$-reduced KP hierarchies \cite{E.Witten1}. The conjecture about $2$-spin intersection numbers was exactly Witten's original statement. The generalised conjecture was proved by C. Faber, S. Shadrin and D. Zvokine using tautological relations \cite{C.Faber}. After then, investigating the problems involved with the $r$-spin intersection numbers become fascinating subjects.

As pointed out by Witten \cite{E.Witten}, that the partition function of the intersection numbers of $\psi$-classes on $\overline{\mathcal {M}}_{g,n}$ is a solution to KdV hierarchies,  and with an additional string equation, completely determine the intersection numbers. Similarly, with an additional string equation, the $r$-reduced KP hierarchy completely determine $r$-spin intersection numbers \cite{E.Witten1}. Besides the string equation, the partition function for $r$-spin intersection numbers also satisfies a dilaton equation and a WDVV equation. All these three equations are called the tautological equations or the universal equations. In principle, the $r$-spin intersection numbers can be obtained through the tautological equations in a recursive way \cite{S.V.Shadrin,K.Liu}. In \cite{K.Liu}, K.Liu and his collaborators express any given $r$-spin intersection number as the sum of products of simpler $r$-spin intersection numbers. By this way, they could obtain all the $r$-spin intersection numbers. But the definitely works for higher genus are highly nontrivial, for the recursive relations would be very complicated.

It is well known that the partition function for $r$-spin intersection numbers also satisfies linear constraints, called as the Virasoro constraint in the $r=2$ case and the $W$-constraint in general cases. In fact, such constraints are equivalent to $r$-reduced KP hierarchies additional with a string equation \cite{M.Fukuma}. Solving the linear constraints is an effective method to calculate the $r$-spin intersection numbers. In the $r=2$ case,  A.Alexandrov gave a cut-and-join type operator representation for this partition function by grading operators \cite{A.Alexandrov1},
\begin{equation}\label{1.1}
\begin{array}{ccc}
Z=e^{A}\cdot1,\\
A=\frac{4}{3}\sum\limits_{k=0}^{\infty}\left(k+\frac{1}{2}\right)\tau_{k}\mathcal {L}_{k},
\end{array}
\end{equation}
where $Z$ is the partition function of $2$-spin intersection numbers and subjects to the constraints

\begin{equation*}
{\mathcal {L}}_{k}Z=0, \ \ \ \ k \geq -1.
\end{equation*}

\noindent where the ${\mathcal {L}_{k}}$ are generators of the Virasoro algebra without central extension

\begin{equation*}
[{\mathcal {L}}_{m},  {\mathcal {L}}_{n}]=(m-n){\mathcal {L}}_{m+n},
\end{equation*}

\noindent For the general case, J. Zhou gave a fermionic representation of the generating function by solving the string equation \cite{Zhou}.
He got a formula as \cite{Zhou1}
\begin{equation}\label{1.2}
Z=\exp\left(\sum\limits_{j=1}^{r-1}A_{j}\right)\cdot1,
\end{equation}
here the operators $\{A_{j},\ j=1,2\cdots r-1\}$ are constructed from $W$-constraints,  and $Z$ is considered as the partition function of the $r$-spin intersection numbers. Unlike the $r=2$ case, \dref{1.2} is correct with the supposing the condition $[A_{i},A_{j}]=0,\ i,j=1,2,\cdots,r-1$. In these
papers,  the operators $A$ in equation \dref{1.1} or $\{A_{i}\}$ in equation \dref{1.2} is not $\hat{gl}(\infty)$ algebra, and the integrability is
obscured.

In fact, the $r$-spin intersection numbers could be calculated from a Generalized Kontsevich Matrix Model (GKMM). The idea was carried out in \cite{B.Brezin}. We use a different processing method in this manuscript. The partition function of a GKMM with monomial potential $V(X)=\frac{\sqrt{-r}X^{r+1}}{r+1}$ is a $r$-reduced KP $\tau$-function, and this $\tau$-function is also subject to a string equation that the partition function of the $r$-intersection numbers satisfies. As stated by the uniqueness property \cite{Liu}, this GKMM partition function is identical with the partition function of the $r$-spin intersection numbers up to a multiple constant. From the integrability of the GKMM, we can get a fermionic representation of it from the method given by S.Kharchev and his collobrators \cite{S.Kharchev}, then expanding the $\tau$-function in terms of the Schur polynomials by the boson-fermion correspondence, in principle, we can get all the $r$-spin intersection numbers.

Besides the partition function of $r$-spin intersection numbers, there are other two well-known generating functions. One is the Hurwitz partition function, there are many interesting results about it. The Hurwitz partition function can be represented in terms of a cut-and-join operator \cite{A.Alexandrov3,A.Mironov2}, this operator is an element of the $\widehat{GL}(\infty)$ group, acting on the space of KP solutions, which guarantees that the Hurwitz partiton function is a KP $\tau$-function. Another one is the Hodge partition function, which is a generating function of linear Hodge integrals. It is well known that these three partition functions are very inherently linked \cite{A.Okounkov,A.Mironov2}. The Ekedahl-Lando-Shapiro-Vainshtein (ELSV) formula connects the Hurwitz partition function to the Hodge partition function, and the relationship between them also can be represented by generators of a Virasoro algebra \cite{M.Kazarian,A.Alexandrov2}. The Hodge partition function is a deformation of the KW $\tau$-function i.e. the partition function of $2$-spin intersection numbers. A.Mironov and A.Morozov had mapped the Hodge partition function to the KW $\tau$-function by a Givental operator, and from this they get the Virasoro constraint for Hodge partition function from the Virasoro constraint of the KW $\tau$-function \cite{A.Mironov1}. However, the Givental operator is not an element of the $\widehat{GL}(\infty)$ group, then the integrable property of the Hodge partition function is obscured in this expression.

The partition function of $r$-spin intersection numbers is identical with a $r$-reduced KP $\tau$-function, while the Hodge partition function is given by a KP $\tau$-function. These facts imply that there should have an operator in the $\widehat{GL}{(\infty)}$ group to match them.  In the case $r=2$, A.Alexandorov has conjectured a form of this operator \cite{A.Alexandrov2}, and recently the conjecture has proved independently by A.Alexandrov  himself\cite{A.Alexandrov8}, X.Liu and G.Wang \cite{X.Liu}\footnote{Our main results had been reported at 'The First Annual Meeting of Vertex Operator Algebras' chaired by C.Dong, at Beijing Institute of Technology, Feb 20, 2014. It is quite sorry that we did not note the article by A.Alexandrov at the first draft of our manuscript, while the paper by Liu\&Wang was appeared when our manuscript is almost completely polished.}. From their result, an intersection number could be expressed as an infinite summation of Hodge integrals. For the partition function of $r$-spin intersection numbers, any element in the $W_{1+\infty}$ constraint is an element of the $\widehat{gl}{(\infty)}$ algebra, so there must be a $W_{1+\infty}$ constraint for the Hodge partition function whose elements are generators of the $\widehat{gl}{(\infty)}$. However, up to now this type of operator and the $W_{1+\infty}$ constraint is unknown yet.

In this paper, we will give a determinantal representation for the partition function of $r$-spin intersection numbers. This expression is equivalent to a fermionic representation of it, with the help of the boson-fermi duality, and we can formulate it as a linear combination of the Schur function on eigenvalues of external field in the GKMM. Then, we can get the Schur polynomials representation by the Miwa transformation.

It is well known that the cut-and-join operator generating the Hurwitz partition function and the Virasoro operators can be expressed as elements of the $\widehat{gl}(\infty)$ algebra. For the two kinds of partition functions, the one for $r$-spin intersection numbers and Hodge partition function, we get a $\widehat{GL}(\infty)$ operator to match them. By boson-fermion correspondence, we get the bosonic version of this operator, and from the expression, a $r$-spin intersection number can be expressed as a finite summation of Hodge integrals. Then, from the $W_{1+\infty}$ constraint for the partition function of $r$-spin intersection numbers, we get the $W_{1+\infty}$ constraint for the Hodge partition function. For the Hodge partition function, the Schur polynomials representation of the Hodge partition function can be got by solving this constraint.

This paper is organized as follows. In section 2, we list basic notations of the partition function of $r$-spin intersection numbers, the Hurwitz partition function and the Hodge partition function. In section 3, it is the material about the $\widehat{GL}(\infty)$ group and KP hierarchy. Section 4 and section 5 are the main parts of this paper, in section 4, we calculate the $r$-spin intersection numbers with the GKMM, in section 5, we build an operator to pair these partition functions, and obtain a $W_{1+\infty}$ constraint for the Hodge partition function. We put detail results of some intersection numbers and a Virasoro constraint for Hurwitz partition function in Appendix A and B, respectively.

\section{$\tau$-Functions for Enumerative Geometry}
\subsection{Partition Function of Witten's $r$-Spin Intersection Numbers}
Let $\overline{\mathcal {M}}_{p,n}$ be the Deligne-Mumford compactification of the moduli space of algebraic curves $X$ with genus $p$ and $n$ marked points $\{x_{1}, x_{2},\cdots,x_{n}\}$. Let us associate a marked point with a line bundle $\mathcal {L}_{i}$ whose fiber at a moduli point $(X; x_{1},\cdots,x_{n})$ is the cotangent space to $X$ at $x_{i}$. The $r$-spin intersection numbers of these holomorphic line bundles are defined by Witten \cite{E.Witten1} as follows:
\begin{equation}\label{2.1}
\langle\tau_{m_{1},a_{1}}\cdots\tau_{m_{n},a_{n}}\rangle_{p}=\int_{\overline{\mathcal {M}}_{p,n}}
c_{W}(a_{1},\cdots,a_{n})\psi(x_{1})^{m_{1}}\cdots\psi(x_{n})^{m_{n}},
\end{equation}
in which $c_{W}(a_{1},\cdots,a_{n})$ is the top Chern class, and $\psi(x_{i})$ is the first Chern class of the bundle $\mathcal {L}_{i}$. The intersection numbers are nonzero if and only if the following selecting rule is satisfied:
\begin{equation}\label{2.2}
(r+1)(2p-2)+rn=r\sum\limits_{i=1}^{n}m_{i}+\sum\limits_{i=1}^{n}a_{i},
\end{equation}
here
\begin{equation}\label{2.3}
a_{i}\in\{0,1,\cdots,r-2\}.
\end{equation}
For a given genus $p$, the intersection numbers satisfy the string equation
\begin{equation}\label{2.4}
\langle\tau_{0,0}\prod\limits_{i=1}^{n}\tau_{m_{i},a_{i}}\rangle_{p}=\sum\limits_{j=1}^{n}
\langle\tau_{m_{j}-1,a_{j}}\cdot\prod\limits_{i=1,i\neq j}^{n}\tau_{m_{i},a_{i}}\rangle_{p},
\end{equation}
and the dilation equation
\begin{equation}\label{2.5}
\langle\tau_{1,0}\prod\limits_{i=1}^{n}\tau_{m_{i},a_{i}}\rangle_{p}=
(2g-2+n)\langle\prod\limits_{i=1}^{n}\tau_{m_{i},a_{i}}\rangle_{p}.
\end{equation}
In the genus 0 case, the following result was obtained \cite{E.Witten1}:
\begin{equation}\label{2.6}
\langle\tau_{0,a_{1}}\tau_{0,a_{2}}\tau_{0,a_{3}}\rangle_{0}=\delta_{a_{1}+a_{2}+a_{3},r-2}.
\end{equation}
For other nontrivial cases, the concrete formula has not yet known until recently.

If we introduce formal variables $t_{m,a}$ corresponding to $\tau_{m,a}$ ($m=0,1,2,\cdots; a=0,1,\cdots,r-2$), then we can define a generating function
named as the free energy
\begin{equation}\label{2.7}
\begin{array}{lll}
F^{\{r\}}_{p}(t)&=&\sum\langle\tau_{m_{1},a_{1}}\cdots\tau_{m_{n},a_{n}}\rangle_{p}\cdot\frac{1}{n!}
\prod\limits_{i=1}^{n}t_{m_{i},a_{i}}\\
&=&\left\langle\exp(\sum\limits_{m=0}^{\infty}\sum\limits_{a=0}^{r-2}t_{m,a}\tau_{m,a})\right\rangle_{p},
\end{array}
\end{equation}
and the total free energy is obtained by summation over all genera
\begin{equation}\label{2.8}
F^{\{r\}}(t;g)=\sum\limits_{p\geq0}g^{2p-2}F^{\{r\}}_{p}(t),
\end{equation}
as well as
\begin{equation}\label{2.9}
Z^{\{r\}}(t;g)=\exp\{F^{\{r\}}(t;g)\}.
\end{equation}
Here, $F^{\{r\}}(t;g)$ is the so called generating function of $r$-spin intersection numbers and $Z^{\{r\}}(t;g)$ is the partition function. In \dref{2.8} and \dref{2.9}, a parameter $g$ is introduced. In fact, we can restore $g$ into $t_{m,a}$ by replacing $t_{m,a}$ with $g^{-\frac{r(m-1)+a}{r+1}}t_{m,a}$. So, in the following we can set $g=1$ without loss of generality. The first few terms of $F^{\{r\}}_{0}$ and $F^{\{r\}}_{1}$ have been already given by Witten \cite{E.Witten1}
\begin{equation}\label{2.10}
\begin{array}{lll}
F^{\{r\}}_{0}(t)=\frac{1}{3!} \sum\limits_{a_{1}+a_{2}+a_{3}=r-2}t_{0,a_{1}}t_{0,a_{2}}t_{0,a_{3}}+\cdots,
\end{array}
\end{equation}
\begin{equation}\label{2.25}
F^{\{r\}}_{1}(t)=\frac{r-1}{24}t_{1,0}+\cdots.
\end{equation}

It is well known that the string equation and dilation equation can be reformulated as the following two differential equations
\begin{equation}\label{2.11}
L^{\{r\}}_{-1}\cdot Z^{\{r\}}=0,
\end{equation}
\begin{equation}\label{2.13}
L^{\{r\}}_{-1}=\frac{\partial}{\partial t_{0,0}}-\sum\limits_{k=1}^{\infty}\sum\limits_{a=0}^{r-2}t_{k,a}\frac{\partial}{\partial t_{k-1,a}}
-\frac{1}{2}\sum\limits_{a=0}^{r-2}t_{0,a}t_{0,r-2-a},
\end{equation}
and
\begin{equation}\label{2.12}
L^{\{r\}}_{0}\cdot Z^{\{r\}}=0,
\end{equation}
\begin{equation}\label{2.14}
L^{\{r\}}_{0}=-\frac{\partial}{\partial t_{1,0}}+\sum\limits_{k=1}^{\infty}\sum\limits_{a=0}^{r-2}\frac{rk+a+1}{r+1}t_{k,a}
\frac{\partial}{\partial t_{k,a}}+\frac{r-1}{24},
\end{equation}
respectively.

Then, the generalized Witten conjecture can be stated as follows: There is a pseudo-differential operator $Q$
\begin{equation}\label{2.15}
Q=\partial^{r}+\sum\limits_{i=0}^{r-2}u_{i}(t)\partial^{i},\ \ \ \ \partial=\frac{\sqrt{-1}}{\sqrt{r}}\frac{\partial}{\partial t_{0,0}},
\end{equation}
such that
\begin{equation}\label{2.16}
\frac{\partial^{2}F^{\{r\}}}{\partial t_{0,0}\partial t_{m,a}}=-c_{m,a}\res(Q^{m+\frac{a+1}{r}}),
\end{equation}
here the constant
\begin{equation}\label{2.17}
c_{m,a}=\frac{(-1)^{m}r^{m+1}}{(a+1)(a+1+r)\cdots(a+1+mr)},
\end{equation}
while
\begin{equation}\label{2.18}
\sqrt{-1}\frac{\partial Q}{\partial t_{m,a}}=\frac{c_{m,a}}{\sqrt{r}}\cdot[(Q^{m+\frac{a+1}{r}})_{+},Q].
\end{equation}
The formula can be simplified by introducing a new set of variables $\{t_{n}\}$, which we name as time variables.
\begin{equation}\label{2.20}
t_{mr+a+1}=\frac{c_{m,a}}{\sqrt{-r}}\cdot t_{m,a}=(-1)^{m}\prod\limits_{j=0}^{m}
\left(j+\frac{a+1}{r}\right)^{-1}\cdot\frac{t_{m,a}}{\sqrt{-r}}.
\end{equation}
Then in terms of the new coordinates $\{t_{1},\cdots,t_{r-1},t_{r+1},\cdots\}$, we can define the Lax operator of KP hierarchy by using the Gelfand-Dickey scheme \cite{L.A.Dickey}. The Lax operator of the hierarchy
$L$ is constructed from the operator $Q$
\begin{equation}\label{2.24}
L=Q^{\frac{1}{r}}.
\end{equation}
Rewrite equations \dref{2.16} and \dref{2.18} with the Lax operator $L$ in the new set of variables $\{t_{k}\}$
\begin{equation}\label{2.21}
\frac{\partial L^{r}}{\partial t_{k}}=[(L^{k})_{+},L^{r}]=[(L^{k})_{+},Q],
\end{equation}
and
\begin{equation}\label{2.22}
\frac{\partial^{2}F^{\{r\}}}{\partial t_{1}\partial t_{k}}=\res(L^{k}),
\end{equation}
in which $\res(P)$ is defined as the coefficient of $\partial^{-1}$ in the pseudo-differential $P$. From equation \dref{2.24} and equation \dref{2.22}, we easily get
\begin{equation}\label{2.101}
\frac{\partial F^{\{r\}}}{\partial t_{kr}}=\text{const}.
\end{equation}
This, together with equation \dref{2.21}, implies that the partition function of $r$-spin intersection numbers is a $r$-reduced KP $\tau$-function.
In terms of the variables $\{t_{k}\}$, the operators $L^{\{r\}}_{-1}$ and $L^{\{r\}}_{0}$ become
\begin{equation}\label{2.23}
L^{\{r\}}_{-1}=-\sqrt{-r}\frac{\partial}{\partial t_{1}}+\sum\limits_{k=r+1}^{\infty}\frac{k}{r}t_{k}\frac{\partial}{\partial t_{k-r}}
+\frac{1}{2r}\sum\limits_{b+c=r}bt_{b}\cdot ct_{c},
\end{equation}
and
\begin{equation}\label{2.24}
L^{\{r\}}_{0}=-\sqrt{-r}\frac{\partial}{\partial t_{r+1}}
+\sum\limits_{k=1}^{\infty}\frac{k}{r}t_{k}\frac{\partial}{\partial t_{k}}+\frac{r^{2}-1}{24r}.
\end{equation}
respectively. The conjecture has been proved by Faber-Shadrin-Zvokine \cite{C.Faber}.
The fact that $Z^{\{r\}}(t)$ is a $r$-reduced KP $\tau$-function with additional the string equation could completely determine the $r$-spin intersection numbers \cite{E.Witten1}.
\subsection{Hurwitz Partition Function}
Hurwitz numbers count ramified coverings of the Riemann sphere. More precisely, the simple Hurwitz number $h(p|m_{1},\cdots,m_{n})$ gives the number of the Riemann sphere coverings with $N$ sheets, $N=\sum_{i=1}^{n}m_{i}$, fixed simple ramification points, and a single point with ramification structure given by $\{m_{i}\}$, a partition of $N$ \cite{A.Hurwitz}. The number of the simple ramification $M$, the genus $p$ of the covering and the partition $\{m_{i}\}$ are related:
\begin{equation}\label{lyp10}
M=(2p-2)+\sum_{i=1}(m_{i}+1)=(2p-2)+N+n.
\end{equation}
One can introduce a generating function of the simple Hurwitz numbers
\begin{equation}\label{lyp11}
H(t_{1},t_{2},\cdots)=\sum_{p=0}^{\infty}g^{2p-2}\sum_{n=1}^{\infty}\sum_{m_{i};M}\dfrac{\beta^{M}}{M!}
h(p|m_{1},\cdots,m_{n})\dfrac{t_{m_{1}}\cdots t_{m_{n}}}{m_{1}\cdots m_{n}},
\end{equation}
in which $g$, $\beta$ are two parameters.
Define an operator
\begin{equation}\label{lyp20}
\hat{W}^{(3)}_{0}=\sum\limits_{i,j\geq1}\left(ijt_{i}t_{j}\dfrac{\partial}{\partial t_{i+j}}+g^{2}(i+j)t_{i+j}\dfrac{\partial^{2}}{\partial t_{i}\partial t_{j}}\right).
\end{equation}
Obviously, the $\hat{W}^{(3)}_{0}$ is a cut-and-join operator. The Hurwitz partition function can be represented from it \cite{A.Alexandrov3,A.Mironov2}
\begin{equation}\label{lyp12}
\begin{array}{lll}
Z_{H}(t_{1},t_{2},\cdots;\beta):&=&\exp\left(H(t_{1},t_{2},\cdots)\right)=
\exp\left(\frac{\beta}{2}\hat{W}_{0}\right)\cdot \exp\left(\dfrac{t_{1}}{g^2}\right)\\
 &=&1+\dfrac{t_{1}}{g^2}+\dfrac{e^{\beta}}{2g^4}\left(\dfrac{t_{1}^{2}}{2}+t_{2}\right)
 +\dfrac{e^{-\beta}}{2g^4}\left(\dfrac{t_{1}^{2}}{2}-t_{2}\right)+\cdots.
\end{array}
\end{equation}
For this partition function, several matrix integral representations are known. For example \cite{A.Morozov}:
\begin{equation}\label{lyp13}
\begin{array}{llll}
Z_{H}(t_{k};\beta)=\disp\int_{N\times N}dM\sqrt{\det\left(\dfrac{\sinh{\left(\dfrac{M\otimes I-I\otimes M}{2}\right)}}{\left(\dfrac{M\otimes I-I\otimes M}{2}\right)}\right)} \exp\left(-\dfrac{1}{2\beta}\text{Tr}M^{2}+\text{Tr}e^{\left(M-Nt/2\right)}\psi\right),
\end{array}
\end{equation}
where $M$ is a $N\times N$ Hermitian matrix and $\psi$ is a $N\times N$ diagonal matrix, meanwhile the times $t_{k}$ are given by the Miwa transform
\begin{equation}\label{lyp14}
t_{k}=\dfrac{1}{k}\text{Tr}\psi^{k}.
\end{equation}

\subsection{Hodge Integrals}

The Hodge integrals, it means intersection numbers of the form
\begin{equation}\label{lyp15}
\left\langle\lambda_{j}\sigma_{k_{1}}\cdots\sigma_{k_{n}}\right\rangle=\int_{\overline{\mathcal {M}}_{p,n}}\lambda_{j}\psi_{1}^{k_{1}}\cdots\psi_{n}^{k_{n}}
\end{equation}
where $\lambda_{j}$ is the $j$th Chern class of the rank $p$ Hodge vector bundle whose fiber is the space of holomorphic one forms and $\psi_{i}$ is the first chern class defined as same as in \eqref{2.1}. Those numbers are well defined whenever the equalities $j+\sum\limits_{i=1}^{n}k_{i}=3p-3+n=\dim\overline{\mathcal {M}}_{p,n}$ holds.

Let us collect the Hodge integrals into the following series in terms of a set of infinite formal variables $\beta,  T_{0},T_{1},\cdots$
\begin{equation}\label{lyp16}
\begin{array}{llll}
\ \ \ \ \overline{G}(\beta;T_{0},T_{1},\cdots)\\
=\sum\limits_{j;k_{0},k_{1},\cdots}\left(-1\right)^{j}\left\langle\lambda_{j}\sigma_{0}^{k_{0}}
\sigma_{1}^{k_{1}}\cdots\right\rangle \beta^{\frac{2j}{3}}\dfrac{T_{0}^{k_{0}}}{k_{0}!}\dfrac{T_{1}^{k_{1}}}{k_{1}!}\cdots,
\end{array}
\end{equation}
where the summation is taken over all possible monomials in the symbols $T_{k}$ and $j\geq 0$. We can introduce the parameter $g$ into the generating function, such that it has the genus expansion. we set up:
\begin{equation}\label{lyp17}
G(g,\beta;T_{0},T_{1},\cdots)=\overline{G}(g\beta;g^{-\frac{2}{3}}T_{0},
\cdots,g^{\frac{2k-2}{3}}T_{k},\cdots).
\end{equation}
The generating function has the following expansion \cite{A.Mironov1}:
\begin{equation}\label{lyp18}
G(g,\beta;T_{0},T_{1},\cdots)=\sum\limits_{p=0}^{\infty}g^{2p-2}G^{(p)}(\beta;T_{0},T_{1},\cdots).
\end{equation}
If set $\beta=0$ in the equation \dref{lyp18}, it is exactly the partition function of $2$-spin intersection numbers \eqref{2.9}, i.e.
\begin{equation}\label{lyp19}
G(g,0;t_{1},t_{3},\cdots,t_{2k+1},\cdots)=Z^{\{2\}}(t;g).
\end{equation}
From equation \dref{lyp17}, one can easily get
\begin{equation}\label{lyp32}
G(g,\beta;T_{0},T_{1},\cdots)=G(1,g\beta;g^{-\frac{2}{3}}T_{0},g^{0}T_{1},\cdots,
g^{\frac{2k-2}{3}}T_{k}\cdots).
\end{equation}
So the parameter $g$ can be restored in the parameters $\beta,T_{0},T_{1}\cdots$. In the following, for convenience we often (not always) consider the $g=1$ case.

The generating function $G(g,\beta;T_{0},T_{1}\cdots)$ is a solution of a KP hierarchy with respect to new variables $q_{i}$'s, which are linear transformations of variables $T_{k}$'s. If we set $g=1$. The transformation are \cite{M.Kazarian}:
\begin{equation}\label{lyp22}
\begin{array}{lll}
T_{0}=\beta^{\frac{4}{3}}q_{1},\\
T_{k+1}=\sum\limits_{m\geq1}\left[\beta^{\frac{2}{3}}mq_{m}+2(m+1)\beta^{\frac{5}{3}}q_{m+1}
+(m+2)\beta^{\frac{8}{3}}q_{m+2}\right]\dfrac{\partial}{\partial q_{m}}T_{k},\ \ \ k\geq0.
\end{array}
\end{equation}
The first few terms of this transformation are
\begin{equation}\label{lyp21}
\begin{array}{lllll}
T_{0}=\beta^{\frac{4}{3}}q_{1},\\
T_{1}=\beta^{2}q_{1}+4\beta^{3}q_{2}+3\beta^{4}q_{3},\\
T_{2}=\beta^{\frac{8}{3}}q_{1}+12\beta^{\frac{11}{3}}q_{2}+36\beta^{\frac{14}{3}}q_{3}
+40\beta^{\frac{17}{3}}q_{4}+15\beta^{\frac{20}{3}}q_{5},\\
T_{3}=\beta^{\frac{10}{3}}q_{1}+28\beta^{\frac{13}{3}}q_{2}+183\beta^{\frac{16}{3}}q_{3}
+496\beta^{\frac{19}{3}} q_{4}+615\beta^{\frac{22}{3}}q_{5}+420\beta^{\frac{25}{3}}q_{6}+105\beta^{\frac{28}{3}}q_{7},\\
\cdots.
\end{array}
\end{equation}
Set the identification of the generating function $G(1,\beta;T_{0}(q),\cdots,T_{k}(q),\cdots)$ with the Hodge free energy
\begin{equation}\label{lyp23}
F_{Hodge}(\beta;q_{1},q_{2},\cdots,q_{k}\cdots)=G(1,\beta;T_{0}(q),T_{1}(q),\cdots,T_{k}(q),\cdots).
\end{equation}
We have the following Hodge partition function
\begin{equation}\label{lyp24}
\begin{array}{lll}
Z_{Hodge}(\beta;q_{1},q_{2},\cdots)=\exp\left(F_{Hodge}(\beta;q_{1},q_{2},\cdots)\right).
\end{array}
\end{equation}
Therefore, $Z_{Hodge}$ is a $\tau$-function of a KP hierarchy with respect to the variables $q_{i}$'s.

\subsection{ELSV Formula}

There are several approaches to investigate the intersection theory of moduli spaces. Among these approaches, the ELSV formula seems to be the most straightforward one \cite{T.Ekedahl}, it expresses the Hurwitz numbers as linear combinations of the Hodge integrals. This formula build a bridge between the Hurwitz partition function and the Hodge partition function \cite{A.Alexandrov2,M.Kazarian}.

Consider two variables $x$ and $z$ related to each other by the following formulas
\begin{equation}\label{lyp27}
\begin{array}{lll}
x=\dfrac{z}{1+\beta z}e^{\frac{\beta z}{1+\beta z}}=z-2\beta z^{2}+\dfrac{7}{2}\beta^{2}z^{3}-\dfrac{17}{3}\beta^{3}z^{4}\cdots,
\end{array}
\end{equation}
and
\begin{equation}\label{new1}
\begin{array}{lllll}
z=\sum\limits_{b\geq1}\frac{b^{b}}{b!}\beta^{b-1}x^{b}=x+2\beta x+\dfrac{9}{2}\beta^{2}x^{3}+\dfrac{32}{3}\beta^{3}x^{4}+\cdots.
\end{array}
\end{equation}
These two formulas provide a linear isomorphism (depending on the parameter $\beta$) between the spaces of formal power series in the variables $x$ and $z$. We set the following correspondence
\begin{equation}\label{lyp28}
t_{b}\longleftrightarrow \dfrac{x^{b}}{b},\ q_{b}\longleftrightarrow \dfrac{z^{b}}{b}.
\end{equation}
Then we can express $t_{b}$ as a linear combination of $q_{m}$, and vice versa.
\begin{equation}\label{lyp29}
\begin{array}{llll}
t_{b}=\sum\limits_{m\geq b}c_{m}^{b}\beta^{m-b}q_{m},\\
q_{m}=\sum\limits_{b\geq m}d_{b}^{m}\beta^{b-m}t_{b}.
\end{array}
\end{equation}
The coefficients $c_{m}^{b}$ and $d_{b}^{m}$ are determined by the following equations
\begin{equation}\label{lyp30}
\begin{array}{llll}
x^{b}=b\sum\limits_{m\geq b}c_{m}^{b}\beta^{m-b}\dfrac{z^{m}}{m},
\end{array}
\end{equation}
and
\begin{equation}\label{new2}
\begin{array}{lllll}
z^{m}=m\sum\limits_{b\geq m}d_{b}^{m}\beta^{b-m}\dfrac{x^{b}}{b},
\end{array}
\end{equation}
respectively.
If we introduce two functions,
\begin{equation}\label{lyp26}
\begin{array}{lll}
H_{0,1}=\sum\limits_{b=1}^{\infty}\dfrac{b^{b-1}}{b!g^{2}}t_{b}\beta^{b-1},
\end{array}
\end{equation}
\begin{equation}\label{new3}
\begin{array}{llll}
H_{0,2}=\dfrac{1}{2}\sum\limits_{b_{1},b_{2}=1}^{\infty}\dfrac{b_{1}^{b_{1}+1}b_{2}^{b_{2}+1}}{(b_{1}
+b_{2})b_{1}!b_{2}!g^{2}}t_{b_{1}}t_{b_{2}}\beta^{b_{1}+b_{2}},
\end{array}
\end{equation}
then the Hurwitz partition function and the Hodge partition function \eqref{lyp11} are linked by the following formula \cite{M.Kazarian}:
\begin{equation}\label{lyp31}
\left(H-H_{0,1}-H_{0,2}\right)|_{t_{n}=t_{n}(q_{m}),g=1}=F_{Hodge}(\beta;q_{1},q_{2},q_{3},\cdots).
\end{equation}
Surely, we can also restore the parameter $g$ into the above formula. From equation \dref{lyp22}, we get
\begin{equation}\label{lyp33}
\begin{array}{llllllllll}
g^{\frac{2k}{3}}T_{k+1}=&&\sum\limits_{m\geq1}\left[\left(g\beta\right)^{\frac{2}{3}}mg^{-m-1}q_{m}
+2(m+1)\left(g\beta\right)^{\frac{5}{3}}g^{-m-2}q_{m+1}\right.\\
&+&\left.(m+2)(g\beta)^{\frac{8}{3}}g^{-m-3}q_{m+2}\right]\dfrac{\partial}{\partial (g^{\frac{m-3}{3}}q_{m})}(g^{\frac{2k-2}{3}}T_{k}).
\end{array}
\end{equation}
If we define the function
\begin{equation}\label{lyp34}
\begin{array}{lllll}
F_{Hodge}(g,\beta;q_{1},q_{2},\cdots)&=&F_{Hodge}((g\beta),g^{-2}q_{1},\cdots,g^{-k-1}q_{k},\cdots)\\
&=&G(g,\beta;T_{0}(q_{k}),T_{1}(q_{k})\cdots),
\end{array}
\end{equation}
then, $F_{Hodge}(g,\beta;q_{1},q_{2},\cdots)$ can also be expanded as
\begin{equation}\label{lyp25}
F_{Hodge}(g,\beta;q_{1},q_{2},\cdots)=\sum\limits_{p=0}^{\infty}g^{2p-2}F_{Hodge}^{(p)}(\beta;q_{1},q_{2},\cdots).
\end{equation}
We can further define the Hodge partition function with parameter $g$
\begin{equation}\label{lyp96}
\begin{array}{llll}
Z_{Hodge}(g,\beta;q_{1},q_{2},\cdots)&=&\exp\left(F_{Hodge}(g,\beta;q_{1},q_{2},\cdots)\right)\\
&=&\exp\left(\sum\limits_{p=0}^{\infty}g^{2p-2}F_{Hodge}^{(p)}(g,\beta;q_{1},q_{2},\cdots)\right).
\end{array}
\end{equation}
If we rewrite the equation \dref{lyp29} as
\begin{equation}\label{lyp35}
\dfrac{t_{b}}{g^{b+1}}=\sum\limits_{m\geq b}c_{m}^{b}(g\beta)^{m-b}\dfrac{q_{m}}{g^{m+1}},
\end{equation}
and substituting $\frac{t_{b}}{g^{b+1}}$ for $t_{b}$, $\frac{q_{m}}{g^{m+1}}$ for $q_{m}$ and $g\beta$ for $\beta$ in equation \dref{lyp31}, respectively, then we get the binding between the Hurwitz partition function and the Hodge partition function with parameter $g$.

We can also relate the Hurwitz partition function with the Hodge partition function by an operator \cite{A.Alexandrov2}. The operator can be realized as a linear combination of certain modes of current algebra.
Consider the bosonic current
\begin{equation}\label{lyp36}
\begin{array}{lllll}
\hat{j}(z)&=&\sum\limits_{k=1}^{\infty}\left(\dfrac{k}{g}t_{k}z^{k-1}+\dfrac{g}{z^{k+1}}
\dfrac{\partial}{\partial t_{k}}\right)\\
&=&\sum\limits_{k\in\mathbb{Z}}\hat{j}_{k}z^{-k-1}.
\end{array}
\end{equation}
From this current, we can get a spin-2 current with central charge $c=1$:
\begin{equation}\label{lyp37}
\hat{L}(z)=\sum\limits_{n=-\infty}^{\infty}\hat{L}_{n}z^{-n-1}=\dfrac{1}{2}:\hat{j}(z)\hat{j}(z):,
\end{equation}
in which $::$ is the normal ordering, which means that the annihilation operators ($\hat{j}_{n},n>-1$) are always moved to the right side.
The explicit forms of $\{\hat{L}_{m}\}$ are
\begin{equation}\label{lyp38}
\begin{array}{lll}
\hat{L}_{m}=\sum\limits_{k=1}^{\infty}kt_{k}\dfrac{\partial}{\partial t_{k+m}}+\dfrac{g^{2}}{2}\sum\limits_{ a+b=m}\dfrac{\partial^{2}}{\partial t_{a}\partial_{b}}+\dfrac{1}{2g^{2}}\sum\limits_{a+b=-m}abt_{a}t_{b},
\end{array}
\end{equation}
and they are subject to the Virasoro algebra relation
\begin{equation}\label{lyp39}
\left[\hat{L}_{m}, \hat{L}_{n}\right]=(m-n)\hat{L}_{m+n}+\dfrac{1}{12}\delta_{m+n,0}(m^{3}-m).
\end{equation}
Furthermore, we can get a spin-$3$ current $W^{(3)}$ from the bosonic current:
\begin{equation}\label{lyp73}
\hat{W}^{(3)}(z)=\dfrac{g}{3}:\hat{j}(z)^{3}:=\sum\limits_{n\in\mathbb{Z}}\hat{W}_{n}z^{-n-3}.
\end{equation}
One of the modes in $\hat{W}^{(3)}(z)$
$$
\hat{W}^{(3)}_{0}=\sum\limits_{i,j\geq1}\left(ijt_{i}t_{j}\dfrac{\partial}{\partial t_{i+j}}
+g^{2}(i+j)t_{i+j}\dfrac{\partial^{2}}{t_{i}t_{j}}\right),
$$
is exactly the cut-and-join operator \dref{lyp20} of the Hurwitz partition function.

The Hodge partition function can be obtained from the Hurwitz partition functions  by the Borel subalgebra action generated by $\hat{L}_{m}$ with $m<0$.
\begin{equation}\label{lyp40}
Z_{Hodge}(g,\beta;t_{k})=\exp\left(\sum\limits_{k=1}^{\infty}a_{k}\beta^{k}\hat{L}_{-k}\right)\cdot
\exp\left(-H_{0,1}\right)\cdot Z_{H}(g,\beta;t_{k}),
\end{equation}
in which $a_{k}$ are constants irrelevant to $g$ or $\beta$. The explicit values of $a_{k}$ are determined by the following equation:
\begin{equation}\label{lyp41}
\exp\left(\sum\limits_{k=1}^{\infty}a_{k}z^{k+1}\dfrac{\partial}{\partial z}\right)\cdot z=\dfrac{z}{1+z}e^{-\frac{z}{1+z}}.
\end{equation}
\section{$\tau$-Functions}
In this section, we outline the expressions of the KP $\tau$-functions \cite{A.Alexandrov7,O.Babelon}.
In the first part, we will represent the KP $\tau$-function in terms of the fermionic correlators parameterised by a set of infinite continuous variables. From the fermionic representation, we can reexpress the $\tau$-function in a specific determinant form. In the second part, we will expand the $\tau$-function with the Schur polynomials, such that we get the explicit form in terms of time variables.

\subsection{$\tau$-Functions in Free Field Representation}

The free fermionic operators $\psi_{n},\psi^{\ast}_{n}, n\in\mathbb{Z}+1/2$, are subjected to the following anti-commutation relation:
\begin{equation}\label{4.1}
\{\psi_{m},\psi_{n}\}=\{\psi^{\ast}_{m},\psi^{\ast}_{n}\}=0,\ \ \ \{\psi_{m},\psi^{\ast}_{n}\}=\delta_{m+n,0}.
\end{equation}
Totally empty vacuum sates $|+\infty\rangle$ and $\langle+\infty|$ are determined by relations
\begin{equation}\label{4.2}
\begin{array}{llll}
\psi_{m}|+\infty\rangle=0,\ \ \ \ m\in\mathbb{Z}+1/2,
\end{array}
\end{equation}
and
\begin{equation}
\langle +\infty|\psi^{\ast}_{n}=0,\ \ \ \ n\in\mathbb{Z}+1/2,
\end{equation}
respectively. The shifted vacuum states $|n\rangle$ and $\langle n|$ are defined as
\begin{equation}\label{4.3}
|n\rangle=\psi^{\ast}_{n+1/2}\psi^{\ast}_{n+3/2}\cdots|+\infty\rangle,
\end{equation}
and
\begin{equation}\label{4.46}
\langle n|=\langle-\infty|\cdots\psi_{-n-3/2}\psi_{-n-1/2}.
\end{equation}
respectively. They satisfy the conditions:
\begin{equation}\label{4.4}
\begin{array}{lll}
\psi^{\ast}_{k}|n\rangle=0,\ \ \text{if}\ k>n;\ \ \ \ \psi_{k}|n\rangle=0,\ \ \text{if}\ k>-n;\\
\langle n|\psi^{\ast}_{k}=0,\ \ \text{if}\ k<n;\ \ \ \ \langle n|\psi_{k}=0,\ \ \text{if}\ k<-n.
\end{array}
\end{equation}
In fact, they can be viewed as definitions of such states.

It is convenient to introduce the free fermionic fields, such that
\begin{equation}\label{4.6}
\psi(z):=\sum\limits_{i\in\mathbb{Z}+1/2}\psi_{i}z^{-i-1/2},\ \ \ \
\psi^{\ast}(z):=\sum\limits_{i\in\mathbb{Z}+1/2}\psi^{\ast}_{i}z^{-i-1/2}.
\end{equation}
It is well known that, they can be expressed in terms of the chiral bosonic filed $\varphi(z)$
\begin{equation}\label{4.7}
\begin{array}{llll}
\varphi(z)=p-iq\log{z}+i\sum\limits_{k\in\mathbb{Z}}\frac{\varphi_{k}}{k}z^{-k},
\end{array}
\end{equation}
\begin{equation}\label{4.8}
[\varphi_{m},\varphi_{n}]=m\delta_{m+n,0},\ \ \ \ [p,q]=i.
\end{equation}
The operator $q$ is charge operator
\begin{equation}\label{4.10}
q|n\rangle=n|n\rangle,\ \ \ \ \langle n|q=n\langle n|
\end{equation}
while $p$ is the conjugate operator of $q$ such that $e^{ip}$ is the shift operator
\begin{equation}\label{4.11}
\begin{array}{lll}
e^{\pm ip}|n\rangle=|n\pm1\rangle,\ \ \ \ \langle n|e^{\pm ip}=\langle n\mp1|.
\end{array}
\end{equation}
The free fermions fields and the chiral bosonic field with the following formulas
\begin{equation}\label{4.9}
\begin{array}{llll}
\psi(z)=:e^{i\varphi(z)}:\equiv e^{ip}z^{q}e^{\varphi _{-}(z)}e^{-\varphi _{+}(z^{-1})},\\
\psi^{\ast}(z)=:e^{-i\varphi(z)}:\equiv e^{-ip}z^{-q}e^{-\varphi _{-}(z)}e^{\varphi _{+}(z^{-1})}.
\end{array}
\end{equation}
where ${\varphi} _{+}(z)=\sum\limits_{k\geq1}\frac{\varphi_{k}}{k}z^{k}$, and  ${\varphi} _{-}(z)=\sum\limits_{k\geq1}\frac{\varphi_{-k}}{k}z^{k}$. In above equation, :(): means the normal ordering for fermionic operators. The only deference with the normal ordering for bosonic operators in equation \dref{lyp37} is that the factor $(-1)$ will be taken into account as two fermionic operators exchanging their positions. For example
\begin{equation}\label{4.5}
:\psi_{m}\psi^{\ast}_{n}:=\left\{
\begin{array}{llll}
\psi_{m}\psi^{\ast}_{n}\ \ \ \ \text{if}\ m<0\\
-\psi^{\ast}_{n}\psi_{m}\ \ \ \ \text{if}\ m>0
\end{array}
\right.
\end{equation}
 Using the definition \dref{4.7}, one can show that
\begin{equation}\label{4.12}
:e^{i\alpha\varphi(z)}::e^{i\beta\varphi(w)}:=(z-w)^{\alpha\beta}:e^{i\alpha\varphi(z)+i\beta\varphi(w)}:.
\end{equation}
Their Operator Product Expansion (OPE) are
\begin{equation}\label{4.13}
\begin{array}{llll}
\psi(z)\psi(w)=(z-w):e^{i\varphi(z)+i\varphi(w)}:,\\
\psi^{\ast}(z)\psi^{\ast}(w)=(z-w):e^{-i\varphi(z)-i\varphi(w)}:,\\
\psi(z)\psi^{\ast}(w)=\frac{1}{z-w}:e^{i\varphi(z)-i\varphi(w)}:
=\frac{1}{z-w}+\cdots,
\end{array}
\end{equation}
respectively. On the other hand, the bosonic free field can be expressed as normal ordering of fermionic fields
\begin{equation}\label{4.14}
\begin{array}{lll}
J(z)=\sum\limits_{k\in\mathbb{Z}}J_{k}z^{-k-1}=i\partial_{z}\varphi(z)
\equiv:\psi(z)\psi^{\ast}(z):.
\end{array}
\end{equation}
Equivalently, the bosonic operators $J_{k}$ can be represented as bilinear combination of the fermionic modes:
\begin{equation}\label{4.15}
J_{k}=\sum\limits_{i\in\mathbb{Z}+1/2}:\psi_{i}\psi^{\ast}_{k-i}:.
\end{equation}
Obviously,
\begin{equation}\label{4.16}
\left\{\begin{array}{lll}
J_{k}|n\rangle=0,\ \ \ k>0\\
\langle n|J_{-k}=0,\ \ \ k>0\\
J_{0}|n\rangle=q|n\rangle=n|n\rangle
\end{array}
\right.
\end{equation}
One should mention that not only the bosonic currents can be represented as bilinear combination of the free fermions, actually, this is true for the Virasoro generators and $W^{(3)}$ generators. The Energy-Momentum tensor $\mathcal {L}(z)$ and $W^{(3)}$ field are defined as
\begin{equation}\label{4.56}
\mathcal {L}(z)\equiv\frac{1}{2}:J(z)J(z):=\sum\limits_{n\in\mathbb{Z}}\mathcal {L}_{n}z^{-k-2}
\end{equation}
and
\begin{equation}\label{4.57}
W^{(3)}(z)\equiv\frac{1}{3}:J(z)J(z)J(z):=\sum\limits_{k\in\mathbb{Z}}W^{(3)}_{k}z^{-k-3},
\end{equation}
respectively. We get the explicit form of these fields by OPE:
\begin{equation}\label{4.58}
\begin{array}{llllll}
J(z)J(w)&=&:\psi(z)\psi^{\ast}(z)::\psi(w)\psi^{\ast}(w):\\
&=&\dfrac{1}{(z-w)^{2}}+:\partial_{w}\psi(w)\psi^{\ast}(w):+:\partial_{w}\psi^{\ast}(w)\psi(w):+\cdots,
\end{array}
\end{equation}
therefore, the Energy-Momentum tensor is
\begin{equation}\label{4.59}
\begin{array}{llll}
\mathcal {L}(z)&=&\frac{1}{2}\left(:\partial_{z}\psi(z)\psi^{\ast}(z):+:\partial_{z}\psi^{\ast}(z)\psi(z):\right)\\
&=&\frac{1}{2}\sum\limits_{a+b=k,k\in\mathbb{Z}}:\psi_{a}\psi^{\ast}_{b}:z^{-k-2}.
\end{array}
\end{equation}
And
\begin{equation}\label{4.60}
\begin{array}{llllllll}
& &J(z):J(w)J(w):\\
&=&:\psi(z)\psi^{\ast}(z):\cdot\left(:\partial_{w}\psi(w)\psi^{\ast}(w):
+:\partial_{w}\psi^{\ast}(w)\psi(w):\right)\\
&=&:\psi(z)\psi^{\ast}(z)\partial_{w}\psi(w)\psi^{\ast}(w):+
:\psi(z)\psi^{\ast}(z)\partial_{w}\psi^{\ast}(w)\psi(w):\\
& &+\dfrac{:\psi(z)\psi^{\ast}(w):}{(z-w)^{2}}
-\dfrac{:\psi^{\ast}(z)\psi(w):}{(z-w)^{2}}+\dfrac{:\psi^{\ast}(z)\partial_{w}\psi(w):}{z-w}
-\dfrac{:\psi(z)\partial_{w}\psi^{\ast}(w):}{z-w},
\end{array}
\end{equation}
So, the $W^{(3)}$ field is
\begin{equation}\label{4.61}
\begin{array}{lll}
W^{(3)}(z)&=&\frac{1}{3}\left(:\partial_{z}\psi^{\ast}(z)\partial_{z}\psi(z):
+\frac{1}{2}:\partial^{2}_{z}\psi(z)\psi^{\ast}(z):\right.\\
&&\left.-:\partial_{z}\psi(z)\partial_{z}\psi^{\ast}(z):-\frac{1}{2}:\partial_{z}^{2}\psi^{\ast}(z)\psi(z):\right).
\end{array}
\end{equation}
In particular, $W^{(3)}_{0}$ can be expressed as
\begin{equation}\label{4.62}
W^{(3)}_{0}=\sum\limits_{r\in\mathbb{Z}+\frac{1}{2}}(r^{2}+\dfrac{1}{12}):\psi_{r}\psi_{-r}^{\ast}:.
\end{equation}

\subsection{Determinantal Representation of $\tau$-Function}
It is well known that, the KP $\tau$-function can be written as the following correlator \cite{A.Alexandrov7,E.Date}.
\begin{equation}\label{4.19}
\begin{array}{lll}
\tau(t)=\frac{\langle0|e^{H(t)}G|0\rangle}{\langle0|G|0\rangle};\\
G=\exp(\sum\limits_{i,j\in\mathbb{Z}+1/2}A_{i,j}:\psi_{i}\psi^{\ast}_{j}:).
\end{array}
\end{equation}
in which $H(t)=\sum\limits_{k=1}^{\infty}t_{k}J_{k}$ is the so called 'Hamiltonian', and $\{t_{k}\}$ is a set of infinite parameters. In order to get the determinantal representation of the $\tau$-function, let us calculate the fermionic correlator $\frac{\langle-N|\psi^{\ast}(\mu_{N})\cdots\psi^{\ast}(\mu_{1})G|0\rangle}{\langle0|G|0\rangle}$ in two ways. On one hand, using the Wick theorem, we can get
\begin{equation}\label{4.20}
\begin{array}{llll}
\frac{\langle-N|\psi^{\ast}(\mu_{N})\cdots\psi^{\ast}(\mu_{1})G|0\rangle}{\langle0|G|0\rangle}
&=&\frac{\langle0|\psi_{1/2}\cdots\psi_{N-1/2}\psi^{\ast}(\mu_{N})\cdots\psi^{\ast}(\mu_{1})G|0\rangle}
{\langle0|G|0\rangle}\\
&=&\det\left(\frac{\langle0|\psi_{i-1/2}\psi^{\ast}(\mu_{j})G|0\rangle}{\langle0|G|0\rangle}\right)_{1\leq i,j\leq N}.
\end{array}
\end{equation}
On the other hand, using the boson-fermion correspondence, we get
\begin{equation}\label{4.21}
\begin{array}{llllll}
\langle-N|\psi^{\ast}(\mu_{N})\cdots\psi^{\ast}(\mu_{1})G|0\rangle
&=&\langle-N|:e^{-i\varphi(\mu_{N})}:\cdots:e^{-i\varphi(\mu_{1})}:G|0\rangle\\
&=&\Delta(\mu)
\langle-N|:\exp\{-i\sum\limits_{j=1}^{N}\varphi(\mu_{j})\}:G|0\rangle\\
&=&\Delta(\mu)\cdot\langle0|\exp\{\sum\limits_{k=1}^{\infty}t_{k}J_{k}\}G|0\rangle,
\end{array}
\end{equation}
where $\Delta(\mu)$ is the VanderMonde determinant, i.e. $\Delta(\mu)=\prod_{i>j}(\mu_{i}-\mu_{j})$, and
\begin{equation}\label{4.22}
t_{k}=\frac{1}{k}\sum\limits_{j=1}^{N}\mu^{-k}_{j}.
\end{equation}
The parametrization \dref{4.22} has been introduced in \cite{T.Miwa1}, so such kind parametrisation is named as the Miwa parametrisation, and $\{\mu_{i}\}$ is called the Miwa variables. Please note that for $N$ is finite, only first $N$ variables, saying $t_{1},\cdots,t_{N}$ are functional independent. So in the following, we will consider the large $N$ case. From \dref{4.20} and \dref{4.21}, we get
\begin{equation}\label{4.23}
\tau(t)=\frac{\det(\phi^{(can)}_{i}(\mu_{j}))}{\Delta(\mu)},
\end{equation}
in which the canonical basis
\begin{equation}\label{4.24}
\begin{array}{lllll}
\phi^{(can)}_{i}(\mu)&=&\frac{\langle0|\psi_{i-1/2}\psi^{\ast}(\mu)G|0\rangle}{\langle0|G|0\rangle}\\
&=&\mu^{i-1}+\sum\limits_{k=0}^{\infty}b_{i,k}\mu^{-k-1},\ \ \ i=1,2,\cdots.
\end{array}
\end{equation}
In which the coefficients $b_{i,k}$ are
\begin{equation}\label{4.55}
b_{i,k}=\frac{\langle0|\psi_{i-1/2}\psi^{\ast}_{k+1/2}G|0\rangle}{\langle0|G|0\rangle}.
\end{equation}
Moreover, the converse statement is also true. Namely, that any function $\tau(t)$ in form \cite{S.Kharchev}
\begin{equation}\label{4.26}
\tau(t)=\frac{\det(\phi_{i}(\mu_{j}))}{\Delta(\mu)},\ \ \ t_{k}=\frac{1}{k}\sum\limits_{i=1}^{N}\mu_{i}^{-k},
\end{equation}
is a KP $\tau$-function, with vectors $\phi_{i}(\mu),(i=1,2,\cdots,N)$ basis have the asymptotic
\begin{equation}\label{4.27}
\phi_{i}(\mu)=\mu^{i-1}(1+O(\frac{1}{\mu})),\ \ \ \ \mu\rightarrow\infty
\end{equation}
From equation \dref{4.26}, it is easy to know that, there are freedoms to change the matrix form, but its determinant is intact, so the vector basis for a KP $\tau$-function is not unique. But the basis in the form as \dref{4.55} is unique. We call the such unique basis as canonical one and denote it as $\{\phi_{i}^{(can)}(\mu)\}$. So, the formula given by equation \dref{4.24} is the canonical basis for the $\tau$-function.

For every $G$ in the form \eqref{4.19}, there is always an element $\widetilde{G}$ in the form
\begin{equation}\label{4.51}
\widetilde{G}=\exp\{\sum\limits_{m,n\geq0}A_{m,n}:\psi_{-m-1/2}\psi^{\ast}_{-n-1/2}:\}
=\exp\{\sum\limits_{m,n\geq0}A_{m,n}\psi_{-m-1/2}\psi^{\ast}_{-n-1/2}\},
\end{equation}
such that
\begin{equation}\label{4.52}
G|0\rangle=\langle0|G|0\rangle\cdot\widetilde{G}|0\rangle.
\end{equation}
The coefficient $A_{m,n}$ in \eqref{4.52} can be got from the canonical basis,
\begin{equation}\label{5.56}
A_{n,m}=\frac{\langle0|\psi_{m+1/2}\psi^{\ast}_{n+1/2}G|0\rangle}{\langle0|G|0\rangle}=b_{m+1,n}
\end{equation}
$G$ may not be an element in certain group, but all the elements in the form of $\widetilde{G}$ form a group. For every KP $\tau$-function, there is an unique element in this group corresponding to it.
From the definition $|0\rangle$, $\widetilde{G}|0\rangle$ can be phrased as
\begin{equation}\label{4.53}
\begin{array}{lllll}
\widetilde{G}|0\rangle&=&\widetilde{G}\psi^{\ast}_{1/2}\psi^{\ast}_{3/2}\cdots\cdots|\infty\rangle\\
&=&(\widetilde{G}\psi^{\ast}_{1/2}\widetilde{G}^{-1})\cdot(\widetilde{G}\psi^{\ast}_{3/2}\widetilde{G}^{-1})
\cdots\cdots|\infty\rangle\\
&=&\widetilde{\psi}^{\ast}_{1/2}\widetilde{\psi}^{\ast}_{3/2}\cdots\cdots|\infty\rangle.
\end{array}
\end{equation}
In this equation,
\begin{equation}\label{4.54}
\begin{array}{lllll}
\widetilde{\psi}^{\ast}_{r+1/2}&=&\widetilde{G}\psi^{\ast}_{r+1/2}\widetilde{G}^{-1}\\
&=&\psi^{\ast}_{r+1/2}-\sum\limits_{n=0}^{\infty}A_{r,n}\psi^{\ast}_{-n-1/2}\\
&=&\psi^{\ast}_{r+1/2}-\sum\limits_{n=0}^{\infty}b_{n+1,r}\psi^{\ast}_{-n-1/2}.
\end{array}
\end{equation}
$\widetilde{G}|0\rangle$ is the fermionic representation of the $\tau$-function, and for simplicity, we write it as
\begin{equation}\label{4.79}
\tau(t)=\widetilde{G}|0\rangle.
\end{equation}

In the case $g=1$, let us denote the current operator $\hat{j}_{k}$ as $\hat{J}_{k}$, and the Virasoro operator $\hat{L}_{k}$ as $\mathcal{L}_{k}$, respectively. That is
\begin{equation}\label{4.49}
\hat{J}_{k}=\left\{\begin{array}{llll}
-kt_{-k},\ \ \ \ \ \ \ \text{for}\  k\leq0;\\
\frac{\partial}{\partial t_{k}}\ \ \ \ \ \ \ \ \text{for}\  k>0,
\end{array}
\right.
\end{equation}
and
\begin{equation}\label{4.31}
\hat{\mathcal {L}_{k}}=\frac{1}{2}\sum\limits_{a+b=k}abt_{-a}t_{-b}+\sum\limits_{a+b=k}at_{-a}\frac{\partial}{\partial t_{b}}+\frac{1}{2}\sum\limits_{a+b=k}\frac{\partial^{2}}{\partial t_{a}\partial t_{b}}.
\end{equation}
Then action of the current operator on the $\tau$-function can be expressed as
\begin{equation}\label{4.50}
\hat{J}_{k}\cdot\tau(t)=\frac{\langle0|e^{H(t)}J_{k}G|0\rangle}{\langle0|G|0\rangle}
\end{equation}
and the action of the Virasoro operator on it is
\begin{equation}\label{4.32}
\hat{\mathcal {L}}_{k}\cdot\tau(t)=\frac{\langle0|e^{H(t)}\mathcal {L}_{k}G|0\rangle}{\langle0|G|0\rangle}.
\end{equation}
The time derivation $\partial\tau/\partial t_{k}$ (i.e. $\hat{J}_{k}\cdot\tau$ for $k>0$) can be also rephrased in the determinant form \dref{4.23} \cite{S.Kharchev1}
\begin{equation}\label{4.28}
\frac{\partial\tau(t)}{\partial t_{k}}=-\frac{1}{\Delta(\mu)}\sum\limits_{m=1}^{N}(B^{k}(\mu_{m})-\mu^{k}_{m})
\det\left(\phi^{(can)}_{i}(\mu_{j})\right),
\end{equation}
in which $B(\mu)$ is a 'shift' operator acting on the vector space $\{\phi^{(can)}_{i}(\mu)\}$
\begin{equation}\label{4.29}
B(\mu)\cdot\phi^{(can)}_{i}(\mu)=\phi^{(can)}_{i+1}(\mu).
\end{equation}
In the case $N\rightarrow\infty$, it is easy to see that
\begin{equation}\label{4.30}
\sum\limits_{m=1}^{\infty}B^{k}(\mu_{m})\cdot\det\left(\phi^{(can)}_{i}(\mu_{j})\right)=0.
\end{equation}
So if the canonical basis $\{\phi_{i}^{(can)}(\mu)\}$ satisfies
\begin{equation}\label{4.48}
\mu^{r}\cdot\phi_{i}^{(can)}(\mu)\in\text{Span}\{\phi_{1}^{(can)}(\mu),\phi_{2}^{(can)}(\mu),\cdots\},
\end{equation}
for all $i>0$, all the derivative of $\tau(t)$ with $t_{rk}$ $(k>0)$ satisfies
\begin{equation}\label{4.63}
\frac{\partial\tau(t)}{\partial t_{kr}}=\text{Const}\cdot\tau(t).
\end{equation}
According to the definition of $r$-reduced KP hierarchies, we know that in this case the KP $\tau$-function is $r$-reduced.

When $N\rightarrow\infty$ action of the Virasoro operators $\hat{\mathcal {L}}_{k}$ on the function can be also reworded in the determinant form:
\begin{equation}\label{4.33}
\hat{\mathcal {L}}_{k}\cdot\tau(t)=-\frac{1}{\Delta(\mu)}\sum\limits_{m=1}^{N}A_{k}(\mu_{m})\det(\phi^{(can)}_{i}(\mu_{j})),
\end{equation}
and here the operator $A_{k}(\mu)$ is
\begin{equation}\label{4.34}
A_{k}(\mu)=\mu^{1+k}\frac{\partial}{\partial\mu}+\frac{1+k}{2}\mu^{k}.
\end{equation}

\subsection{Schur Polynomials Expansion}
At first, some notations should be introduced. A partition $\lambda=(\lambda_{1},\lambda_{2},\cdots\,\lambda_{l})$ is a sequence of positive integers $\lambda_{i}$ such that $\lambda_{1}\geq\lambda_{2}\geq\cdots\geq\lambda_{l}>0$. The numbers of nonzero $\lambda_{i}$ in $\lambda$, denoted by $l=l(\lambda)$, is called the length of the partition. A partition can be naturally graphed by a Young diagram. A Young diagram of $\lambda$ is a table whose $j$-th row (counting from the top) consists of $\lambda_{j}$ boxes (see Fig.1). We will identify a partition with a Young diagram in the following. The total number of boxes in the diagram $\lambda$ is $|\lambda|=\sum_{i=1}^{l}\lambda_{i}$, and the empty diagram is denoted by $\emptyset$. The conjugate of a partition $\lambda$ is the partition $\lambda^{\prime}$ whose diagram is the transpose Young diagram $\lambda$, i.e. $\lambda^{\prime}_{j}$ is the height of the $j$-th column of $\lambda$. We shall denote the set of partitions of $m$ by $\mathscr{P}(m)$, and the set of all partitions by $\mathscr{P}$.

For a given partition $\lambda=(\lambda_{1},\lambda_{2}\,\cdots,\lambda_{l})$ with $l=l(\lambda)$ nonzero rows, due to Frobenius, there is another notation for the diagram $\lambda$, saying  $(\vec{\alpha}|\vec{\beta})=(\alpha_{1},\cdots,\alpha_{d}|\beta_{1},\cdots,\beta_{d})$, here $d=d(\lambda)$, $d(\lambda)$ is the number of boxes in the diagonal of Young diagram $\lambda$, and $\alpha_{i}=\lambda_{i}-i$, $\beta_{j}=\lambda^{\prime}_{j}-j$ which are the arm-length of square $s=(i,i)$ and the leg-length of square $s=(j,j)$, respectively. Clearly, if $\lambda=(\vec{\alpha}|\vec{\beta})$, $\alpha_{1}>\alpha_{2}>\cdots>\alpha_{d}$ and $\beta_{1}>\beta_{2}>\cdots>\alpha_{d}$, and $\lambda^{\prime}=(\vec{\beta}|\vec{\alpha})$. Note that
\begin{equation}\label{4.43}
\sum\limits_{i=1}^{d}(\alpha_{i}+\beta_{i})+d=|\lambda|.
\end{equation}
Sometimes, it is convenient to use a notation which indicates the number of times each integer occurs as a part:
\begin{equation}\label{4.44}
\lambda=(1^{m_1}2^{m_2}\cdots r^{m_r}\cdots).
\end{equation}
\begin{equation}\label{4.45}
m_{i}=m_{i}(\lambda)=\text{Card}\{j:\lambda_{j}=i\}.
\end{equation}
The number $m_{i}$ is called the multiplicity of $i$ in $\lambda$.

\begin{figure} [htp] \centering
{\includegraphics[width=5.5cm]{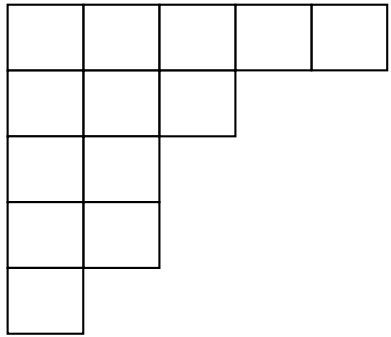}\label{Fig3.1}}
\caption{partition $\lambda=(5,3,2,2,1)=(4,1|4,2)$}
\end{figure}

If $\mathcal {B}=\mathbb{C}[[t_{1},t_{2},\cdots]]$ is the space of formal power series, there exists an additive basis of this apace. In order to introduce an additive basis for $\mathcal {B}$ which is indexed by partitions, we first need to define the elementary Schur polynomials $S_{k}(t)$. They are defined by the following generating function
\begin{equation}\label{4.35}
\sum\limits_{k\in\mathbb{Z}}S_{k}(t)z^{k}=\exp\left(\sum\limits_{k=0}^{\infty}t_{k}z^{k}\right).
\end{equation}
Thus
\begin{equation}\label{4.36}
\begin{array}{lllll}
S_{k}=0\ \ \text{for}\ k<0,\\
S_{0}=1,\\
S_{k}=\sum\limits_{k_{1}+2k_{2}+\cdots=k}\frac{t_{1}^{k_{1}}}{k_{1}!}\frac{t_{2}^{k_{2}}}{k_{2}!}\cdots \ \text{for} \ \ k>0.
\end{array}
\end{equation}
For a partition $\lambda$, we associate it a Schur polynomial $S_{\lambda}(t)$ defined by the following $k\times k$ determinant
\begin{equation}\label{4.37}
\begin{array}{lll}
S_{\lambda}(t)=\det\left(\begin{array}{llllll}
S_{\lambda_{1}}\  \ \ \ \ S_{\lambda_{1}+1}\ \ \ S_{\lambda_{1}+2} \cdots\cdots\\
S_{\lambda_{2}-1}\ \ \ \ S_{\lambda_{2}}\ \ \ \ S_{\lambda_{2}+1} \cdots\cdots\\
S_{\lambda_{3}-2}\ \ S_{\lambda_{3}-1}\ \ \ \ S_{\lambda_{3}} \cdots\cdots\\
\cdots  \cdots  \cdots  \cdots\cdots\cdots\cdots\cdots\cdots\\
\cdots  \cdots  \cdots  \cdots\cdots\cdots\cdots\cdots\cdots
\end{array}\right)\\
=\det\left(S_{\lambda_{i}+j-i}\right).
\end{array}
\end{equation}
Examples are:
\begin{equation}\label{4.38}
\begin{array}{llll}
S_{(1,1)}=\frac{t_{1}^{2}}{2}-t_{2},\\
S_{(2,1)}=\frac{t_{1}^{3}}{3}-t_{3},\\
S_{(2,2)}=\frac{t_{1}^{4}}{12}-t_{1}t_{3}+t_{2}^{2}.
\end{array}
\end{equation}
The set $\{S_{\lambda}(t)|\lambda\in\mathscr{P}\}$ is just the an additive basis of $\mathcal {B}=\mathbb{C}[[t_{1},t_{2},\cdots]]$. For more about the Schur polynomials, please refer to \cite{Macdonald}.

The Schur polynomials can be formulated in terms of fermionic operators, we define two basic states $|\lambda,n\rangle$ and $\langle\lambda,n|$ with respect to a partition $\lambda=(\lambda_{1},\cdots,\lambda_{l})=(\vec{\alpha}|\vec{\beta})$ \cite{A.Alexandrov7}
\begin{equation}\label{4.39}
\begin{array}{lllll}
|\lambda,n\rangle\equiv\psi^{\ast}_{n-\beta_{1}-1/2}\cdots\psi^{\ast}_{n-\beta_{d(\lambda)}-1/2}
\psi_{-n-\alpha_{d(\lambda)}-1/2}\cdots\psi_{-n-\alpha_{1}-1/2}|n\rangle,
\end{array}
\end{equation}
and
\begin{equation}\label{4.47}
\begin{array}{llll}
\langle\lambda,n|\equiv\langle n|\psi^{\ast}_{n+\alpha_{1}+1/2}\cdots\psi^{\ast}_{n+\alpha_{d(\lambda)}+1/2}
\psi_{-n+\beta_{d(\lambda)}+1/2}\cdots\psi_{-n+\beta_{1}+1/2},
\end{array}
\end{equation}
respectively. It is well known that the basic vectors \dref{4.39} and \dref{4.47} are orthonormal, i.e.
\begin{equation}\label{4.40}
\langle \lambda,n|\mu,m\rangle=\delta_{\lambda,\mu}\delta_{n,m}.
\end{equation}
For the given 'Hamiltonian' $H(t)$, we can expand $\langle n|e^{H(t)}$ in the basis vectors $\langle\lambda,n|$
\begin{equation}\label{4.41}
\langle n|e^{H(t)}=\sum\limits_{\lambda}(-1)^{b(\lambda)}S_{\lambda}(t)\langle\lambda,n|,
\end{equation}
here the summation runs over all Young diagrams as well as the empty one and $b(\lambda)=\sum_{i=1}^{d(\lambda)}(\beta_{i}+1)$. From \dref{4.19}, we can expand the KP $\tau$-function in Schur polynomials
\begin{equation}\label{4.42}
\tau(t)=\sum\limits_{\lambda}c_{\lambda}S_{\lambda}(t).
\end{equation}
In this equation, $c_{\lambda}=(-1)^{b(\lambda)}\langle\lambda,0|G|0\rangle$ is the Pl\"{u}cker coordinate.

\section{The $r$-Spin Intersection Numbers with GKMM}
The aim of this section is to calculate the $r$-spin intersection numbers. The partition function of the $r$-spin intersection numbers has matrix integral representations \cite{R.Dijkgraaf}. At first, we overview the generalized Kontsevich matrix model, whose partition function is a KP $\tau$-function. Then, we calculate explicitly the matrix model corresponding to the partition function of the $r$-spin intersection numbers.

\subsection{Generalised Kontsevich Matrix Model}
The main subject which we will investigate below is one-matrix integral depending on an external field $N\times N$ Hermitian matrix $M$ \cite{S.Kharchev,S.Kharchev1,S.Kharchev2}:
\begin{equation}\label{5.1}
Z_{N}^{\{V\}}[M]=\frac{\int dYe^{-S(M,Y)}}{\int dYe^{-S_{2}(M,Y)}},
\end{equation}
where the measure $dY$ is the Haar measure of Hermitian matrix space
\begin{equation}\label{5.2}
dY=\prod\limits_{i=1}^{N}dY_{ii}\prod\limits_{i<j}dReY_{ij}dImY_{ij}.
\end{equation}
For any Taylor series of $V(Y)$, we set, by definition,
\begin{equation}\label{5.3}
S(M,Y)=\text{Tr}[V(M+Y)-V(M)-V^{\prime}(M)Y],
\end{equation}
and $S_{2}(M,Y)$ is the quadratic part of $S(M,Y)$,
\begin{equation}\label{5.4}
S_{2}(M,Y)=\lim_{\epsilon\rightarrow0}\frac{1}{\epsilon^{2}}S(M,\epsilon Y).
\end{equation}
It is clear that the partition function (\ref{5.1}) depends only on the eigenvalues $\mu_{1},\mu_{2},\cdots,\mu_{N}$ of the external field $M$. The partition function defined by equation \dref{5.1} is often called as a Generalised Kontsevich Matrix Model (GKMM), while for $V(Y)=\frac{Y^{3}}{3}$, it is exactly the original matrix model defined by Kontsevich \cite{M.Kontsevich}.

In order to investigate the GKMM \dref{5.1}, we first study the following matrix integral
\begin{equation}\label{5.5}
\mathcal {F}_{N}^{\{V\}}[\Lambda]=\int dXe^{-\text{Tr}[V(X)-\Lambda X]},
\end{equation}
in which the potential $V(X)$ is a polynomial, $\Lambda$ is a Hermitian matrix and it can be diagonalised by unitary matrix $U_{0}$, i.e. $U_{0}^{-1}\Lambda U_{0}=\text{diag}\{\lambda_1,\cdots,\lambda_{N}\}$.  Because the Haar measure \dref{5.2} is invariant under the action of unitary group $U(N)$, we get
\begin{equation}\label{5.18}
\begin{array}{lllll}
\mathcal {F}_{N}^{\{V\}}[\Lambda]&=&\int dXe^{-\text{Tr}[V(X)-U_{0}\cdot\text{diag}\{\lambda_{1},\cdots,\lambda_{N}\}\cdot U_{0}^{-1}X]}\\
&=&\int dXe^{-\text{Tr}[V(X)-\text{diag}\{\lambda_{1},\cdots,\lambda_{N}\}\cdot X]}\\
&=&\mathcal {F}_{N}^{\{V\}}[\text{diag}\{\lambda_{1},\cdots,\lambda_{N}\}].
\end{array}
\end{equation}
Therefore, it is convenient to take the external field as a diagonal matrix, i.e. $\Lambda=\text{diag}\{\lambda_{1},\cdots,\lambda_{N}\}$. For a Hermitian matrix $X$, there is a unitary matrix $U$, such that
\begin{equation}\label{5.19}
\begin{array}{lllll}
X&=&U\cdot\text{diag}\{x_{1},x_{2},\cdots,x_{N}\}\cdot U^{-1}\\
&=&U\hat{X} U^{-1}.
\end{array}
\end{equation}
In the second identity, we have used $\hat{X}$ to represent the diagonal form $\text{diag}\{x_{1},x_{2},\cdots,x_{N}\}$. So, the matrix integral \dref{5.5} can be written as
\begin{equation}\label{5.6}
\begin{array}{lllll}
\mathcal {F}_{N}^{\{V\}}&=&\int\prod_{i=1}^{N}dx_{i}\int[dU]\Delta^{2}(X)e^{-\text{Tr}[V(\hat{X})-\Lambda U\hat{X}U^{\dag}]}\\
&=&\frac{N!V_{N}}{\Delta(\Lambda)}\prod\limits_{i=1}^{N}\int dx_{i}e^{-V(x_{i})+\lambda_{i}x_{i}}\Delta(X)\\
&\sim&\frac{\det(F_{i}(\lambda_{j}))_{1\leq i,j\leq N}}{\Delta(\Lambda)},
\end{array}
\end{equation}
where $[dU]$ is the Haar measure of the group $U(N)$ and $V_{N}$ is its volume. In the above performing, the well-known Itzyskon-Zuber fomula
\begin{equation}\label{5.7}
\int [dU]e^{\text{Tr}\Lambda X}=V_{N}\frac{\det(e^{\lambda_{i}x_{j}})_{1\leq i,j\leq N}}{\Delta(X)\Delta(\Lambda)},
\end{equation}
is used, and the functions $F_{i}(\lambda)$ in this equation is
\begin{equation}\label{5.8}
\begin{array}{llll}
F_{i}(\lambda)&=&\int dxx^{i-1}e^{-V(x)+\lambda x}\\
&=&\left(\frac{\partial}{\partial\lambda}\right)^{i-1}F_{1}(\lambda).
\end{array}
\end{equation}
The goal in the following is to get a determinantal representation of the GKMM \dref{5.1}. We will deal with the numerator and denominator in two different ways, for convenience, we denote them as $N_{K}$ and $D_{K}$, respectively. So,
\begin{equation}
N_{K}=\int dYe^{-\text{Tr}[V(M+Y)-V(M)-V^{\prime}(M)Y]}
\end{equation}
and
\begin{equation}
D_{K}=\int dYe^{-S_{2}(M,Y)}
\end{equation}

We first deal with the numerator. After shifting the integration variable
\begin{equation}\label{5.9}
X=Y+M,
\end{equation}
the numerator can be expressed as
\begin{equation}\label{5.10}
\begin{array}{lllll}
N_{K}&=&e^{\text{Tr}[V(M)-MV^{\prime}(M)]}\cdot\mathcal {F}_{N}^{\{V\}}[V^{\prime}(M)]\\
&=&\prod\limits_{i=1}^{N}e^{V(\mu_{i})-\mu_{i}V^{\prime}(\mu_{i})}
\frac{\det(\widetilde{\Phi}^{V}_{i}(\mu_{j}))_{1\leq i,j\leq N}}{\Delta(V^{\prime}(M))},
\end{array}
\end{equation}
in which
\begin{equation}\label{5.11}
\widetilde{\Phi}_{i}^{\{V\}}(\mu)=F_{i}(V^{\prime}(\mu)).
\end{equation}

Then, let us convey the denominator $D_{K}$. If the potential can be represented as a formal series
\begin{equation}\label{5.12}
V(X)=\sum\limits_{n=1}^{\infty}\frac{V_{n}X^{n}}{n}.
\end{equation}
and it is supposed to be analytic in $X$ at $X=0$, the equation (\ref{5.4}) implies that,
\begin{equation}\label{5.13}
S_{2}(M,Y)=\frac{1}{2}\sum\limits_{n=2}^{\infty}V_{n}\sum\limits_{a+b=n-2}M^{a}YM^{b}Y,
\end{equation}
then the denominator can be identical with
\begin{equation}\label{5.14}
D_{K}\sim\frac{\Delta(M)}{\Delta(V^{\prime}(M))}\prod\limits_{i=1}^{N}
\left[V^{\prime\prime}(\mu_{i})\right]^{-\frac{1}{2}}.
\end{equation}
Therefore, we get the determinantal representation of \dref{5.1}
\begin{equation}\label{5.15}
\begin{array}{llll}
Z_{N}^{\{V\}}[M]\sim\frac{\prod\limits_{k=1}^{N}s(\mu_{k})
\det(\widetilde{\Phi}_{i}^{\{V\}}(\mu_{j}))}{\Delta(M)},
\end{array}
\end{equation}
in which
\begin{equation}\label{5.20}
s(\mu)=\left[V^{\prime\prime}(\mu)\right]^{\frac{1}{2}}e^{V(\mu)-\mu V^{\prime}(\mu)},
\end{equation}
and $s(\mu)\widetilde{\Phi}_{i}^{\{V\}}(\mu)$ has the following asymptotic
\begin{equation}\label{5.21}
\begin{array}{lllll}
s(\mu)\widetilde{\Phi}_{i}^{\{V\}}(\mu)\sim\mu^{i-1}\left(1+O(\frac{1}{\mu})\right).
\end{array}
\end{equation}
as $\mu\rightarrow\infty$. If we set
\begin{equation}\label{5.16}
\begin{array}{llll}
\Phi_{i}^{\{V\}}(\mu)&=&\frac{1}{\sqrt{2\pi}}s(\mu)\widetilde{\Phi}_{i}^{\{V\}}(\mu)\\
&=&\frac{1}{\sqrt{2\pi}}\left[V^{\prime\prime}(\mu)\right]^{\frac{1}{2}}e^{V(\mu)-\mu V^{\prime}(\mu)}\Phi_{i}^{\{V\}}(\mu)\\
&=&A^{\{V\}}(\mu)\Phi_{i-1}^{\{V\}}(\mu)
\end{array}
\end{equation}
where $A^{\{V\}}(\mu)$ is a first-order differential operator in special form
\begin{equation}\label{5.17}
\begin{array}{llll}
A^{\{V\}}(\mu)&=&\left[V^{\prime\prime}(\mu)\right]^{\frac{1}{2}}e^{V(\mu)-\mu V^{\prime}(\mu)}\frac{\partial}{\partial(V^{\prime}(\mu))}\left[V^{\prime\prime}(\mu)\right]^{-\frac{1}{2}}
e^{-V(\mu)+\mu V^{\prime}(\mu)}\\
&=&\frac{1}{V^{\prime\prime}(\mu)}\frac{\partial}{\partial\mu}+\mu-
\frac{V^{\prime\prime\prime}(\mu)}{2[V^{\prime\prime}(\mu)]^{2}}
\end{array}
\end{equation}
The operator $A^{\{V\}}$ is viewed as a Kac-Schwarz operator, this kind of operator in $V(x)=\frac{x^{3}}{3}$ case was obtained in \cite{V.Kac}. In the next section, we will take a specific potential $V(Y)$, such that \dref{5.1} satisfies \dref{2.23}.

It is remarkable to use another parametrisation of the partition function $Z_{N}^{\{V\}}$ by treating it as a function of the time variables $t_{k}$ defined by \dref{4.22}. From \dref{4.26} and \dref{4.27}, we know that the partition function of the GKMM \dref{5.1} is a KP $\tau$-function in the time variables $\{t_{k}\}$

\subsection{Calculate $r$-spin Intersection Numbers with GKMM}
In section 4.1, we consider the GKMM with very general potential. If we restrict the potential function to a concrete form, more details could be to carry out.

If the potential function in equation \dref{5.1} is chosen as
\begin{equation}\label{6.1}
V(Y)=\sqrt{-r}\frac{Y^{r+1}}{r+1}
\end{equation}
From now on, we denote the partition function $Z_{N}^{\{V\}}[M]$ in \dref{5.1} by $Z_{N}^{\{r\}}[M]$, meanwhile $\Phi^{\{V\}}_{i}(\mu)$ by $\Phi_{i}(\mu)$. From the definition of $\Phi_{i}(\mu)$ and equation \dref{5.21}, it is easy to know
\begin{equation}\label{6.2}
\Phi_{i}(\mu)=\mu^{i-1}(1+O(\frac{1}{\mu})),\ \ \ \ \mu\rightarrow\infty,
\end{equation}
and
\begin{equation}\label{6.3}
\begin{array}{lllll}
\mu^{r}\cdot\Phi_{i}(\mu)&=&\frac{1}{\sqrt{2\pi}}\mu^{r}\cdot s(\mu) \int dxx^{i-1}e^{-\sqrt{-r}\frac{x^{r+1}}{r+1}+\sqrt{-r}\mu^{r}x}\\
&\in&\text{Span}\{\Phi_{1}(\mu),\Phi_{2}(\mu),\cdots\}.
\end{array}
\end{equation}
The canonical basis $\Phi_{i}^{(can)}(\mu)$ is a linear combination of $\Phi_{j}(\mu)$'s, so it is also satisfies \dref{6.3}, i.e.
\begin{equation}\label{6.24}
\mu^{r}\cdot\Phi_{i}^{(can)}(\mu)\in\text{Span}\{\Phi_{1}^{can}(\mu),\Phi_{2}^{can}(\mu),\cdots\}.
\end{equation}
According to \dref{4.48}, as $N\rightarrow\infty$, the partition function $Z_{\infty}^{\{r\}}[M]$ of the GKMM with the potential \dref{6.1} $Z_{\infty}^{\{r\}}[M]$ is a $r$-reduced KP $\tau$-function, that is to say
\begin{equation}\label{4.66}
\frac{\partial Z^{\{r\}}_{\infty}[M]}{\partial t_{kr}}=\text{Const}\cdot Z^{\{r\}}_{\infty}[M].
\end{equation}
For potential \dref{6.1}, the recursive relation \dref{5.17} for $\Phi_{i}(\mu)$ becomes explicitly
\begin{equation}\label{6.4}
\Phi_{i+1}(\mu)=\{\frac{1}{\sqrt{-r}r\mu^{r-1}}\frac{\partial}{\partial\mu}+\mu
-\frac{r-1}{2r\sqrt{-r}\mu^{r}}\}\cdot\Phi_{i}(\mu).
\end{equation}
With the same reason as \dref{6.24}, there is the constraint for the canonical basis $\Phi_{i}^{(can)}(\mu)$
\begin{equation}\label{6.15}
\{\frac{1}{r\mu^{r-1}}\frac{\partial}{\partial\mu}+\sqrt{-r}\mu
-\frac{r-1}{2r\mu^{r}}\}\cdot\Phi^{(can)}_{i}(\mu)\in \text{Span}\{\Phi^{(can)}_{1}(\mu),\Phi^{(can)}_{2}(\mu),\cdots\}.
\end{equation}
From equations \dref{4.29} and \dref{4.33}, equation \dref{6.15} is equivalent to a constraint on $Z_{\infty}^{\{r\}}[M]$
\begin{equation}\label{6.19}
\{-\sqrt{-r}\frac{\partial}{\partial t_{1}}+\frac{1}{r}\hat{\mathcal {L}}_{-r}\}\cdot Z^{\{r\}}_{\infty}[M]=\text{Const}\cdot Z^{\{r\}}_{\infty}[M].
\end{equation}
The operator $-\sqrt{-r}\frac{\partial}{\partial t_{1}}+\frac{1}{r}\hat{\mathcal {L}}_{-r}$ is exactly the operator $L^{\{r\}}_{-1}$ defined in equation \dref{2.13}. Equations \eqref{4.66} and \eqref{6.15} are two additional constraints for $Z_{\infty}^{\{r\}}[M]$, more precisely the constants on the right hand of these constraints are identical to zero. In the following, we will prove this fact. For simplicity, we denote
\begin{equation}\label{4.64}
a_{W}^{\{r\}}=\{\frac{1}{r\mu^{r-1}}\frac{\partial}{\partial\mu}+\sqrt{-r}\mu
-\frac{r-1}{2r\mu^{r}}\},
\end{equation}
and
\begin{equation}\label{4.65}
b_{W}^{\{r\}}=\mu^{r},
\end{equation}
respectively. In fact, both $a_{W}^{\{r\}}$ and $b_{W}^{\{r\}}$ are Kac-Schwarz operators in the Grassmannian notation \cite{A.Alexandrov8}. From the operators $a_{W}^{\{r\}}$ and $b_{W}^{\{r\}}$, we construct operators $l_{k}$ for $k\geq-1$
\begin{equation}\label{4.67}
l_{k}=\frac{1}{2}\{(b_{W}^{\{r\}})^{k+1},a_{W}^{\{r\}}\}=\frac{1}{r}\mu^{kr+1}\frac{\partial}{\partial\mu}
+\frac{1+kr}{2r}\mu^{kr}
+\sqrt{-r}\mu^{(k+1)r+1},
\end{equation}
where $\{ , \}$ is the anticommutator for differential operators.
From the definition of $l_{k}$ and equations \dref{6.24} and \dref{6.15}, it is obvious that
\begin{equation}\label{4.68}
l_{k}\cdot\Phi^{(can)}_{i}(\mu)\in\text{Span}\{\Phi_{1}^{(can)}(\mu),\Phi_{2}^{(can)}(\mu),\cdots\}.
\end{equation}
It is equivalent to
\begin{equation}\label{4.69}
(\sqrt{-r}\frac{\partial}{\partial t_{(k+1)r+1}}-\frac{1}{r}\hat{\mathcal {L}}_{kr})\cdot Z_{\infty}^{\{r\}}[M]
=\text{Const}\cdot Z_{\infty}^{\{r\}}[M].
\end{equation}
In order to investigate the relations of operators $\sqrt{-r}\frac{\partial}{\partial t_{(k+1)r+1}}-\frac{1}{r}\hat{\mathcal {L}}_{kr}$, we denote\\
$L^{\{r\}}_{k}=-\sqrt{-r}\frac{\partial}{\partial t_{(k+1)r+1}}+\frac{1}{r}\hat{\mathcal {L}}_{kr}+\delta_{k,0}\frac{(r+1)(r-1)}{24r}$ for simplicity. The operators $L^{\{r\}}_{k},\ k\geq-1$ and $\hat{J}_{nr},\ n\geq1$ are subject to the following relations
\begin{equation}\label{4.70}
\begin{array}{lllll}
[\hat{J}_{mr},\hat{J}_{nr}]=0;
\end{array}
\end{equation}
\begin{equation}\label{4.71}
[L^{\{r\}}_{k},\hat{J}_{mr}]=-m\hat{J}_{(m+k)r};
\end{equation}
and
\begin{equation}\label{4.72}
[L^{\{r\}}_{m},L^{\{r\}}_{n}]=(m-n)L^{\{r\}}_{m+n}.
\end{equation}
As each operator can be expressed as the commutator of other two operators, we get the following constraints for partition function $Z_{\infty}^{\{r\}}[M]$
\begin{equation}\label{4.73}
\hat{J}_{kr}\cdot Z_{\infty}^{\{r\}}[M]=\frac{\partial}{\partial t_{kr}}Z_{\infty}^{\{r\}}[M]=0,\ \ \text{for}\
k\geq1,
\end{equation}
and
\begin{equation}\label{4.74}
L^{\{r\}}_{k}\cdot Z_{\infty}^{\{r\}}[M]=-(\sqrt{-r}\frac{\partial}{\partial t_{(k+1)r+1}}-\frac{1}{r}\hat{\mathcal {L}}_{kr})\cdot Z_{\infty}^{\{r\}}[M]=0,\ \ \text{for}\ k\geq-1.
\end{equation}
The fact that $Z^{\{r\}}_{\infty}[M]$ is a $r$-reduced KP $\tau$-function, and with \dref{6.19} determines that the partition function $Z^{\{r\}}_{\infty}[M]$ satisfies a formal vacuum condition of $W_{1+\infty}$ algebra. Furthermore, it satisfies the vacuum condition of $W_{r}$ algebra \cite{M.Fukuma,J.Goeree}. The $W_{r}$-constraints can uniquely determine the partition function up to a constant factor \cite{Liu}. So, up to a constant, the partition function defined in \dref{2.9} and $Z^{\{r\}}_{\infty}[M]$ here are the same matter. More specifically, as the constant terms in the expansions of $Z(t)$ and $Z^{\{r\}}_{\infty}[M]$ both are one, so they are the same one, i.e.
\begin{equation}\label{6.20}
Z(t)= Z^{\{r\}}_{\infty}[M]
\end{equation}

Our next aim is to calculate the expression of $\Phi_{i}(\mu)$ and carry out $\Phi_{j}^{(can)}(\mu)$, and the fermionic representation of $Z_{\infty}^{\{r\}}[M]$. From the definition of $\Phi_{i}(\mu)$, it is easy to see that all $\Phi_{i}(\mu)$ can be written in the following form
\begin{equation}\label{6.5}
\Phi_{i}(\mu)=\mu^{i-1}+\sum\limits_{j=1}^{\infty}a^{\{r\}}_{i,j}\mu^{i-j-1},\ \ \ i=1,2,\cdots,
\end{equation}
the coefficients $a^{\{r\}}_{i,j}$ can be obtained directly from $\dref{5.16}$ by performing the integral. In the $r=2$ case, these coefficients have obtained in the way \cite{A.Alexandrov8}. Before giving the explicit form of $a_{i,j}$, we introduce some notations. For $m>0$, we define
\begin{equation}\label{6.7}
\mathscr{P}^{r+1}(m)=\{\lambda\in\mathscr{P}(m)|3\leq\lambda_{i}\leq r+1,\ \ \text{for}\ 1\leq i\leq l(\lambda)\},
\end{equation}
and
\begin{equation}\label{6.23}
\mathscr{P}^{r+1}(0)=\{\emptyset\}.
\end{equation}
At first, we calculate $\Phi_{1}(\mu)$
\begin{equation}\label{6.6}
\begin{array}{llll}
\Phi_{1}(\mu)&=&\frac{1}{\sqrt{2\pi}}(\sqrt{-r}r\mu^{r-1})^{\frac{1}{2}}e^{-\frac{\sqrt{-r}r}{r+1}\mu^{r+1}}
\int dxe^{-\frac{x^{r+1}}{r+1}+\sqrt{-r}\mu^{r}x}\\
&=&\frac{1}{\sqrt{2\pi}}(\sqrt{-r}\mu^{r-1})^{\frac{1}{2}}\int dx\exp\{-\frac{\sqrt{-r}r}{r+1}\sum\limits_{k=2}^{r+1}C_{r+1}^{k}x^{k}\mu^{r+1-k}\}.
\end{array}
\end{equation}
By derivation the integral \dref{6.6}, we get the explicit form of $\{a^{\{r\}}_{1,j}\}$
\begin{equation}\label{6.8}
\begin{array}{lll}
a^{\{r\}}_{1,j}=0,\ \ \ \text{if}\ j\neq0(\text{mod}(r+1)),\\
a^{\{r\}}_{1,k(r+1)}=\sum\limits_{m=k}^{3k}\sum\limits_{\lambda\in{\it Par}^{r+1}(2m)}\left(\prod\limits_{l=1}^{l(\lambda)}C_{r+1}^{\lambda_{l}}\right)
\frac{(-1)^{l(\lambda)}(\sqrt{-r})^{l(\lambda)-m}(2m-1)!!}{(r+1)^{l(\lambda)}r^{m}
\prod_{i=1}^{\infty}m_{i}(\lambda)!}
\delta_{l(\lambda)-m+k,0}.
\end{array}
\end{equation}
For other $a^{\{r\}}_{i,j},(i>1)$, it can be calculated by recursion derived from \dref{6.4}
\begin{equation}\label{6.9}
\left\{
\begin{array}{lllll}
a^{\{r\}}_{i+1,j}=0,\ \ \ \ \text{if}\ j\neq0(\text{mod}(r+1));\\
a^{\{r\}}_{i+1,r+1}=a^{\{r\}}_{i,r+1}+\frac{2i-1-r}{2r\sqrt{-r}};\\
a^{\{r\}}_{i+1,(k+1)(r+1)}=a^{\{r\}}_{i,(k+1)(r+1)}-\frac{(2k+1)r+(2k-2i+1)}{2r\sqrt{-r}}a^{\{r\}}_{i,k(r+1)}.
\end{array}
\right.
\end{equation}
In the next, we proceed to get the canonical basis $\Phi^{(can)}_{i}(\mu)$ which is a linear combination of $\Phi_{i}(\mu)$'s. Recall that in the expression of $\Phi_{i}^{(can)}(\mu)$, there is only one term $\mu$ with non-negative power, i.e.
\begin{equation}\label{6.10}
\begin{array}{llll}
\Phi^{(can)}_{i}(\mu)=\mu^{i-1}+O(\frac{1}{\mu}),\ \ \ \mu\rightarrow\infty,\\
\det(\Phi_{i}(\mu_{j}))=\det(\Phi^{(can)}_{i}(\mu_{j})).
\end{array}
\end{equation}
Obviously, the identities between $\Phi_{i}(\mu)$ and $\Phi_{i}^{(can)}(\mu)$ are
\begin{equation}\label{6.11}
\begin{array}{lll}
\Phi^{(can)}_{1}(\mu)=\Phi_{1}(\mu),\\
\Phi^{(can)}_{i}(\mu)=\Phi_{i}(\mu)-\sum\limits_{k=1}^{i-1}a^{\{r\}}_{i,k}\cdot\Phi^{(can)}_{i-k}(\mu).
\end{array}
\end{equation}
Then, the explicit form of $\Phi^{(can)}_{i}$ is
\begin{equation}\label{6.12}
\Phi^{(can)}_{i}(\mu)=\mu^{i-1}+\sum\limits_{j=1}^{\infty}b^{\{r\}}_{i,j}\cdot\mu^{-j},
\end{equation}
in which
\begin{equation}\label{6.13}
b^{\{r\}}_{i,j}=a^{\{r\}}_{i,i+j-1}-\sum\limits_{k=1}^{i-1}
a^{\{r\}}_{i,k}\cdot b^{\{r\}}_{i-k,j}.
\end{equation}
From the explicit expression of $a^{\{r\}}_{i,j}$, it is easy to known that $b^{\{r\}}_{i,j}\neq0$ if and only if $i+j=1(\text{mod}(r+1))$.

Reversing the progress what we have done in section 3, we will construct the fermionic representation of this GKMM from its determinantal representation. As accounted in section 3, there is a unique $\widetilde{G}^{\{r\}}$ in the form
\begin{equation}\label{4.75}
\widetilde{G}^{\{r\}}=\exp\{\sum\limits_{m,n\geq0}A^{\{r\}}_{m,n}\psi_{-m-1/2}\psi^{\ast}_{-n-1/2}\},
\end{equation}
such that
\begin{equation}\label{4.76}
Z^{\{r\}}(t)=Z^{\{r\}}_{\infty}[M]=\langle0|e^{H(t)}\widetilde{G}^{\{r\}}|0\rangle.
\end{equation}
From equations \eqref{4.55} and \eqref{6.12},The coefficient $A^{\{r\}}_{m,n}$ is
\begin{equation}\label{4.77}
A^{\{r\}}_{m,n}=b^{\{r\}}_{n+1,m+1}.
\end{equation}
In the next subsection, we will expand the partition function $Z^{\{r\}}_{\infty}[M]$ in a additive basis of $\mathcal {B}=\mathbb{C}[[t_{1},t_{2},\cdots]]$, and we get the coefficients before the monomials of $t_{k}$ that is we really need.

\subsection{$Z^{\{r\}}_{\infty}[M]$ in terms of Schur polynomials}
We have known that, the GKMM partition function with potential $V(Y)=\sqrt{-r}\frac{Y^{r+1}}{r+1}$ is a $r$-reduced KP
$\tau$-functiion. According to equation \dref{4.42}
\begin{equation}\label{6.21}
\begin{array}{lllllll}
Z^{\{r\}}_{\infty}[M]&=&\sum\limits_{\lambda}c_{\lambda}\cdot S_{\lambda}(t)\\
\end{array}
\end{equation}
here the pl\"{u}cker coordinates $c_{\lambda}$ are
\begin{equation}\label{6.22}
\begin{array}{llll}
c_{\lambda}&=&(-1)^{b(\lambda)}\langle\lambda,0|G_{0}|0\rangle\\
&=&(-1)^{b(\lambda)+d(\lambda)}\cdot\det(b_{\beta_{i}+1,\alpha_{j}+1})_{1\leq i,j\leq d(\lambda)},
\end{array}
\end{equation}
and as usual, $(\vec{\alpha}|\vec{\beta})$ is the Frobenius notation for a partition $\lambda$. After simple algebra, we find that $c_{\lambda}\neq0$ if and only if $|\lambda|=0\ (\text{mod}\ (r+1))$, that is,
\begin{equation}\label{4.78}
\begin{array}{llllll}
Z^{\{r\}}_{\infty}[M]&=&1+\sum\limits_{k=1}^{\infty}\sum\limits_{|\lambda|=k(r+1)}c_{\lambda}S_{\lambda}(t)\\
&=&1+b^{\{r\}}_{1,r+1}\cdot S_{(r+1)}(t)-b^{\{r\}}_{2,r}\cdot S_{(r,1)}(t)+(-1)^{r+1}b^{\{r\}}_{r+1,1}\cdot S_{(1^{r+1})}(t)+\cdots\\
&&+b^{\{r\}}_{1,2r+2}\cdot S_{(2r+2)}(t)-b^{\{r\}}_{2,2r+1}\cdot S_{(2r+1,1)}(t)+(-1)^{2r+2}b^{\{r\}}_{2r+2,1}
\cdot S_{(1^{2r+2})}(t)+\cdots\\
&&+\cdots\cdots\cdots.
\end{array}
\end{equation}

In equation \dref{6.21}, we have expanded the partition function $Z_{\infty}^{\{r\}}$ in terms of the Schur functions $\{S_{\lambda}(t)|\lambda\in\mathscr{P}\}$. By definition of the Schur polynomials, we can express this
partition function in the basis $\{\prod_{k=0}^{\infty}t_{k}^{m_{k}}|m_{k}\geq0 ,\  \text{and only finite}\  m_{k} \ \text{are non-zero}\}$. By the transformation of variables \dref{2.20},  the partition function $Z_{\infty}^{\{r\}}[M]$ can be expanded as a Taylor series in the variable $\{t_{m,a}|m\geq0; 0\leq a\leq r-2\}$, and from these expressions, we can obtain the partition function of $r$-spin intersection number. After taking the logarithm of $Z_{\infty}^{\{r\}}[M]$ and resplacing the parameter $\lambda$ by $t_{m,a}\rightarrow\lambda^{\frac{r(m-1)}{(r+1)}+\frac{a}{r+1}}t_{m,a}$, we can get the genus expansion of the free energy which is the generating function of the $r$-spin intersection numbers. Unlike the recursive method given by K.Liu, R.Vakil and H.Xu \cite{K.Liu}, we completely solve a generating function of $r$-spin intersection numbers. From this generating function, we can read all the $r$-spin intersection numbers directly. We will give some examples in more details in the Appendix A.

\section{The Partition Functions $Z^{\{r\}}(t)$ and $Z_{Hodge}(t)$}
\subsection{Fermionic Representations of $Z_{H}(t)$ and $Z_{Hodge}(t)$}
As mentioned in equation \dref{lyp21}, the Hurwitz partition function can be represented in terms of the cut-and-join operator $\hat{W}_{0}$ (i.e.\dref{lyp20}). From the fermionic representation of $\hat{W}^{(3)}_{0}$ , i.e. $W^{(3)}_{0}$ \eqref{4.61}, the Hurwitz partition function can be represented as
\begin{equation}\label{lyp84}
Z_{H}=\frac{\exp(\dfrac{\beta}{2}\sum\limits_{r\in\mathbb{Z}+\frac{1}{2}}(r^{2}
+\frac{1}{12}):\psi_{r}\psi^{\ast}_{-r}:)\exp(\sum\limits_{r+s=-1}
:\psi_{r}\psi^{\ast}_{s}:)|0\rangle}{\langle0|\exp(\dfrac{\beta}{2}\sum\limits_{r\in\mathbb{Z}+\frac{1}{2}}(r^{2}
+\frac{1}{12}):\psi_{r}\psi^{\ast}_{-r}:)\exp(\sum\limits_{r+s=-1}
:\psi_{r}\psi^{\ast}_{s}:)|0\rangle}.
\end{equation}
It has be proved that the Hurwitz partition function is a KP $\tau$-function. For the operators $\hat{L}_{k}$ are generators of $\widehat{gl}(\infty)$, the Hodge partition function is also a KP $\tau$-function. Expanding \eqref{lyp84}, we get the fermionic representation of the Hurwitz partition function
\begin{equation}\label{li11}
\begin{array}{lllllll}
Z_{H}&=&\left(\sum\limits_{i_{1}=0}^{\infty}\dfrac{(-1)^{i_{1}}}{i_{1}!}\exp\frac{\beta}{2}[-(-i_{1}
+\frac{1}{2})^{2}+\frac{1}{4}]\cdot\psi^{\ast}_{\frac{1}{2}-i_{1}}\right)\cdot\\
& &\cdot\left(\sum\limits_{i_{2}=0}^{\infty}\dfrac{(-1)^{i_{2}}}{i_{2}!}\exp\frac{\beta}{2}[-(-i_{2}
+\frac{3}{2})^{2}+\frac{9}{4}]\cdot\psi^{\ast}_{\frac{3}{2}-i_{2}}\right)\cdots\cdot\\
& &\cdot\left(\sum\limits_{i_{k}=0}^{\infty}\dfrac{(-1)^{i_{k}}}{i_{k}!}\exp\frac{\beta}{2}[-(-i_{k}
+k+\frac{1}{2})^{2}+(k-\frac{1}{2})]\cdot\psi^{\ast}_{k-\frac{1}{2}-i_{k}}\right)\cdots|\infty\rangle\\
&=&\widetilde{\psi}^{H}_{1/2}\widetilde{\psi}^{H}_{3/2}\cdots\widetilde{\psi}^{H}_{r+1/2}\cdots|\infty\rangle.
\end{array}
\end{equation}
In this equation,
\begin{equation}\label{5.22}
\widetilde{\psi}^{H}_{r+1/2}=\psi^{\ast}_{r+\frac{1}{2}}+\sum\limits_{i=1}^{\infty}a^{H}_{r,u}\psi^{\ast}_{r+1/2-i}
\end{equation}
and
\begin{equation}\label{5.23}
a^{H}_{r,u}=\frac{(-1)^{u}}{u!}\exp\{\frac{\beta}{2}[(r+1/2)^{2}-(r-u+1/2)^{2}]\}.
\end{equation}
$\widetilde{\psi}^{H}_{r+1/2}$ corresponds to the basis $\{\phi_{i}(\mu)\}$ in the notation \eqref{4.27}, and from $\widetilde{\psi}^{H}_{r+1/2}$, we can construct another operator $\Psi^{H}_{r+1/2}$ which corresponds to the canonical basis \eqref{4.24}
\begin{equation}\label{5.24}
\begin{array}{lllll}
\Psi^{H}_{1/2}=\widetilde{\psi}^{H}_{1/2};\\
\Psi^{H}_{r+1/2}=\widetilde{\psi}^{H}_{r+1/2}-\sum\limits_{i=1}^{r}a^{H}_{r,i}\Psi^{H}_{r-i+1/2}.
\end{array}
\end{equation}
From the definition of $\Psi^{H}_{r+1/2}$, we can written it as
\begin{equation}\label{5.25}
\Psi^{H}_{r+1/2}=\psi^{\ast}_{r+1/2}+\sum\limits_{i=1}^{\infty}b^{H}_{r,i}\psi^{\ast}_{-i+1/2},
\end{equation}
and the coefficients are
\begin{equation}\label{5.26}
b^{H}_{r,i}=a^{H}_{r,i}-\sum\limits_{j=1}^{r}a^{H}_{r,j}b^{H}_{r-j,i}.
\end{equation}
From equation \eqref{4.54}, we can find a fermionic representation of the Hurwitz partition function $Z_{H}(t)$
\begin{equation}\label{5.27}
\begin{array}{lllll}
Z_{H}&=&\exp\{\sum\limits_{m,n\geq0}b^{H}_{m,n+1}\psi_{-m-1/2}\psi^{\ast}_{-n-1/2}\}|0\rangle
&=&\widetilde{G}^{H}|0\rangle.
\end{array}
\end{equation}

Using the boson-fermion correspondence, one can rewrite $Z_{H}$ in terms of the Schur polynomials,
\begin{equation}\label{li12}
\begin{array}{llllllll}
Z_{H}&=&1+S_{(1)}(t)+\frac{1}{2}e^{\beta}S_{(2)}(t)+e^{-\beta}S_{(1,1)}(t)+\frac{1}{3!}
e^{3\beta}S_{(3)}(t)+\frac{1}{2}S_{(2,1)}(t)+e^{-3\beta}S_{(1,1,1)}(t)\\
&&+\frac{1}{4!}e^{6\beta}S_{(4)}(t)+\frac{1}{3!}e^{2\beta}S_{(3,1)}(t)+\frac{1}{2!2!}S_{(2,2)}(t)
+\frac{1}{2!}e^{-2\beta}S_{(2,1,1)}(t)+e^{6\beta}S_{(1,1,1,1)}(t)\\
&&+\frac{1}{5!}e^{10\beta}S_{(5)}(t)+\frac{1}{4!}e^{5\beta}S_{(4,1)}(t)+\frac{1}{3!2!}e^{2\beta}
S_{(3,2)}(t)+\frac{1}{3!}S_{(3,1,1)}(t)+\dfrac{1}{2!}e^{-2\beta}S_{(2,2,1)}(t)\\
&&+\frac{1}{2!}e^{-5\beta}S_{(2,1,1,1)}(t)+e^{-10\beta}S_{(1,1,1,1,1)}(t)\cdots\cdots\\
&=&\sum\limits_{(\lambda_{1},\cdots,\lambda_{k})\in
\mathscr{P}}\frac{1}{\lambda_{1}!\lambda_{2}!\cdots\lambda_{k}!}\exp\left[\dfrac{\beta}{2}
\sum\limits_{i=1}^{k}\lambda_{i}(\lambda_{i}-2i+1)\right]\cdot S_{(\lambda_{1},\cdots,\lambda_{k})}(t).
\end{array}
\end{equation}

In the following, we will give the fermionic representation of the Hodge partition function. At first, we rewrite the ELSV formula for the Hodge partition function
$$
Z_{Hodge}=\exp\left(\sum\limits_{k=1}^{\infty}a_{k}\beta^{k}\hat{L}_{-k}\right)\cdot
\exp\left(-\sum\limits_{b=1}^{\infty}\frac{b^{b-2}\beta^{b-1}\hat{J}_{-b}}{b!}\right)\cdot Z_{H},
$$
where the coefficients $a_{k}$ are some constants that are determined by equation \dref{lyp41}
$$
\exp\left(\sum\limits_{k=1}^{\infty}a_{k}z^{k+1}\dfrac{\partial}{\partial z}\right)\cdot z=\dfrac{z}{1+z}e^{-\frac{z}{1+z}}.
$$

We will derive the fermionic representation of the Hodge partition function in two steps. In first step, we construct a partition function $Z_{b}$, which bridges the Hodge partition function and the Hurwitz Partition function. In next step, we give the fermionic representation of the Hodge partition function from $Z_{b}$. At first, we define the partition function $Z_{b}$ as
\begin{equation}\label{5.30}
Z_{b}=\exp\{-\sum\limits_{b=1}^{\infty}\frac{b^{b-2}\beta^{b-1}\hat{J}_{-b}}{b!}\}\cdot Z_{H}.
\end{equation}
From equation \eqref{4.15}, $Z_{b}$ can be represented in terms of fermions
\begin{equation}\label{5.31}
\begin{array}{lllllll}
Z_{b}&=&\exp\{-\sum\limits_{b=1}^{\infty}\frac{b^{b-2}\beta^{b-1}}{b!}\sum\limits_{i+j=-b}:\psi_{u}\psi^{\ast}_{v}:\}
\widetilde{G}^{H}|0\rangle\\
&=&\widetilde{\psi}^{1}_{1/2}\widetilde{\psi}^{1}_{3/2}\cdots\widetilde{\psi}^{1}_{r+1/2}\cdots|\infty\rangle.
\end{array}
\end{equation}
In this equation, the operator $\widetilde{\psi}^{1}_{r+1/2}$ is gotten from $\widetilde{\psi}^{H}_{r+1/2}$ in the following way
\begin{equation}\label{5.32}
\widetilde{\psi}^{1}_{r+1/2}=\exp\{-\sum\limits_{b=1}^{\infty}\frac{b^{b-2}\beta^{b-1}}{b!}
\sum\limits_{i+j=-b}:\psi_{u}\psi^{\ast}_{v}:\}\cdot\widetilde{\psi}^{H}_{r+1/2}\cdot
\exp\{\sum\limits_{b=1}^{\infty}\frac{b^{b-2}\beta^{b-1}}{b!}\sum\limits_{i+j=-b}:\psi_{u}\psi^{\ast}_{v}:\}
\end{equation}
From equation \eqref{4.15}, it is easy to get $[J_{k},\psi^{\ast}_{r+1/2}]=-\psi^{\ast}_{k+r+1/2}$. Then from the explicit form of $\widetilde{\psi}^{H}_{r+1/2}$, we get the explicit expression of $\widetilde{\psi}^{1}_{r+1/2}$
\begin{equation}\label{5.33}
\widetilde{\psi}^{1}_{r+1/2}=\psi^{\ast}_{r+1/2}+\sum\limits_{i=1}^{\infty}a^{1}_{r,i}\psi^{\ast}_{r-i+1/2}
\end{equation}
where
\begin{equation}\label{5.34}
a^{1}_{r,i}=a^{H}_{r,i}+\sum\limits_{j=1}^{i}\left(\sum\limits_{l=1}^{j}\frac{\beta^{j-l}}{l!}\sum_{
k_{1}+\cdots +k_{l}=u}\frac{k_{1}^{k_1-2}\cdots k_{l}^{k_{l}-2}}{k_{1}!\cdots k_{l}!}\right)a^{H}_{r,i-j}.
\end{equation}
From the definition of $Z_{b}$, it is obvious that
\begin{equation}\label{5.35}
Z_{Ho}=\exp\left(\sum\limits_{k=1}^{\infty}a_{k}\beta^{k}\hat{L}_{-k}\right)\cdot Z_{b},
\end{equation}
The operator $\hat{L}_{k}$ is a operator being in $\hat{gl}(\infty)$, we have gotten its fermionic representation in equation \eqref{4.59}. So the Hodge partition function can be represented as
\begin{equation}\label{5.36}
\begin{array}{llllll}
Z_{Ho}&=&\exp\left(\frac{1}{2}\sum\limits_{k=1}^{\infty}a_{k}\beta^{k}\sum\limits_{a+b=-k}(b-a)
:\psi_{a}\psi^{\ast}_{b}:\right)\cdot Z_{b}\\
&=&\widetilde{\psi}^{Ho}_{1/2}\widetilde{\psi}^{Ho}_{3/2}\cdots\widetilde{\psi}^{Ho}_{r+1/2}\cdots|\infty\rangle.
\end{array}
\end{equation}
In this equation,
\begin{equation}\label{5.37}
\begin{array}{lllll}
\widetilde{\psi}^{Ho}_{r+1/2}&=&\exp\left(\frac{1}{2}\sum\limits_{k=1}^{\infty}a_{k}\beta^{k}
\sum\limits_{a+b=-k}(b-a)
:\psi_{a}\psi^{\ast}_{b}:\right)\cdot\widetilde{\psi}^{1}_{r+1/2}\\
& &\cdot\exp\left(-\frac{1}{2}\sum\limits_{k=1}^{\infty}a_{k}\beta^{k}\sum\limits_{a+b=-k}(b-a)
:\psi_{a}\psi^{\ast}_{b}:\right).
\end{array}
\end{equation}
As $L_{k}=\frac{1}{2}\sum_{a+b=k}(b-a):\psi_{a}\psi^{\ast}_{b}:$, the commutator of $L_{k}$ and $\psi^{\ast}_{r+1/2}$ is
\begin{equation}\label{5.38}
[L_{k},\psi^{\ast}_{r+1/2}]=\frac{k+2r+1}{2}\psi^{\ast}_{r+k+1/2}.
\end{equation}
Therefore, the explicit expression of $\widetilde{\psi}^{Ho}_{r+1/2}$ is
\begin{equation}\label{5.39}
\widetilde{\psi}^{Ho}_{r+1/2}=\psi^{\ast}_{r+1/2}+\sum\limits_{i=1}^{\infty}a^{Ho}_{r,i}\psi^{\ast}_{r-i+1/2},
\end{equation}
where the coefficients $a^{Ho}_{r,i}$ are
\begin{equation}\label{5.40}
\begin{array}{lllll}
a^{Ho}_{r,i}&=&a^{1}_{r,i}+\sum\limits_{j=1}^{i}a^{1}_{r,i-j}\sum\limits_{l=1}^{j}\frac{\beta^{j}}{2^{l}l!}
\sum\limits_{k_{1}+\cdots+k_{l}=u}a_{k_{1}}\cdots a_{k_{l}}(-k_{1}+2r+2j-2i+1)\\
& &\cdot(-k_{2}+2r+2j-2k_{1}-2i+1)\cdots
(-k_{l}+2r+2j-2k_{l-1}-2i+1).
\end{array}
\end{equation}
We repeat the approach for $Z_{H}$, and construct $\Psi^{Ho}_{r+1/2}$ from $\widetilde{\psi}^{Ho}_{r+1/2}$
\begin{equation}\label{5.41}
\begin{array}{lllll}
\Psi^{Ho}_{1/2}=\widetilde{\psi}^{Ho}_{1/2};\\
\Psi^{Ho}_{r+1/2}=\widetilde{\psi}^{Ho}_{r+1/2}-\sum\limits_{i=1}^{r}a^{H}_{r,i}\Psi^{Ho}_{r-i+1/2}.
\end{array}
\end{equation}
Similarly,
\begin{equation}\label{5.42}
\Psi^{Ho}_{r+1/2}=\psi^{\ast}_{r+1/2}+\sum\limits_{i=1}^{\infty}b^{Ho}_{r,i}\psi^{\ast}_{-i+1/2},
\end{equation}
and the coefficients are
\begin{equation}\label{5.43}
b^{Ho}_{r,i}=a^{H}_{r,i}-\sum\limits_{j=1}^{r}a^{Ho}_{r,j}b^{Ho}_{r-j,i}.
\end{equation}
Finally, the fermionic representation of $Z_{Ho}$ is
\begin{equation}\label{5.44}
\begin{array}{lllll}
Z_{Ho}&=&\exp\{\sum\limits_{m,n\geq0}b^{Ho}_{m,n+1}\psi_{-m-1/2}\psi^{\ast}_{-n-1/2}\}|0\rangle
&=&\widetilde{G}^{Ho}|0\rangle.
\end{array}
\end{equation}

\subsection{Connection Between $Z^{\{r\}}(t)$ and $Z_{Hodge}(t)$}
In above sections, we have expressed the partition function of $r$-spin intersection numbers, the Hurwitz partition function and the Hodge partition function in terms of fermionic fields. We could consider the linkage between any two by a $\widehat{GL}(\infty)$ operator.
Denote the operators
\begin{equation}\label{lyp86}
\begin{array}{lllll}
U_{HW}&=&\widetilde{G}^{H}\cdot(\widetilde{G}^{r})^{-1}\\
&=&\exp\{\sum\limits_{m,n\geq0}(b^{Ho}_{m,n+1}-b^{r}_{m+1,n+1})\psi_{-m-1/2}\psi^{\ast}_{-n-1/2}\},
\end{array}
\end{equation}
and
\begin{equation}\label{5.45}
\begin{array}{llll}
U_{HoW}&=&\widetilde{G}^{Ho}(\widetilde{G}^{r})^{-1}\\
&=&\exp\{\sum\limits_{m,n\geq0}(b^{Ho}_{m,n+1}-b^{r}_{m+1,n+1})\psi_{-m-1/2}\psi^{\ast}_{-n-1/2}\},
\end{array}
\end{equation}
respectively. Then, the partition function of $r$-spin intersection numbers $Z^{\{r\}}$ and the Hurwitz partition function $Z_{H}$ can be bridged by $U_{HW}$:
\begin{equation}\label{lyp87}
Z_{H}=U_{HW}\cdot Z^{\{r\}}.
\end{equation}
And, the partition function of $r$-spin intersection numbers $Z^{\{r\}}$ and the Hodge partition functin $Z_{Ho}$ can be connected by $U_{HoW}$
\begin{equation}\label{5.46}
Z_{Ho}=U_{HoW}\cdot Z^{\{r\}}.
\end{equation}

As the boson-fermion correspondence, we can express $\psi_{r+1/2}$ and $\psi^{\ast}_{r+1/2}$ with $J_{k}$. From equation \eqref{4.19},
\begin{equation}\label{5.47}
\begin{array}{llllll}
\psi_{-m-1/2}&=&\text{res}(z^{-m-1}e^{ip}\exp\{\sum\limits_{k>0}\frac{J_{-k}}{k}z^{k}\}
\exp\{-\sum\limits_{l>0}\frac{J_{k}}{k}z^{-k}\})\\
&=&e^{ip}\sum\limits_{u=0}^{\infty}S_{u+m}(\frac{J_{-k}}{k})S_{k}(-\frac{J_{k}}{k}),
\end{array}
\end{equation}
and
\begin{equation}\label{5.48}
\begin{array}{lllll}
\psi^{\ast}_{-n-1/2}&=&\text{res}(z^{-n-1}e^{-ip}\exp\{-\sum\limits_{k>0}\frac{J_{-k}}{k}z^{k}\}
\exp\{\sum\limits_{l>0}\frac{J_{k}}{k}z^{-k}\})\\
&=&e^{-ip}\sum\limits_{u=0}^{\infty}S_{u+n}(-\frac{J_{-k}}{k})S_{k}(\frac{J_{k}}{k}).
\end{array}
\end{equation}
Here, $S_{u}$ is the elementary Schur function defined in \eqref{4.35}.
Therefore, equation \eqref{5.46} is equivalent to
\begin{equation}\label{5.49}
\begin{array}{lllll}
Z_{Ho}&=&\exp\{\sum\limits_{m,n\geq0}(b^{Ho}_{m,n+1}-b^{r}_{m+1,n+1})\sum\limits_{u,v\geq0}
S_{m+u}(\frac{J_{-k}}{k})S_{u}(-\frac{J_{k}}{k})S_{v+n}(-\frac{J_{-k}}{k})S_{v}(\frac{J_{k}}{k})\}\cdot Z^{\{r\}}\\
&=&\exp\{\sum\limits_{m,n\geq0}(b^{Ho}_{m,n+1}-b^{r}_{m+1,n+1})\sum\limits_{u,v\geq0}
(S_{m+u}(\frac{J_{-k}}{k})S_{v+n}(-\frac{J_{-k}}{k})S_{u}(-\frac{J_{k}}{k})S_{v}(\frac{J_{k}}{k})\\
& &+S_{m+u}(\frac{J_{-k}}{k})[S_{u}(-\frac{J_{k}}{k}),S_{v+n}(-\frac{J_{-k}}{k})]S_{v}(\frac{J_{k}}{k}))\}
\cdot Z^{\{r\}}.
\end{array}
\end{equation}
From equation \eqref{4.50}, equation \eqref{5.49} can also be rephrased in terms of time coordinate $\{t_{k}\}$
\begin{equation}\label{5.61}
\begin{array}{llllll}
Z_{Ho}&=&\exp\{\sum\limits_{m,n\geq0}(b^{Ho}_{m,n+1}-b^{r}_{m+1,n+1})\sum\limits_{u,v\geq0}
S_{m+u}(t_{k})S_{u}(-\frac{1}{k}\frac{\partial}{\partial t_{k}})S_{v+n}(-t_{k})S_{v}(\frac{1}{k}\frac{\partial}{\partial t_{k}})\}\cdot Z^{\{r\}}\\
&=&\exp\{\sum\limits_{m,n\geq0}(b^{Ho}_{m,n+1}-b^{r}_{m+1,n+1})\sum\limits_{u,v\geq0}
(S_{m+u}(t_{k})S_{v+n}(-t_{k})S_{u}(-\frac{1}{k}\frac{\partial}{\partial t_{k}})S_{v}(\frac{1}{k}\frac{\partial}{\partial t_{k}})\\
& &+S_{m+u}(t_{k})[S_{u}(-\frac{1}{k}\frac{\partial}{\partial t_{k}}),S_{v+n}(-\frac{1}{k}\frac{\partial}{\partial t_{k}})]S_{v}(\frac{1}{k}\frac{\partial}{\partial t_{k}}))\}
\cdot Z^{\{r\}}.
\end{array}
\end{equation}
Equation \eqref{5.61} gives a relationship between the partition function of $r$-spin intersection numbers and the Hodge partition function. By comparing the coefficients in both side of equation \eqref{5.49}, one get the formula between the $r$-spin intersection numbers and the Hodge integrals. The $r$-spin intersection numbers and the Hodge integrals both are invariants in the moduli space of curves, and they play important roles in both mathematics and physics. We expect equation \eqref{5.61} should have significant roles to investigating the topics.

\subsection{$W_{1+\infty}$ constraint for the Hodge Partition Function}
In section 4, we have given two Kac-Schwarz operators $a^{\{r\}}_{W}$ and $b^{\{r\}}_{W}$ for the partition function of $r$-spin intersection numbers. From these two operators, one can construct a $W_{1+\infty}$ constraint for the partition function of $r$-spin intersection numbers. In fermionic fields, they can be expressed as
\begin{equation}\label{5.50}
\mathscr{W}^{(k)}_{n}=\frac{(-1)^{k-1}}{2}\text{res}_{z}
\left(:\psi(z)\{(b^{\{r\}}_{W})^{n+k-1},(a^{\{r\}}_{W})^{k}\}\psi^{\ast}(z):\right)+\delta_{n,0}c_{k},
 \ \ \ k\geq1,\ n>-k,
\end{equation}
where $c_{k}$ are constants which can be determined by the commutators of these operators. These operators satisfy the equation
\begin{equation}\label{5.51}
\mathscr{W}^{(k)}_{n}\cdot Z^{\{r\}}=0\ \ \ \ \text{for} \ k\geq1,\ n>-k.
\end{equation}
We list the first few examples in the following:
\begin{equation}\label{5.52}
\begin{array}{llllllllll}
\mathscr{W}^{(1)}_{n}&=&J_{nr}=\sum\limits_{a+b=nr}:\psi_{a}\psi^{\ast}_{b}:,\\
\mathscr{W}^{(2)}_{n}&=&-\frac{1}{2}\text{res}_{z}\left(:\psi(z)\{(b^{\{r\}}_{W})^{n+1},a^{\{r\}}_{W}\}
\psi^{\ast}(z):\right)+\delta_{n,0}\frac{r^{2}-1}{24r}\\
&=&\frac{1}{2r}\sum\limits_{a+b=nr}(b-a):\psi_{a}\psi^{\ast}_{b}:-
\sqrt{-r}\sum\limits_{a+b=(n+1)r+1}:\psi_{a}\psi^{\ast}_{b}:+\delta_{n,0}\frac{r^{2}-1}{24r}\\
&=&\frac{1}{r}\mathcal {L}_{nr}-\sqrt{-r}J_{(n+1)r+1}+\delta_{n,0}\frac{r^{2}-1}{24r},\\
\mathscr{W}^{(3)}_{n}&=&\frac{1}{2}\text{res}_{z}\left(:\psi(z)\{(b^{\{r\}}_{W})^{n+2},(a^{\{r\}}_{W})^{2}\}
\psi^{\ast}(z):\right)\\
&=&\sum\limits_{a+b=nr}\frac{(2a+r)(2a+3r)+(2b+r)(2b+3r)}{8r^{2}}:\psi_{a}\psi^{\ast}_{b}:\\
& &+\sum\limits_{a+b=(n+1)r+1}(a-b):\psi_{a}\psi^{\ast}_{b}:
-r\sum\limits_{a+b=(n+2)r+2}:\psi_{a}\psi^{\ast}_{b}:
\end{array}
\end{equation}
Equation \eqref{5.45} gives an operator in term of fermions connecting the partition function of $r$-spin intersection numbers and the Hodge partition function. From it and equation \eqref{5.50}, we can construct a $W_{1+\infty}$ constraint for the Hodge partition function
\begin{equation}\label{5.55}
\begin{array}{lllll}
\mathcal {W}^{(k)}_{n}&=&U_{HoW}\cdot\mathscr{W}^{(k)}_{n}\cdot U^{-1}_{HoW}\\
&=&\frac{(-1)^{k-1}}{2}\text{res}_{z}\left(\sum\limits_{a,b\in\mathbb{Z}+1/2}
U_{HoW}:\psi_{a}\psi^{\ast}_{b}:U^{-1}_{HoW}z^{-a-1/2}\{(b^{\{r\}}_{W})^{n+k-1},(a^{\{r\}}_{W})^{k}\}
z^{-b-1/2}\right).
\end{array}
\end{equation}
These operators satisfy the equation
\begin{equation}\label{5.57}
\mathcal {W}^{(k)}_{n}\cdot Z_{Ho}=0,\ \ \ \ \text{for}\ k\geq1,\ n>-k.
\end{equation}
As examples, we give the explicit expressions of $\mathcal {W}^{(1)}_{n}$ and $\mathcal {W}^{(2)}_{-1}$ which are the generator of the whole $W_{1+\infty}$ constraint,
\begin{equation}\label{5.58}
\begin{array}{llllllllll}
\mathcal {W}^{(1)}_{n}&=&\sum\limits_{a+b=nr}:\psi_{a}\psi^{\ast}_{b}:\\
& &+\sum\limits_{0\leq k<kr}\sum\limits_{l\geq0}(b^{Ho}_{l,nr-k}-b^{r}_{l+1,nr-k})\psi^{\ast}_{-l-1/2}\psi_{k+1/2}\\
& &-\sum\limits_{k\geq0}\sum\limits_{l\geq0}(b^{Ho}_{k,l+1}-b^{r}_{k+1,l+1})\psi^{\ast}_{nr-k-1/2}\psi_{-l-1/2}\\
& &-\sum\limits_{k\geq0}\sum\limits_{l\geq0}(b^{Ho}_{l,nr+k+1}-b^{r}_{l+1,nr+k+1})
\psi_{-k-1/2}\psi^{\ast}_{-l-1/2}\\
& &+\sum\limits_{m,l\geq0}\sum\limits_{0\leq k<nr}(b^{Ho}_{l,nr-k}-b^{r}_{l+1,nr-k})(b^{Ho}_{k,m+1}-b^{r}_{k+1,m+1})
\psi^{\ast}_{-l-1/2}\psi_{-m-1/2},
\end{array}
\end{equation}
and
\begin{equation}\label{5.59}
\begin{array}{lllllllll}
\mathcal {W}^{(2)}_{-1}&=&\frac{1}{2r}\sum\limits_{a+b=-r}:\psi_{a}\psi^{\ast}_{b}:-\sqrt{-r}\sum\limits_{a+b=1}
:\psi_{a}\psi^{\ast}_{b}:\\
& &+\frac{1}{2r}\sum\limits_{k\geq0}\sum\limits_{m\geq0}(b^{Ho}_{k,m+1}-b^{r}_{k+1,m+1})(r+2k+1)
\psi^{\ast}_{-r-k-1/2}\psi_{-m-1/2}\\
& &+\frac{1}{2r}\sum\limits_{k\geq r}\sum\limits_{n\geq0}(b^{Ho}_{n,k-r+1}-b^{r}_{n+1,k-r+1})(r-2k-1)
\psi_{-k-1/2}\psi^{\ast}_{-n-1/2}\\
& &+\sqrt{-r}\sum\limits_{k\geq0}\sum\limits_{m\geq0}(b^{Ho}_{k,m+1}-b^{r}_{k+1,m+1})
\psi^{\ast}_{1/2-k}\psi_{-m-1/2}\\
& &-\sqrt{-r}\sum\limits_{n\geq0}(b^{Ho}_{n,1}-b^{r}_{n+1,1})\psi^{\ast}_{-n-1/2}\psi_{1/2}\\
& &+\sqrt{-r}\sum\limits_{k\geq0}\sum\limits_{n\geq0}(b^{Ho}_{n,k+2}-b^{r}_{n+1,k+2})
\psi_{-k-1/2}\psi^{\ast}_{-n-1/2}\\
& &-\sqrt{-r}\sum\limits_{m\geq0}\sum\limits_{n\geq0}(b^{Ho}_{n,1}-b^{r}_{n+1,1})(b^{Ho}_{0,m+1}-b^{r}_{1,m+1})
\psi^{\ast}_{-n-1/2}\psi_{-m-1/2}.
\end{array}
\end{equation}
As operators $\mathscr{W}^{(1)}_{1}$ and $\mathscr{W}^{(2)}_{-1}$ are another operator representations of the Kac-Schwarz operators $b^{\{r\}}_{W}$ and $a^{\{r\}}_{W}$, the operators $\mathcal {W}^{(1)}_{1}$ and $\mathcal {W}^{(2)}_{-1}$ are equivalent to two Kac-Schwarz operators for the Hodge partition function, denoted $b^{Ho}$ and $a^{Ho}$. These two operators are subject to relation
$$
[a^{Ho},b^{Ho}]=1,
$$
where $[ , ]$ is the commutator for differential operators.
The operators $\mathcal {W}^{(1)}_{1}$ and $\mathcal {W}^{(2)}_{-1}$ are generators of the $W_{1+\infty}$ constraint for the Hodge partition function and the Hodge partition function can be determined by these two operators.
\section{Conclusion}
In this paper, we want to tryout the $r$-spin intersection numbers through the GKMM. In the paper \cite{M.Kontsevich}, Kontsevich carry out his model to the intersection numbers by the Feynman diagram techniques. Unlike his way, we unite them by $W$-constraint. The partition function of the GKMM with a given potential $V(M)$ and the generating function for the $r$-spin intersection numbers satisfy the same $W$-constraint. As the uniqueness of the solution to the constraint, the
two kind functions should be coincide with each other. So, if we represent the partition function for the GKMM in
determinant form, we can get the fermionic representation, and the Schur polynomials representation for the partition function of the $r$-spin intersection numbers.

We constructed a $\widehat{GL}(\infty)$ operator that makes up a connection between the partition function of  $r$-spin intersection numbers and the Hodge partition function. We expressed this operator in both fermionic language and bosonic language. Unlike the results got by A.Alexandrov \cite{A.Alexandrov2,A.Alexandrov8}, X.Liu and G.Wang \cite{X.Liu}, a Hodge integral can be expressed as a finite summation of $r$-spin intersection numbers. The Hodge integrals and $r$-spin intersection numbers are both geometric invariants in the moduli space of curves, and this operator build a bridge between them. The operator must have certain geometric meaning, and it is a problem we want to investigate in near future.

From the operator between the partition function of $r$-spin intersection numbers and the Hodge partition function, we construct the $W_{1+\infty}$ constraint for the Hodge partition function. This constraint completely determines the Hodge partition function. We also give two Kac-Schwarz operators for the Hodge partition function which are the generators of the $W_{1+\infty}$ constraint. It is not only a $W_{1+\infty}$ constraint for the partition function of $r$-spin intersection numbers, but is also a $W_{r}$-constraint for it. There may be an operator consisting of $W_{r}$ operators that connect the $r$-spin intersection numbers and Hodge integrals, like the operator given in A.Alexandrov's conjecture \cite{A.Alexandrov2}.

\section*{Acknowledgments}
The financial supports from the National Natural Science Foundation
of China (Grant No. 11375258) are gratefully
acknowledged from one of the author (Ding).

\section*{Appendix A. Examples of $r$-Spin Intersection Numbers}
We will give some particular examples for the potential $V(Y)=\sqrt{-r}\frac{Y^{r+1}}{r+1}$ with given $r$. We calculate the following results with the help of Matlab progrom. If anyone is interested in the source program,
please email us to receive it.
\subsection*{A.1 $r=2$}
In this case, from \dref{6.8}, we get
\begin{equation}\label{7.1}
a_{1,3m}=\left(-\frac{\sqrt{-2}}{144}\right)^{m}\frac{(6m-1)!!}{(2m)!}.
\end{equation}
By the recursion formula \dref{6.9}
\begin{equation}
\begin{array}{llllllllll}
a_{2,3m}=-\left(-\frac{\sqrt{-2}}{144}\right)^{m}\frac{(6m-3)!!}{(2m)!}(6m+1),\\
a_{3,3m}=\left(-\frac{\sqrt{-2}}{144}\right)^{m}\frac{(6m-1)!!}{(2m)!},\\
a_{4,3}=-\frac{\sqrt{-2}}{144}\cdot\frac{123}{2},\\
a_{4,3(m+1)}=\left(-\frac{\sqrt{-2}}{144}\right)^{m+1}\frac{(6m-1)!!}{(2m+2)!}(6m+3)(41-36m^{2})\nonumber
\end{array}
\end{equation}
\begin{equation}\label{7.2}
\begin{array}{llllllllll}
a_{5,3}=-\frac{\sqrt{-2}}{144}\cdot\frac{303}{2},\\
a_{5,3(m+1)}=\left(-\frac{\sqrt{-2}}{144}\right)^{m+1}\frac{(6m-3)!!}{(2m+2)!}(6m+3)(6m+1)
(36m^{2}-72m-101),\\
a_{6,3}=-\frac{\sqrt{-2}}{144}\cdot\frac{555}{2},\\
a_{6,3(m+1)}=\left(-\frac{\sqrt{-2}}{144}\right)^{m+1}\frac{(6m-1)!!}{(2m+2)!}(2m+1)
(555+432m-108m^{2}),
\end{array}
\end{equation}
then, from equation \dref{6.13},we get the explicit form of $b_{i,j}$
\begin{equation}\label{7.3}
\begin{array}{llllllll}
b_{1,3m}=a_{1,3m}=\left(-\frac{\sqrt{-2}}{144}\right)^{m}\frac{(6m-1)!!}{(2m)!},\\
b_{2,3m-1}=a_{2,3m}=\left(-\frac{\sqrt{-2}}{144}\right)^{m}\frac{(6m-1)!!}{(2m)!},\\
b_{3,3m-2}=a_{3,3m}=\left(-\frac{\sqrt{-2}}{144}\right)^{m}\frac{(6m-1)!!}{(2m)!},\\
b_{4,3m}=-\left(-\frac{\sqrt{-2}}{144}\right)^{m+1}\frac{(6m-1)!!}{(2m+2)!}(2m+1)(108m^{2}+123m),\\
b_{5,3m-1}=\left(-\frac{\sqrt{-2}}{144}\right)^{m+1}\frac{(6m-3)!!}{(2m+2)!}(6m+1)(2m+1)
(108m^{2}+87m),\\
b_{6,3m-2}=-\left(-\frac{\sqrt{-2}}{144}\right)^{m+1}\frac{(6m-1)!!}{(2m+2)!}(2m+1)(108m^{2}+123m),\\
\cdots\cdots\cdots
\end{array}
\end{equation}
Expanding the partition function $Z$ in the Schur polynomials,
\begin{equation}\label{7.4}
\begin{array}{llllll}
Z=1-\frac{\sqrt{-2}}{96}\left(5\cdot S_{(3)}(t)+7\cdot S_{(2,1)}(t)+5\cdot S_{(1^{3})}(t)\right)\\
-\frac{1}{9216}\left(385\cdot S_{(6)}(t)+455\cdot S_{(5,1)}(t)+0\cdot S_{(4,2)}(t)
+385\cdot S_{(4,1^{2})}(t)\right.\\
\left.-70\cdot S_{(3,3)}(t)-50\cdot S_{(3,2,1)}(t)-70\cdot S_{(2,2,2)}(t)+385\cdot S_{(3,1^{3})}(t)\right.\\
\left.+455\cdot S_{(2,1^{4})}(t)+385\cdot S_{(1^{6})}(t)\right)+\cdots\\
=1+(16t_1^6 + 600t_1^3T_3 + 720t_5t_1+ 225t_3^2)/4608(\sqrt{-2})^2+(64t_1^9 + 7056t_1^6t_3 + 60480t_5t_1^4 \\ + 132300t_1^3t_3^2 + 181440t_7t_1^2 + 317520t_5t_1t_3 + 33075t_3^3 + 68040t_9)/663552(\sqrt{-2})^3
\end{array}
\end{equation}
These results are consistent with the formulas got by Zhou \cite{Zhou} by solving the Virasoro constraint of the Kontsevich matrix model. And A. Alexandrov also got the same result from other method \cite{A.Alexandrov1}.

\subsection*{A.2 $r=3$}
In this case
\begin{equation}\label{7.5}
\begin{array}{lllllllllll}
b_{1,4}=a_{1,4}=\frac{7}{36\sqrt{-3}},\ \ \ \ \ \
b_{1,8}=a_{1,8}=\frac{140}{(36\sqrt{-3})^{2}},\\
b_{2,3}=a_{2,4}=-\frac{5}{36\sqrt{-3}},\ \ \ \ \ \
b_{2,7}=a_{2,8}=-\frac{4480}{(36\sqrt{-3})^{2}},\\
b_{3,2}=a_{3,4}=-\frac{5}{36\sqrt{-3}},\ \ \ \ \ \
b_{3,6}=a_{3,8}=-\frac{6760}{(36\sqrt{-3})^{2}},\\
b_{4,1}=a_{4,4}=\frac{7}{36\sqrt{-3}},\ \ \ \ \ \
b_{4,5}=a_{4,8}=-\frac{7420}{(36\sqrt{-3})^{2}},\cdots\\
\vdots
\end{array}
\end{equation}
Then, the partition function can be expressed as
\begin{equation}\label{7.6}
\begin{array}{lllllllllllllll}
Z=&&1+\frac{7}{36\sqrt{-3}}\cdot S_{(4)}(t)+\frac{5}{36\sqrt{-3}}\cdot S_{(3,1)}(t)\\
&-&\frac{5}{36\sqrt{-3}}\cdot S_{(2,1^{2})}(t)-\frac{7}{36\sqrt{-3}}\cdot S_{(1^{4})}(t)
+0\cdot S_{(2,2)}(t)+\\
&+&\frac{1}{2592(\sqrt{-3})^{2}}[385\cdot S_{(8)}(t)+455\cdot S_{(7,1)}(t)+0\cdot S_{(6,2)}(t)+25\cdot S_{(6,1^{2})(t)}\\
&+&0\cdot S_{(5,3)}(t)+0\cdot S_{(5,2,1)}(t)-385\cdot S_{(5,1^{3})}(t)-70\cdot S_{(4,4)}(t)\\
&+&70\cdot S_{(4,3,1)}(t)+0\cdot S_{(4,2^{2})}(t)+98\cdot S_{(4,2,1^{2})}(t)-385\cdot S_{(4,1^{4})}(t)\\
&+&50\cdot S_{(3^{2},2)}(t)+0\cdot S_{(3^{2},1^{2})}(t)+70\cdot S_{(3,2^{2},1)}(t)+0\cdot S_{(3,2,1^{3})}(t)\\
&+&25\cdot S_{(3,1^{5})}(t)-70\cdot S_{(2^{4})}(t)+0\cdot S_{(2^{3},1^{2})}(t)+0\cdot S_{(2^{2},1^{2})}(t)\\
&+&455\cdot S_{(2,1^{6})}(t)+385\cdot S_{(1^{8})}(t)]+\cdots\\
=&&1+\frac{4}{36\sqrt{-3}}t_{4}+\frac{24}{36\sqrt{-3}}\cdot\frac{1}{2}t^{2}_{1}t_{2}+\frac{1}{162(\sqrt{-3})^{2}}
[9t_{1}^{4}t_{2}^{2}+30t_{1}^{3}t_{5}\\
&+&78t_{1}^{2}t_{2}t_{4}+42t_{1}t_{7}-12t_{2}^{4}+13t_{4}^{2}]+\cdots
\end{array}
\end{equation}
Replacing $t_{3n+a+1}$ in \dref{7.6} by $\frac{t_{n,a}\cdot c_{n,a}}{\sqrt{-3}}$, we get
\begin{equation}\label{7.7}
\begin{array}{lllll}
Z=&&1+\frac{1}{2}t^{2}_{0,0}t_{0,1}+\frac{1}{12}t_{1,0}+\frac{1}{8}t_{0,0}^{4}t_{0,1}^{2}
+\frac{1}{6}t_{0,0}^{3}t_{1,1}\\
&+&\frac{13}{24}t_{0,0}^{2}t_{0,1}t_{1,0}+\frac{1}{12}t_{0,0}t_{2,0}+\frac{1}{72}t_{0,1}^{4}
+\frac{13}{288}t_{1,0}^{2}
+\cdots
\end{array}
\end{equation}
The second term in the RHS of \dref{7.7} is exactly the $F_{0}$ in \dref{2.10} obtained by E.Witten \cite{E.Witten1}, while the third term is given by \dref{2.25}. It seems that there is no practical effort to write the first few 3-spin intersection numbers out. In them, some results coincide with those in \cite{Zhou1},
\begin{equation}\label{7.8}
\begin{array}{llll}
\langle\tau_{0,0}^{3}\tau_{1,1}\rangle_{0}=1;\ \ \ \ \ \ \langle\tau_{0,0}\tau_{2,0}\rangle_{1}=\frac{1}{12}\\
\langle\tau_{0,1}^{4}\rangle_{0}=\frac{1}{3}
\end{array}
\end{equation}
However, the following results are different,
\begin{equation}\label{7.8}
\begin{array}{llll}
\langle\tau_{1,0}^{2}\rangle_{1}=\frac{13}{144};\ \ \ \ \langle\tau_{0,0}^{2}\tau_{0,1}\tau_{1,0}\rangle_{0}=\frac{13}{12}
\end{array}
\end{equation}

\subsection*{A.3 $r$=4}
In this case, the partition function
\begin{equation}\label{7.9}
\begin{array}{lllllllllllllllllllllllllll}
Z=&&1+\frac{9}{32\sqrt{-4}}\cdot S_{(5)}(t)+\frac{3}{32\sqrt{-4}}\cdot S_{(4,1)}(t)+0\cdot S_{(3,2)}(t)\\
&-&\frac{7}{32\sqrt{-4}}\cdot S_{(3,1^{2})}(t)+0\cdot S_{(2^{2},1)}(t)+\frac{3}{32\sqrt{-4}}(t)+\frac{9}{32\sqrt{-4}}\cdot S_{(1^{5})}(t)\\
&+&\frac{1}{2048(\sqrt{-4})^{2}}[441\cdot S_{(10)}(t)+495\cdot S_{(9,1)}(t)+0\cdot S_{(8,2)}(t)-231\cdot S_{(8,1^{2})}(t)\\
&+&0\cdot S_{(7,3)}(t)+0\cdot S_{(7,2,1)}(t)-273\cdot S_{(7,1^{3})}(t)+0\cdot S_{(6,4)}(t)\\
&+&0\cdot S_{(6,3,1)}(t)+0\cdot S_{(6,2^{2})}(t)+0\cdot S_{(6,2,1^{2})}(t)+441\cdot S_{(6,1^{4})}(t)\\
&-&54\cdot S_{(5^{2})}(t)+126\cdot S_{(5,4,1)}(t)+0\cdot S_{(5,3,2)}(t)-54\cdot S_{(5,3,1^{2})}(t)\nonumber
\end{array}
\end{equation}
\begin{equation}
\begin{array}{lllllllllllllllllllllllllll}
&+&0\cdot S_{(5,2^{2},1)}(t)-162\cdot S_{(5,2,1^{3})}(t)+441\cdot S_{(5,1^{5})}(t)+42\cdot S_{(4^{2},2)}(t)\\
&+&0\cdot S_{(4^{2},1^{2})}(t)+0\cdot S_{(4,3^{2})}(t)-18\cdot S_{(4,3,2,1)}(t)+0\cdot S_{(4,3,1^{3})}(t)\\
&+&0\cdot S_{(4,2^{3})}(t)-54\cdot S_{(4,2^{2},1^{2})}(t)+0\cdot S_{(4,2,1^{4})}(t)-273\cdot S_{(4,1^{6})}(t)\\
&+&0\cdot S_{(3^{3},1)}(t)+42\cdot S_{(3^{2},2^{2})}(t)+0\cdot S_{(3,3,2,1^{2})}(t)+0\cdot S_{(3^{2},1^{4})}(t)\\
&+&126\cdot S_{(3,2^{3})}(t)+0\cdot S_{(3,2^{2},1)}(t)+0\cdot S_{(3,2,1^{5})}(t)-231\cdot S_{(3,1^{7})}(t)\\
&-&54\cdot S_{(2^{5})}(t)+0\cdot S_{(2^{4},1^{2})}(t)+0\cdot S_{(2^{3},1^{4})}(t)+0\cdot S_{(2^{2},1^{6})}(t)\\
&+&495\cdot S_{(2,1^{8})}(t)+441\cdot S_{(1^{10})}(t)]+\cdots\\
=&&1+\frac{1}{32\sqrt{-4}}[12t_{1}^{2}t_{3}+16t_{1}t_{2}^{2}+5t_{5}]+[144t_1^4t_3^2 + 384t_1^3t_2^2t_3+ 448t_7t_1^3\\ &+&256t_1^2t_2^4
+1536t_6t_1^2t_2+ 1080t_1^2t_3t_5+ 1440t_1t_2^2t_5+ 720t_9t_1- 1152t_2^2t_3^2 \\&-&336t_7t_3+ 225t_5^2]/2048(\sqrt{-4})^2
+\cdots\cdots
\end{array}
\end{equation}
Substituting $\frac{t_{n,a}\cdot c_{n,a}}{\sqrt{-4}}$ for $t_{4n+a+1}$ in above equation, we get
\begin{equation}\label{7.10}
\begin{array}{lllllllllllllllllllllllllll}
Z&=&1+\frac{1}{2}t_{0,0}^{2}t_{0,2}+\frac{1}{2}t_{0,0}t_{0,1}^{2}+\frac{1}{8}t_{1,0}+\frac{9}{32}t_{0,0}^{4}t_{0,2}^2+\frac{1}{4}t_{0,0}^{3}t_{0,1}^2t_{0,2}
+\frac{1}{6}t_{0,0}^{3}t_{1,2}\\&+&\frac{1}{8}t_{0,0}^{2}t_{0,1}^4+\frac{1}{2}t_{0,0}^{2}t_{0,1}t_{1,1}+\frac{9}{16}t_{0,0}^{2}t_{0,2}t_{1,0}+\frac{9}{16}t_{0,1}^{2}t_{0,0}t_{1,0}
+\frac{1}{8}t_{0,0}t_{2,0}\\&+&\frac{1}{16}t_{0,1}^{2}t_{0,2}^2+\frac{1}{96}t_{0,2}t_{1,2}+\frac{9}{128}t_{1,0}^{2}+\cdots\cdots
\end{array}
\end{equation}
The second and the third term of the RHS in equation \dref{7.9} are terms in the $F_{0}$ in \dref{2.10}, and the fourth term is term in the $F_{1}$, respectively. From equation \dref{7.10}, we get the following 4-spin intersection numbers
$$
\begin{array}{llllll}
<\tau_{0,0}^{2}\tau_{0,2}>=1,\ \ \ \ \ \ \ \ \ \ <\tau_{0,0}\tau_{0,1}^{2}>=1\\
<\tau_{1,0}>=\frac{1}{8},\ \ \ \ \ \ \ \ \ \ <\tau_{0,2}\tau_{1,2}>=\frac{1}{96}
\end{array}
$$
The above results are coincided with the given ones in\cite{K.Liu,E.Witten1}. We have also obtained some intersection numbers which are not listed in the references yet.
$$
\begin{array}{llllll}
<\tau_{0,0}^{3}\tau_{0,1}^2t_{0,2}>=3,\ \ \ \ \ \ \ \ \ \ <\tau_{0,0}^{3}\tau_{1,2}>=1\\
<\tau_{0,0}^{2}\tau_{0,1}^4>=6,\ \ \ \ \ \ \ \ \ \ <\tau_{0,0}^{2}\tau_{0,1}\tau_{1,1}>=1\\
<\tau_{0,0}^{2}\tau_{0,2}\tau_{1,0}>=\frac{9}{8},\ \ \ \ \ \ \ \ \ \ <\tau_{0,0}\tau_{0,1}^{2}\tau_{1,0}>=\frac{9}{8}\\
<\tau_{0,0}\tau_{2,0}>=\frac{1}{8},\ \ \ \ \ \ \ \ \ \ <\tau_{0,1}^{2}\tau_{0,2}^{2}>=\frac{1}{4}\\
<\tau_{0,0}^{4}\tau_{0,2}^2>=\frac{27}{2},\ \ \ \ \ \ \ \ \ \ <\tau_{1,0}^{2}>=\frac{9}{64}
\end{array}
$$

\subsection*{A.4 $r=5$}
In this case, the partition function
\begin{equation}\label{7.11}
\begin{array}{lllllllllllllllllllllllllllllllllllllllllllllllllllllllllllllllllllll}
Z=&&1+\frac{1}{30\sqrt{-5}}[11\cdot S_{(6)}(t)+1\cdot S_{(5,1)}(t)+0\cdot S_{(4,2)}(t)\\
&-&7\cdot S_{(4,1^{2})}(t)+0\cdot S_{(3^{2})}(t)+0\cdot S_{(3,2,1)}(t)+7\cdot S_{(3,1^{3})}(t)\\
&+&0\cdot S_{(2^{3})}(t)+0\cdot S_{(2^{2},1^{2})}(t)-1\cdot S_{(2,1^{4})}(t)-11\cdot S_{(1^{6})}(t)]\\
&+&\frac{1}{1800(\sqrt{-5})^{2}}[517\cdot S_{(12)}(t)+539\cdot S_{(11,1)}(t)+0\cdot S_{(10,2)}(t)-455\cdot S_{(10,1^{2})}(t)\\
&+&0\cdot S_{(9,3)}(t)+0\cdot S_{(9,2,1)}(t)-49\cdot S_{(9,1^{3})}(t)+0\cdot S_{(8,4)}(t)\\
\end{array}
\end{equation}
$$
\begin{array}{lllllllllllllllllllllllll}
&+&0\cdot S_{(8,3,1)}(t)+0\cdot S_{(8,2^{2})}(t)+0\cdot S_{(8,2,1^{2})}(t)+469\cdot S_{(8,1^{4})}(t)\\
&+&0\cdot S_{(7,5)}(t)+0\cdot S_{(7,4,1)}(t)+0\cdot S_{(7,3,2)}(t)+0\cdot S_{(7,3,1^{2})}(t)\\
&+&0\cdot S_{(7,2^{2},1)}(t)+0\cdot S_{(7,2,1^{3})}(t)-517\cdot S_{(7,1^{5})}(t)-22\cdot S_{(6^{2})}(t)\\
&+&154\cdot S_{(6,5,1)}(t)+0\cdot S_{(6,4,2)}(t)-154\cdot S_{(6,4,1^{2})}(t)+0\cdot S_{(6,3^{2})}(t)\\
&+&0\cdot S_{(6,3,2,1)}(t)+22\cdot S_{(6,3,1^{3})}(t)+0\cdot S_{(6,2^{3})}(t)+0\cdot S_{(6,2^{2},1^{2})}(t)\\
&+&242\cdot S_{(6,2,1^{4})}(t)-517\cdot S_{(6,1^{6})}(t)+14\cdot S_{(5^{2},2)}(t)+0\cdot S_{(5^{2},1^{2})}(t)\\
&+&0\cdot S_{(5,4,3)}(t)-14\cdot S_{(5,4,2,1)}(t)+0\cdot S_{5,4,1^{3}}(t)+0\cdot S_{(5,3^{2},1)}(t)\\
&+&0\cdot S_{(5,3,2^{2})}(t)+2\cdot S_{(5,3,2,1^{2})}(t)+0\cdot S_{(5,3,1^{4})}(t)+0\cdot S_{(5,2^{3},1)}(t)\\
&+&22\cdot S_{(5,2^{2},1^{3})}(t)+0\cdot S_{(5,2,1^{5})}(t)+469\cdot S_{(5,1^{7})}(t)+0\cdot S_{(4^{3})}(t)\\
&+&0\cdot S_{(4^{2},3,1)}(t)+98\cdot S_{(4^{2},2^{2})}(t)+0\cdot S_{(4^{2},2,1^{2})}(t)+0\cdot S_{4^{2},1^{4}}(t)\\
&+&0\cdot S_{(4,3^{2},2)}(t)+0\cdot S_{(4,3^{2},1^{2})}(t)-14\cdot S_{(4,3,2^{2},1)}(t)+0\cdot S_{(4,3,2,1^{3})}(t)\\
&+&0\cdot S_{(4,3,1^{5})}(t)+0\cdot S_{(4,2^{4})}(t)-154\cdot S_{(4,2^{3},1^{2})}(t)+0\cdot S_{(4,2^{2},1^{4})}(t)\\
&+&0\cdot S_{(4,2,1^{6})}(t)-49\cdot S_{(4,1^{8})}(t)+0\cdot S_{(3^{4})}(t)+0\cdot S_{(3^{2},2,1)}(t)\\
&+&0\cdot S_{(3^{3},1^{3})}(t)+14\cdot S_{(3^{2},2^{3})}(t)+0\cdot S_{(3^{2},2^{2},1^{2})}(t)+0\cdot S_{(3^{2},2,1^{4})}(t)\\
&+&0\cdot S_{(3^{2},1^{6})}(t)+154\cdot S_{(3,2^{4},1)}(t)+0\cdot S_{(3,2^{3},1^{3})}(t)+0\cdot S_{(3,2^{2},1^{5})}(t)\\
&+&0\cdot S_{(3,2,1^{7})}(t)-455\cdot S_{(3,1^{9})}(t)-22\cdot S_{(2^{6})}(t)+0\cdot S_{(2^{5},1^{2})}(t)\\
&+&0\cdot S_{(2^{4},1^{4})}(t)+0\cdot S_{(2^{3},1^{6})}(t)+0\cdot S_{(2^{2}),1^(8)}(t)+539\cdot S_{(2,1^{10})}(t)\\
&+&517\cdot S_{(1^{12})}(t)]+\cdots\cdots\\
\end{array}
$$
By the definition of Schur polynomials, we expand the partition function as a Taylor series of $\{t_{k}\}$
\begin{equation}
\begin{array}{lllllllllllllllllllllll}
Z=&&1+\frac{1}{15\sqrt{-5}}[6\cdot t_{1}^{2}t_{4}+18\cdot t_{1}t_{2}t_{3}+4\cdot t_{2}^{3}+3\cdot t_{6}]\\
&+&\frac{1}{900(\sqrt{-5})^{2}}[72\cdot t_{1}^{4}t_{4}^{2}+432\cdot t_{1}^{3}t_{2}t_{3}t_{4}+216\cdot t_{1}^{3}t_{9}+96\cdot t_{1}^{2}t_{2}^{3}t_{4}\\
&+&648\cdot t_{1}^{2}t_{2}^{2}t_{3}^{2}+864\cdot t_{1}^{2}t_{2}t_{8}+756\cdot t_{1}^{2}t_{3}t_{7}+504\cdot t_{1}^{2}t_{4}t_{6}+288\cdot t_{1}t_{2}^{4}t_{3}\\
&+&1008\cdot t_{1}t_{2}^{2}t_{7}+1512\cdot t_{1}t_{2}t_{3}t_{6}+396\cdot t_{1}t_{11}+32\cdot t_{2}^{6}+336\cdot t_{2}^{3}t_{6}\\
&-&576\cdot t_{2}^{2}t_{4}^{2}-1296t_{2}t_{3}^{2}t_{4}-243\cdot t_{3}^{4}-324\cdot t_{3}t_{9}-288\cdot t_{4}t_{8}+126\cdot t_{6}^{2}]\\
&+&\frac{1}{202500(\sqrt{-5})^3}[2160\cdot t_{1}^{6}t_{4}^{3} + 19440\cdot t_{1}^{5}t_{2}t_{3}t_{4}^{2} +19440\cdot t_{9}t_{1}^{5}t_{4}+4320\cdot t_{1}^{4}t_{2}^{3}t_{4}^{2}\\
&+&58320\cdot t_{1}^{4}t_{2}^{2}t_{3}^{2}t_{4}+58320\cdot t_{9}t_{1}^{4}t_{2}t_{3}+77760\cdot t_{1}^{4}t_{2}t_{4}t_{8}+68040\cdot t_{1}^{4}t_{3}t_{4}t_{7}\\
&+&42120\cdot t_{1}^{4}t_{4}^{2}t_{6}+34020\cdot t_{14}t_{1}^{4}+25920\cdot t_{1}^{3}t_{2}^{4}t_{3}t_{4}+58320\cdot t_{1}^{3}t_{2}^{3}t_{3}^{3}\\
&+&12960\cdot t_{9}t_{1}^{3}t_{2}^{3}+233280\cdot t_{1}^{3}t_{2}^{2}t_{3}t_{8}+90720\cdot t_{1}^{3}t_{2}^{2}t_{4}t_{7}+204120\cdot t_{1}^{3}t_{2}t_{3}^{2}t_{7}\\
&+&252720\cdot t_{1}^{3}t_{2}t_{3}t_{4}t_{6}+168480\cdot t_{13}t_{1}^{3}t_{2}+136080\cdot t_{12}t_{1}^{3}t_{3}+106920t_{11}t_{1}^{3}t_{4}\\
&+&126360\cdot t_{9}t_{1}^{3}t_{6}+181440\cdot t_{1}^{3}t_{7}t_{8}+ 2880\cdot t_{1}^{2}t_{2}^{6}t_{4}+38880\cdot t_{1}^{2}t_{2}^{5}t_{3}^{2}\\
&+&51840\cdot t_{1}^{2}t_{2}^{4}t_{8}+317520\cdot t_{1}^{2}t_{2}^{3}t_{3}t_{7}+56160\cdot t_{1}^{2}t_{2}^{3}t_{4}t_{6}+379080\cdot t_{1}^{2}t_{2}^{2}t_{3}^{2}t_{6}\\
&-&51840\cdot t_{1}^{2}t_{2}^{2}t_{4}^{3}+272160\cdot t_{12}t_{1}^{2}t_{2}^{2}-116640\cdot t_{1}^{2}t_{2}t_{3}^{2}t_{4}^{2}+427680\cdot t_{11}t_{1}^{2}t_{2}t_{3}\\
&+&505440\cdot t_{1}^{2}t_{2}t_{6}t_{8}+317520\cdot t_{1}^{2}t_{2}t_{7}^{2}-21870\cdot t_{1}^{2}t_{3}^{4}t_{4}-29160\cdot t_{9}t_{1}^{2}t_{3}t_{4}\\
\end{array}
\end{equation}
$$
\begin{array}{llllllllllllllllllllllllll}
&+&442260\cdot t_{1}^{2}t_{3}t_{6}t_{7}-25920\cdot t_{1}^{2}t_{4}^{2}t_{8}+147420\cdot t_{1}^{2}t_{4}t_{6}^{2}+142560\cdot t_{16}t_{1}^{2}\\
&+&8640\cdot t_{1}t_{2}^{7}t_{3}+60480\cdot t_{1}t_{2}^{5}t_{7}+ 168480\cdot t_{1}t_{2}^{4}t_{3}t_{6}-155520\cdot t_{1}t_{2}^{3}t_{3}t_{4}^{2}\\
&+&166320\cdot t_{11}t_{1}t_{2}^{3}-349920\cdot t_{1}t_{2}^{2}t_{3}^{3}t_{4}-466560\cdot t_{9}t_{1}t_{2}^{2}t_{4}+589680\cdot t_{1}t_{2}^{2}t_{6}t_{7}\\
&-&65610\cdot t_{1}t_{2}t_{3}^{5}-612360\cdot t_{9}t_{1}t_{2}t_{3}^{2}-1010880\cdot t_{1}t_{2}t_{3}t_{4}t_{8}+442260\cdot t_{1}t_{2}t_{3}t_{6}^{2}\\
&-&362880\cdot t_{1}t_{2}t_{4}^{2}t_{7}-349920\cdot t_{1}t_{3}^{3}t_{8}-408240\cdot t_{1}t_{3}^{2}t_{4}t_{7}-204120\cdot t_{14}t_{1}t_{3}\\
&-&168480\cdot t_{13}t_{1}t_{4}+ 231660\cdot t_{11}t_{1}t_{6}-233280\cdot t_{9}t_{1}t_{8}+640\cdot t_{2}^{9}\\
&+&18720\cdot t_{2}^{6}t_{6}-34560\cdot t_{2}^{5}t_{4}^{2}-77760\cdot t_{2}^{4}t_{3}^{2}t_{4}-14580\cdot t_{2}^{3}t_{3}^{4}\\
&-&252720\cdot t_{9}t_{2}^{3}t_{3}-432000\cdot t_{2}^{3}t_{4}t_{8}+98280\cdot t_{2}^{3}t_{6}^{2}- 699840\cdot t_{2}^{2}t_{3}^{2}t_{8}\\
&-&1088640\cdot t_{2}^{2}t_{3}t_{4}t_{7}-336960\cdot t_{2}^{2}t_{4}^{2}t_{6}-272160\cdot t_{14}t_{2}^{2}-612360\cdot t_{2}t_{3}^{3}t_{7}\\
&-&758160\cdot t_{2}t_{3}^{2}t_{4}t_{6}-758160\cdot t_{13}t_{2}t_{3}-544320\cdot t_{12}t_{2}t_{4}-408240t_{9}t_{2}t_{7}\\
&-&311040\cdot t_{2}t_{8}^{2}-142155\cdot t_{3}^{4}t_{6}+155520\cdot t_{3}^{2}t_{4}^{3}-408240\cdot t_{12}t_{3}^{2}\\
&-&213840\cdot t_{11}t_{3}t_{4}-189540\cdot t_{9}t_{3}t_{6}-544320\cdot t_{3}t_{7}t_{8}-168480\cdot t_{4}t_{6}t_{8}\\
&-&158760\cdot t_{4}t_{7}^{2}+24570\cdot t_{6}^{3}-138996\cdot t_{18}]+\cdots\cdots
\end{array}
$$
After substituting $\frac{c_{n,a}\cdot t_{n,a}}{\sqrt{-5}}$ for $t_{5n+a+1}$ in above equation, we get the following 5-spin intersection numbers
\begin{equation}\label{7.12}
\begin{array}{lllllllllllllllllllllllllllllllllll}
\langle\tau_{1,0}\rangle_{1}=\dfrac{1}{6},\ \ \ \ \ \ \ \ \ \ \ \ \ \ \ \ \ \ \ \ \ \ \ \ \ \langle\tau_{0,2}\tau_{1,3}\rangle_{1}=\frac{1}{60},\\
\langle\tau_{0,3}\tau_{1,2}\rangle_{1}=\frac{1}{60}, \ \ \ \ \ \ \ \ \ \ \ \ \ \ \ \ \ \ \ \  \langle\tau_{0,2}\tau_{0,3}\tau_{2,0}\rangle_{1}=\frac{1}{30},\\
\langle\tau_{0,1}^{2}\tau_{2,3}\rangle_{1}=\frac{1}{30},\ \ \ \ \ \ \ \ \ \ \ \ \ \ \ \ \ \ \ \langle\tau_{0,1}\tau_{0,2}\tau_{2,2}\rangle_{1}=\frac{1}{20},\\
\langle\tau_{3,2}\rangle_{2}=\frac{11}{3600},\ \ \ \ \ \ \ \ \ \ \ \ \ \ \ \ \ \ \ \ \ \ \ \ \ \langle\tau_{2,1}^{2}\rangle_{2}=\frac{9}{400},\\
\langle\tau_{1,1}\tau_{3,1}\rangle_{2}=\frac{17}{1200},\ \ \ \ \ \ \ \ \ \ \ \ \ \ \ \ \ \ \ \
\langle\tau_{1,2}\tau_{3,0}\rangle_{2}=\frac{47}{3600},\\
\langle\tau_{2,0}\tau_{2,2}\rangle_{2}=\frac{59}{3600}
\end{array}
\end{equation}
The listed results in \dref{7.12} are coincided with the ones which have also been computed by K.Liu and his collaborators\cite{K.Liu}. However, if we want to derive certain intersection numbers listed in \cite{K.Liu}, we need to compute more schur polynomials. The calculation is beyond the capability of our computers. On the other side, in our approach, while it is easy to compute certain intersection numbers which is not listed in \cite{K.Liu}.
\begin{equation}
\begin{array}{lllllllllllllllllllllllllllllllllll}
\langle\tau_{0,0}^{2}\tau_{0,3}\rangle_{0}=1,\ \ \ \ \ \ \ \ \ \ \ \ \ \ \ \ \ \ \ \ \ \langle\tau_{0,0}\tau_{0,1}\tau_{0,2}\rangle_{0}=1,\\
\langle\tau_{0,1}^{3}\rangle_{0}=1,\ \ \ \ \ \ \ \ \ \ \ \ \ \ \ \ \ \ \ \ \ \ \ \ \ \ \langle\tau_{0,0}^{3}\tau_{1,3}\rangle_{0}=\frac{1}{6},\\
\langle\tau_{0,0}^{2}\tau_{0,3}\tau_{1,0}\rangle_{0}=\frac{7}{6},\ \ \ \ \ \ \ \ \ \ \ \ \ \ \ \langle\tau_{0,0}\tau_{0,1}^{2}\tau_{1,1}\rangle_{0}=2,\\
\langle\tau_{0,0}\tau_{0,1}\tau_{0,2}\tau_{1,0}\rangle_{0}=\frac{7}{6},\ \ \ \ \ \ \ \ \ \ \ \ \ \ \ \ \ \ \langle\tau_{0,1}^{3}\tau_{1,0}\rangle_{0}=\frac{7}{6},\\
\langle\tau_{0,1}^{2}\tau_{0,3}^{2}\rangle_{0}=\frac{1}{20},\ \ \ \ \ \ \ \ \ \ \ \ \ \ \ \ \ \ \langle\tau_{0,1}\tau_{0,2}^{2}\tau_{0,3}\rangle_{0}=\frac{1}{5},\\
\langle\tau_{0,2}^{4}\rangle_{0}=\frac{2}{5},\ \ \ \ \ \ \ \ \ \ \ \ \ \ \ \ \ \ \ \ \ \ \ \ \ \langle\tau_{0,0}\tau_{2,0}\rangle_{1}=\frac{1}{6},\\
\langle\tau_{0,0}\tau_{1,3}\tau_{1,2}\rangle_{1}=\frac{1}{30},\ \ \ \ \ \ \ \ \ \ \ \ \ \ \ \ \ \ \ \ \ \ \ \langle\tau_{1,0}^{2}\rangle_{1}=\frac{7}{36},\\
\langle\tau_{0,0}^{2}\tau_{3,0}\rangle_{1}=\frac{1}{12},\ \ \ \ \ \ \ \ \ \ \ \ \ \ \ \ \ \ \ \ \ \langle\tau_{0,0}\tau_{0,2}\tau_{2,3}\rangle_{1}=\frac{1}{60},\\
\langle\tau_{0,0}\tau_{0,3}\tau_{2,2}\rangle_{1}=\frac{1}{60},\ \ \ \ \ \ \ \ \ \ \ \ \ \ \ \ \ \ \langle\tau_{0,0}\tau_{1,0}\tau_{2,0}\rangle_{1}=\frac{1}{3},\nonumber
\end{array}
\end{equation}
\begin{equation}
\begin{array}{lllllllllllllllllllllllllllllllllll}
\langle\tau_{0,1}\tau_{1,1}\tau_{1,3}\rangle_{1}=\frac{1}{20},\ \ \ \ \ \ \ \ \ \ \ \ \ \ \ \ \ \ \
\langle\tau_{0,3}\tau_{1,1}^{2}\rangle_{1}=\frac{1}{40},\\
\langle\tau_{1,0}^{3}\rangle_{1}=\frac{1}{18},\ \ \ \ \ \ \ \ \ \ \ \ \ \ \ \ \ \ \ \ \ \ \ \ \ \ \ \
\langle\tau_{1,0}\tau_{3,2}\rangle_{2}=\frac{11}{1200},\\
\cdots\cdots\nonumber
\end{array}
\end{equation}

\subsection*{A.5 $r=7$}
In this case, the partition function
\begin{equation}\label{7.13}
\begin{array}{llllllllllllllllllllllllllllllllllllllllllllllll}
Z=&&1+\frac{1}{28\sqrt{-7}}[15\cdot S_{(8)}(t)-3\cdot S_{(7,1)}(t)+0\cdot S_{(6,2)}(t)-5\cdot S_{(6,1^{2})}(t)\\
&+&0\cdot S_{(5,3)}(t)+0\cdot S_{(5,2,1)}(t)+9\cdot S_{(5,1^{3})}(t)+0\cdot S_{(4^{2})}(t)\\
&+&0\cdot S_{(4,3,1)}(t)+0\cdot S_{(4,2^{2})}(t)+0\cdot S_{(4,2,1^{2})}(t)-9\cdot S_{(4,1^{4})}(t)\\
&+&0\cdot S_{(3^{2},2)}(t)+0\cdot S_{(3^{2},1^{2})}(t)+0\cdot S_{(3,2^{2},1)}(t)+0\cdot S_{(3,2,1^{3})}(t)\\
&+&5\cdot S_{(3,1^{5})}(t)+0\cdot S_{(2^{4})}(t)+0\cdot S_{(2^{3},1^{2})}(t)+0\cdot S_{(2^{2},1^{4})}(t)\\
&+&3\cdot S_{(2,1^{6})}(t)-15\cdot S_{(1^{8})}(t)]\\
&+&\frac{1}{1568(\sqrt(-7))^{2}}[705\cdot S_{(16)}(t)+615\cdot S_{(15,1)}(t)+0\cdot S_{(14,2)}(t)-855\cdot S_{(14,1^{2})}(t)\\
&+&0\cdot S_{(13,3)}(t)+0\cdot S_{(13,2,1)}(t)+495\cdot S_{(13,1^{3})}(t)+\cdots]\\
&+&\cdots\cdots\\
=&&1+\frac{1}{7\sqrt{-7}}[3\cdot t_{6}t_{1}^{2}+10\cdot t_{5}t_{1}t_{2}+12\cdot t_{4}t_{1}t_{3}+ 8\cdot t_{4}t_{2}^{2}+9\cdot t_{2}t_{3}^{2}+2\cdot t_{8}]\\
&+&\frac{1}{98(\sqrt{-7})^{2}}[9\cdot t_{1}^{4}t_{6}^{2}+60\cdot t_{1}^{3}t_{2}t_{5}t_{6}+ 72\cdot t_{1}^{3}t_{3}t_{4}t_{6}\\
&+&26\cdot t_{13}t_{1}^{3}+48\cdot t_{1}^{2}t_{2}^{2}t_{4}t_{6}+100\cdot t_{1}^{2}t_{2}^{2}t_{5}^{2}+ 54\cdot t_{1}^{2}t_{2}t_{3}^{2}t_{6}\\
&+&240\cdot t_{1}^{2}t_{2}t_{3}t_{4}t_{5}+120\cdot t_{12}t_{1}^{2}t_{2}+144\cdot t_{1}^{2}t_{3}^{2}t_{4}^{2}+132\cdot t_{11}t_{1}^{2}t_{3}\\
&+&120\cdot t_{10}t_{1}^{2}t_{4}+90\cdot t_{9}t_{1}^{2}t_{5}+60\cdot t_{1}^{2}t_{6}t_{8}+ 160\cdot t_{1}t_{2}^{3}t_{4}t_{5}\\
&+&180\cdot t_{1}t_{2}^{2}t_{3}^{2}t_{5}+192\cdot t_{1}t_{2}^{2}t_{3}t_{4}^{2}+176\cdot t_{11}t_{1}t_{2}^{2}\\
&+&216\cdot t_{1}t_{2}t_{3}^{3}t_{4}+360\cdot t_{10}t_{1}t_{2}t_{3}+288\cdot t_{9}t_{1}t_{2}t_{4}+ 200\cdot t_{1}t_{2}t_{5}t_{8}\\
&+&162\cdot t_{9}t_{1}t_{3}^{2}+240\cdot t_{1}t_{3}t_{4}t_{8}+60\cdot t_{15}t_{1}+64\cdot t_{2}^{4}t_{4}^{2}\\
&+&144\cdot t_{2}^{3}t_{3}^{2}t_{4}+80\cdot t_{10}t_{2}^{3}+81\cdot t_{2}^{2}t_{3}^{4}+216\cdot t_{9}t_{2}^{2}t_{3}\\
&+&160\cdot t_{2}^{2}t_{4}t_{8}-72\cdot t_{2}^{2}t_{6}^{2}+180\cdot t_{2}t_{3}^{2}t_{8}-360\cdot t_{2}t_{3}t_{5}t_{6}\\
&-&192\cdot t_{2}t_{4}^{2}t_{6}-200\cdot t_{2}t_{4}t_{5}^{2}-216\cdot t_{3}^{2}t_{4}t_{6}-225\cdot t_{3}^{2}t_{5}^{2}\\
&-&480\cdot t_{3}t_{4}^{2}t_{5}-78\cdot t_{13}t_{3}-64\cdot t_{4}^{4}-120\cdot t_{12}t_{4}\\
&-&110\cdot t_{11}t_{5}-60\cdot t_{10}t_{6}+20\cdot t_{8}^{2}]\\
&+&(262567831872062961684419606434295856611174463307776000(\sqrt{-7})^{3})^{-1}\cdot \\
&\cdot&[-585610470501831386847174923971534490692561120788480000\cdot t_24\\
&+&45930232980535795046837248938943881622945970257920000\cdot t_{8}^{3}\\
&-&1591482572775565298372910675734405498235077869436928000\cdot t_{18}t_{3}^{2}\\
&-&2181686066575450264724769324599834377089933587251200000\cdot t_{19}t_{2}t_{3}+\cdots]\\
&+&\cdots\cdots
\end{array}
\end{equation}
So, the intersection numbers at $r=7$ are
\begin{equation}\label{7.14}
\begin{array}{llllllllllllllllllllllllllll}
\langle\tau_{0,5}\tau_{0,0}^{2}\rangle_{0}=1;\ \ \ \ \ \ \ \ \ \ \ \ \ \ \ \ \ \ \ \ \ \ \ \
\langle\tau_{0,1}\tau_{0,4}\tau_{0,0}\rangle_{0}=1\\
\langle\tau_{0,0}\tau_{0,2}\tau_{0,3}\rangle_{0}=1;\ \ \ \ \ \ \ \ \ \ \ \ \ \ \ \ \ \ \ \
\langle\tau_{0,1}\tau_{0,2}^{2}\rangle_{0}=1\\
\langle\tau_{1,2}\tau_{0,1}^{3}\rangle_{0}=1;\ \ \ \ \ \ \ \ \ \ \ \ \ \ \ \ \ \ \ \ \ \ \ \ \langle\tau_{0,5}^{2}\tau_{0,1}^{2}\rangle_{0}=\frac{1}{7}\\
\langle\tau_{0,1}\tau_{0,2}\tau_{0,4}\tau_{0,5}\rangle_{0}=\frac{1}{7};\ \ \ \ \ \ \ \ \ \ \ \ \ \ \ \ \langle\tau_{0,2}^{2}\tau_{0,3}\tau_{0,5}\rangle_{0}=\frac{1}{7}\\
\langle\tau_{1,0}\rangle_{1}=\frac{1}{4};\ \ \ \ \ \ \ \ \ \ \ \ \ \ \ \ \ \ \ \ \ \ \ \ \ \ \ \ \ \langle\tau_{0,0}\tau_{2,0}\rangle_{0}=\frac{1}{4}\\
\langle\tau_{0,2}\tau_{1,5}\rangle_{1}=\frac{1}{42};\ \ \ \ \ \ \ \ \ \ \ \ \ \ \ \ \ \ \ \ \ \ \ \ \langle\tau_{0,5}\tau_{1,2}\rangle_{1}=\frac{1}{42}\\
\langle\tau_{0,4}\tau_{1,3}\rangle_{1}=\frac{1}{28};\ \ \ \ \ \ \ \ \ \ \ \ \ \ \ \ \ \ \ \ \ \ \ \ \langle\tau_{1,0}^{2}\rangle_{1}=\frac{5}{16}\\
\langle\tau_{0,1}^{2}\tau_{2,5}\rangle_{1}=\frac{1}{28};\ \ \ \ \ \ \ \ \ \ \ \ \ \ \ \ \ \ \ \ \ \ \ \
\langle\tau_{0,1}\tau_{0,2}\tau_{2,4}\rangle_{1}=\frac{5}{84}\\
\langle\tau_{0,2}^{2}\tau_{2,3}\rangle_{1}=\frac{1}{12};\ \ \ \ \ \ \ \ \ \ \ \ \ \ \ \ \ \ \ \ \ \ \ \
\langle\tau_{1,0}^{3}\rangle_{1}=\frac{1}{2}\\
\langle\tau_{3,2}\rangle_{2}=\frac{1}{112}\ \ \ \ \ \ \ \ \ \cdots\cdots
\end{array}
\end{equation}

In \dref{7.14}, there are nontrivial terms of 7-spin intersection number. In fact, we have gotten much more terms than these, but for the length of this article, we have not write down them completely here. Again, it seems that, there is no practical tryout to give the intersection numbers in this case.

\section*{Appendix B. A Virasoro Constraint for Hurwitz Partition function}
As an example, we will give a Virasoro constraint for Hurwitz partition function in this Appendix. Our starting point is the partition function of $2$-spin intersection numbers, i.e. Kontsevich-Witten $\tau$-function (KW $\tau$-function). From section 4, it is easy to get the coefficients $a^{\{2\}}){n,m}$
$$
\begin{array}{llll}
a^{\{2\}}_{n,i}=0,\ \ \ \ \ \ \ \ \text{if}\ i\neq0(\text{mod}~3),\\
a^{\{2\}}_{n,3m}=\sqrt{-2}\left(\sum\limits_{k+l=m}C^{2k}_{n}\frac{(6l+2k-1)!!}{(2l)!2^{k}(72)^{l}}-
\sum\limits_{k+l=m-1}C^{2k+1}_{n}\frac{(6l+2k+3)!!}{(2l+1)!12\cdot2^{k}(72)^{l}}\right),
\end{array}
$$
and the coefficients $b^{\{2\}}_{n,m}$ are
$$
b^{\{2\}}_{n,m}=a^{\{2\}}_{n,n+m-1}-\sum\limits_{k=1}^{n-1}
a^{\{2\}}_{n,k}\cdot b^{\{r\}}_{n-k,m}.
$$
It is easy to know that $b^{\{2\}}_{n,m}\neq0$ only if $m+n=1(\text{mod}~3)$. J.Zhou got another expression for $b^{\{2\}}_{n,m}$ by solving the Virasoro constraint of KW $\tau$-function in \cite{Zhou}. The fact that KW $\tau$-function is a KdV $\tau$-function, together with its Virasoro constraint, determines the explicit form of KW $\tau$-function. The Virasoro constraint for KW $\tau$-function is
$$
\hat{\mathcal{L}}_{m}\cdot Z^{\{2\}}(t)=0,\ \ \ \ \ m\geq-1,
$$

$$
\begin{array}{llll}
\hat{\mathcal {L}}_{m}&=&-\sqrt{-2}\dfrac{\partial}{\partial t_{2m+3}}+\sum\limits_{k=1}^{\infty}\left(k+\dfrac{1}{2}\right)t_{2k+1}\dfrac{\partial}{\partial t_{2k+2m+1}}\\
&&+\dfrac{1}{4}\sum\limits_{k=1}^{m-1}\dfrac{\partial^2}{\partial t_{2k+1}\partial t_{2m-2k-1}}+\dfrac{t_{1}^2}{4}\delta_{m,-1}+\dfrac{1}{16}\delta_{m,0}, \  m\geq-1.
\end{array}
$$
The fermionic representation of Virasoro operators are
$$
\begin{array}{llll}
\mathcal {L}_{n}=-\sqrt{-2}\sum\limits_{r+s=2n+3}:\psi_{r}\psi_{s}^{\ast}:
+\dfrac{1}{4}\sum\limits_{r+s=2n}(s-r):\psi_{r}\psi_{s}^{\ast}:+\dfrac{1}{16}\delta_{n,0},\ n\geq-1.
\end{array}
$$
In particular, $\mathcal{L}_{-1}$ is
$$
\begin{array}{llll}
\mathcal {L}_{-1}&=&-\sqrt{-2}\sum\limits_{r+s=1}:\psi_{r}\psi_{s}^{\ast}:
+\dfrac{1}{4}\sum\limits_{r+s=-2}:\psi_{r}\psi_{s}^{\ast}:\\
&=&-\sqrt{-2}\left(\psi_{\frac{1}{2}}\psi_{\frac{1}{2}}^{\ast}
+\sum\limits_{k=0}^{\infty}\left(\psi_{-k-\frac{1}{2}}\psi_{k+\frac{3}{2}}^{\ast}
-\psi_{-k-\frac{1}{2}}^{\ast}\psi_{k+\frac{3}{2}}\right)\right)\\
&&+\dfrac{1}{4}\left(\psi_{-\frac{3}{2}}\psi_{-\frac{1}{2}}^{\ast}-\psi_{-\frac{1}{2}}
\psi_{-\frac{3}{2}}^{\ast}+\sum\limits_{l=0}^{\infty}(2l+3)\left(
\psi_{-l-\frac{5}{2}}\psi_{l+\frac{1}{2}}^{\ast}+\psi_{-l-\frac{5}{2}}^{\ast}
\psi_{l+\frac{1}{2}}\right)\right).
\end{array}
$$
From section 4, it is straightforward to get the fermionic representation of KW $\tau$-function
$$
Z^{\{2\}}(t)=\text{exp}\{\sum\limits_{m,n\geq0}b^{\{2\}}_{n+1,m+1}\psi_{-m-1/2}\psi^{\ast}_{-n-1/2}\}|0\rangle
=\widetilde{G}^{\{2\}}|0\rangle.
$$
From equation \dref{lyp86}, it is easy to know that the operator $U_{HW}$ bridging the Hurwitz partition function and KW $\tau$-function can be expressed as
$$
U_{HW}=\text{exp}(\frac{\beta}{2}W^{(3)}_{0})\cdot\text{exp}(-J_{-1})\cdot(\widetilde{G}^{\{2\}})^{-1}.
$$
So, we construct a Virasoro constraint for Hurwitz partition function as following
$$
\mathfrak{L}_{n}=U_{HW}\cdot\mathcal{L}_{n}\cdot(U_{HW})^{-1},\ \ \ \ \ n\geq-1.
$$
The Hurwitz partition function satisfies the following equations
$$
\mathfrak{L}_{n}\cdot Z_{H}(t)=0,\ \ \ \ \ \ n\geq-1.
$$

Now, we want to get the explicit expression of operators $\mathfrak{L}_{n}$. At first, we calculate the "gauge transformation"
$$
\begin{array}{lllllllll}
&&(\widetilde{G}^{\{2\}})^{-1}\mathcal {L}_{n}\widetilde{G}^{\{2\}}\\
&=&-\sqrt{-2}\sum\limits_{k=0}^{2n+2}\psi_{2n-k+\frac{5}{2}}\psi^{\ast}_{k+\frac{1}{2}}
+\dfrac{1}{4}\sum\limits_{k=0}^{2n-1}(2k-2n+1)\psi_{2n-k-\frac{1}{2}}\psi^{\ast}_{k+\frac{1}{2}}\\
&&-\sqrt{-2}\sum\limits_{k=0}^{\infty}\left(\psi_{-k-\frac{1}{2}}\psi^{\ast}_{2n+k+\frac{7}{2}}
-\psi^{\ast}_{-k-\frac{1}{2}}\psi_{2n+k+\frac{7}{2}}\right)\\
\end{array}
$$
$$
\begin{array}{llllllllllllllll}
&&+\dfrac{1}{4}\sum\limits_{k=0}^{\infty}(2k+2n+1)\left(
\psi_{-k-\frac{1}{2}}\psi^{\ast}_{2n+k+\frac{1}{2}}+\psi^{\ast}_{-k-\frac{1}{2}}
\psi_{2n+k+\frac{1}{2}}\right)\\
&&+\sqrt{-2}\sum\limits_{k=0}^{2n+2}\sum\limits_{l=0}^{\infty}\left\{b^{\{2\}}_{2n+3-k,l+1}
\psi_{-l-\frac{1}{2}}\psi^{\ast}_{k+\frac{1}{2}}+b^{\{2\}}_{l+1,2n+3-k}\psi^{\ast}_{-l-\frac{1}{2}}
\psi_{k+\frac{1}{2}}\right\}\\
&&-\dfrac{1}{4}\sum\limits_{k=0}^{2n-1}\sum\limits_{l=0}^{\infty}(2k-2n+1)\left\{b^{\{2\}}_{2n-k,l+1}
\psi_{-l-\frac{1}{2}}\psi^{\ast}_{k+\frac{1}{2}}+b^{\{2\}}_{l+1,2n-k}\psi^{\ast}_{-l-\frac{1}{2}}
\psi_{k+\frac{1}{2}}\right\}\\
&&-\sqrt{-2}\sum\limits_{k,l=0}^{\infty}\left\{b^{\{2\}}_{l+1,2n+k+4}\psi_{-k-\frac{1}{2}}
\psi^{\ast}_{-l-\frac{1}{2}}+b^{\{2\}}_{2n+k+4,l+1}\psi^{\ast}_{-k-\frac{1}{2}}\psi_{-l-\frac{1}{2}}\right\}\\
&&+\dfrac{1}{4}\sum\limits_{k,l=0}^{\infty}(2k+2n+1)\left\{b^{\{2\}}_{l+1,2n+k+1}\psi_{-k-\frac{1}{2}}
\psi^{\ast}_{-l-\frac{1}{2}}-b^{\{2\}}_{2n+k+1,l+1}\psi^{\ast}_{-k-\frac{1}{2}}\psi_{-l-\frac{1}{2}}\right\}\\
&&+\sqrt{-2}\sum\limits_{k=0}^{2n+2}\sum\limits_{l,s=0}^{\infty}\left\{b^{\{2\}}_{2n+3-k,l+1}b^{\{2\}}_{s+1,k+1}
\psi_{-l-\frac{1}{2}}\psi^{\ast}_{-s-\frac{1}{2}}\right.\\
&&\left.+b^{\{2\}}_{l+1,k+1}b^{\{2\}}_{2n+3-k,s+1}\psi_{-s-\frac{1}{2}}
\psi^{\ast}_{-l-\frac{1}{2}}\right\}\\
&&-\dfrac{1}{4}\sum\limits_{k=0}^{2n-1}\sum\limits_{l,s=0}^{\infty}(2k-2n+1)
\left\{b^{\{2\}}_{s+1,k+1}b^{\{2\}}_{2n-k,l+1}\psi_{-l-\frac{1}{2}}\psi^{\ast}_{-s-\frac{1}{2}}\right.\\
&&\left.-b^{\{2\}}_{l+1,2n-k}b^{\{2\}}_{k+1,s+1}\psi_{-s-\frac{1}{2}}\psi^{\ast}_{-l-\frac{1}{2}}\right\}\\
&&-\sqrt{-2}\sum\limits_{k=0}^{2n+2}b^{\{2\}}_{k+1,2n+3-k}+\dfrac{1}{4}\sum\limits_{k=0}^{2n-1}(2k-2n+1)b^{\{2\}}_{k+1,2n-k}
+\dfrac{1}{16}\delta_{n,0}.
\end{array}
$$
The Virasoro constraint for KW $\tau$-function can also be written as
$$
\begin{array}{llll}
0=(\widetilde{G}^{\{2\}})^{-1}\cdot\mathcal {L}_{n}\cdot Z^{\{2\}}(t)=(\widetilde{G}^{\{2\}})^{-1}\cdot\mathcal {L}_{n}\cdot\widetilde{G}^{\{2\}}|0\rangle  \ \ \ \ n\geq-1,
\end{array}
$$
so, the summation of terms is zero if they consist of only creation operators ($\psi_{r+1/2},\psi^{\ast}_{r+1/2},\ r>0$) only, i.e.
$$
\begin{array}{llllllll}
0&=&-\sqrt{-2}\sum\limits_{k,l=0}^{\infty}\left\{b^{\{2\}}_{l+1,2n+k+4}\psi_{-k-\frac{1}{2}}
\psi^{\ast}_{-l-\frac{1}{2}}+b^{\{2\}}_{2n+k+4,l+1}\psi^{\ast}_{-k-\frac{1}{2}}\psi_{-l-\frac{1}{2}}\right\}\\
&&+\dfrac{1}{4}\sum\limits_{k,l=0}^{\infty}(2k+2n+1)\left\{b^{\{2\}}_{l+1,2n+k+1}\psi_{-k-\frac{1}{2}}
\psi^{\ast}_{-l-\frac{1}{2}}-b^{\{2\}}_{2n+k+1,l+1}\psi^{\ast}_{-k-\frac{1}{2}}\psi_{-l-\frac{1}{2}}\right\}\\
&&+\sqrt{-2}\sum\limits_{k=0}^{2n+2}\sum\limits_{l,s=0}^{\infty}\left\{b^{\{2\}}_{2n+3-k,l+1}b^{\{2\}}_{s+1,k+1}
\psi_{-l-\frac{1}{2}}\psi^{\ast}_{-s-\frac{1}{2}}\right.\\
&&\left.+b^{\{2\}}_{l+1,k+1}b^{\{2\}}_{2n+3-k,s+1}\psi_{-s-\frac{1}{2}}
\psi^{\ast}_{-l-\frac{1}{2}}\right\}\\
&&-\dfrac{1}{4}\sum\limits_{k=0}^{2n-1}\sum\limits_{l,s=0}^{\infty}(2k-2n+1)
\left\{b^{\{2\}}_{s+1,k+1}b^{\{2\}}_{2n-k,l+1}\psi_{-l-\frac{1}{2}}\psi^{\ast}_{-s-\frac{1}{2}}\right.\\
&&\left.-b^{\{2\}}_{l+1,2n-k}b^{\{2\}}_{k+1,s+1}\psi_{-s-\frac{1}{2}}\psi^{\ast}_{-l-\frac{1}{2}}\right\}\\
&&-\sqrt{-2}\sum\limits_{k=0}^{2n+2}b^{\{2\}}_{k+1,2n+3-k}+\dfrac{1}{4}\sum\limits_{k=0}^{2n-1}(2k-2n+1)
b^{\{2\}}_{k+1,2n-k}
+\dfrac{1}{16}\delta_{n,0}.
\end{array}
$$
With certain assumptions, in $r=2$ case, J.Zhou got the fermionic representation of KW $\tau$-function by solving above equation directly. Therefore, $(\widetilde{G}^{\{2\}})^{-1}\mathcal {L}_{n}\widetilde{G}^{\{2\}}$ can be rewritten as
$$
\begin{array}{lllllllllllll}
&&(\widetilde{G}^{\{2\}})^{-1}\mathcal {L}_{n}\widetilde{G}^{\{2\}}\\
&=&-\sqrt{-2}\sum\limits_{k=0}^{2n+2}\psi_{2n-k+\frac{5}{2}}\psi^{\ast}_{k+\frac{1}{2}}
+\dfrac{1}{4}\sum\limits_{k=0}^{2n-1}(2k-2n+1)\psi_{2n-k-\frac{1}{2}}\psi^{\ast}_{k+\frac{1}{2}}\\
&&-\sqrt{-2}\sum\limits_{k=0}^{\infty}\left(\psi_{-k-\frac{1}{2}}\psi^{\ast}_{2n+k+\frac{7}{2}}
-\psi^{\ast}_{-k-\frac{1}{2}}\psi_{2n+k+\frac{7}{2}}\right)\\
&&+\dfrac{1}{4}\sum\limits_{k=0}^{\infty}(2k+2n+1)\left(
\psi_{-k-\frac{1}{2}}\psi^{\ast}_{2n+k+\frac{1}{2}}+\psi^{\ast}_{-k-\frac{1}{2}}
\psi_{2n+k+\frac{1}{2}}\right)\\
&&+\sqrt{-2}\sum\limits_{k=0}^{2n+2}\sum\limits_{l=0}^{\infty}\left\{b^{\{2\}}_{2n+3-k,l+1}
\psi_{-l-\frac{1}{2}}\psi^{\ast}_{k+\frac{1}{2}}+b^{\{2\}}_{l+1,2n+3-k}\psi^{\ast}_{-l-\frac{1}{2}}
\psi_{k+\frac{1}{2}}\right\}\\
&&-\dfrac{1}{4}\sum\limits_{k=0}^{2n-1}\sum\limits_{l=0}^{\infty}(2k-2n+1)\left\{b^{\{2\}}_{2n-k,l+1}
\psi_{-l-\frac{1}{2}}\psi^{\ast}_{k+\frac{1}{2}}+b^{\{2\}}_{l+1,2n-k}\psi^{\ast}_{-l-\frac{1}{2}}
\psi_{k+\frac{1}{2}}\right\}.
\end{array}
$$
In particular, $(\widetilde{G}^{\{2\}})^{-1}\mathcal{L}_{-1}\widetilde{G}^{\{2\}}$ is
$$
\begin{array}{lllllllllll}
&&(\widetilde{G}^{\{2\}})^{-1}\mathcal {L}_{-1}\widetilde{G}^{\{2\}}\\
&=&-\sqrt{-2}\psi_{\frac{1}{2}}\psi^{\ast}_{\frac{1}{2}}+\sqrt{-2}
\sum\limits_{k=0}^{\infty}\left(b^{\{2\}}_{1,k+1}\psi_{-k-\frac{1}{2}}\psi^{\ast}_{\frac{1}{2}}
+b^{\{2\}}_{k+1,1}\psi^{\ast}_{-k-\frac{1}{2}}\psi_{\frac{1}{2}}\right)\\
&&-\sqrt{-2}\sum\limits_{k=0}^{\infty}\left(\psi_{-k-\frac{1}{2}}\psi^{\ast}_{k+\frac{3}{2}}
-\psi^{\ast}_{-k-\frac{1}{2}}\psi_{k+\frac{3}{2}}\right)\\
&&+\dfrac{1}{2}
\sum\limits_{k=0}^{\infty}(2k+3)\left(\psi_{-\frac{5}{2}}\psi^{\ast}_{k+\frac{1}{2}}
+\psi^{\ast}_{-k-\frac{5}{2}}\psi_{k+\frac{1}{2}}\right).
\end{array}
$$
Finally, we get a set of Virasoro operators $\{\mathfrak{L}_{n},n\geq-1\}$ which is a Virasoro constraint for the Hurwitz partition function
$$
\begin{array}{lllllllllll}
\mathfrak{L}_{n}
&=&\dfrac{1}{2}\sum\limits_{u,v=0}^{2n+2}\dfrac{(-1)^{v}}{u!v!}\left\{-2\sqrt{-2}\sum\limits_{k=0}^{\infty}
\exp\dfrac{\beta}{2}\left[(2n-k+\frac{5}{2}-u)^{2}-(k+\frac{1}{2}-v)^{2}\right]\right.\\
&&\left.\cdot
\psi_{2n-k+\frac{5}{2}-u}\psi^{\ast}_{k+\frac{1}{2}-v}
+\dfrac{1}{2}\sum\limits_{k=0}^{2n-1}(2k-2n+1)\exp\dfrac{\beta}{2}
\left[(2n-k-\frac{1}{2}-u)^{2}\right.\right.\\
&&\left.\left.-(k+\frac{1}{2}-v)^{2}\right]\cdot\psi_{2n-k-\frac{1}{2}-u}
\psi^{\ast}_{k+\frac{1}{2}-v}\right.\\
&&\left.+2\sqrt{-2}\sum\limits_{k=0}^{\infty}\left[\exp\dfrac{\beta}{2}\left[(-k-\frac{1}{2}-u)^{2}
-(2n+k+\frac{7}{2}-v)^{2}\right]\right.\right.\\
&&\left.\left.\cdot\psi_{-k-\frac{1}{2}-u}\psi^{\ast}_{2n+k-\frac{7}{2}-v}
-\exp\dfrac{\beta}{2}\left[(2n+k+\frac{7}{2}-u)^{2}-(-k-\frac{1}{2}-v)^{2}\right]\right.\right.\\
&&\left.\left.\cdot\psi^{\ast}_{-k-\frac{1}{2}}\psi_{2n+k+\frac{7}{2}-u}\right]\right.\\
&&\left.+\dfrac{1}{2}\sum\limits_{k=0}^{\infty}(2n+2k+1)\left[\exp\dfrac{\beta}{2}
\left[(-k-\frac{1}{2}-u)^{2}-(2n+k+\frac{1}{2}-v)^{2}\right]\right.\right.\\
&&\left.\left.\cdot\psi_{-k-\frac{1}{2}-u}
\psi^{\ast}_{2n+k+\frac{1}{2}-v}
+\exp\dfrac{\beta}{2}\left[(2n+k+\frac{1}{2}-u)^{2}-(-k-\frac{1}{2}-v)^{2}\right]\right.\right.\\
&&\left.\left.\cdot\psi^{\ast}_{-k-\frac{1}{2}-v}\psi_{2n+k+\frac{1}{2}-u}\right]\right.\\
\end{array}
$$
$$
\begin{array}{llllllllllllll}
&&\left.+2\sqrt{-2}\sum\limits_{k=0}^{2n+2}\sum\limits_{l=0}^{\infty}\left[\exp\dfrac{\beta}{2}
\left[(-l-\frac{1}{2}-u)^{2}-(k+\frac{1}{2}-v)^{2}\right]\cdot b^{\{2\}}_{2n+3-k,l+1}\right.\right.\\
&&\left.\left.\cdot\psi_{-l-\frac{1}{2}-u}\psi^{\ast}_{k+\frac{1}{2}-v}
+\exp\dfrac{\beta}{2}\left[(k+\frac{1}{2}-u)^{2}-(-l-\frac{1}{2}-v)^{2}\right]\cdot b^{\{2\}}_{l+1,2n+3-k}\right.\right.\\
&&\left.\left.\cdot\psi^{\ast}_{-l-\frac{1}{2}-v}\psi_{k+\frac{1}{2}-u}\right]\right.\\
&&\left.-\dfrac{1}{2}\sum\limits_{k=0}^{2n-1}(2k-2n+1)\sum\limits_{l=0}^{\infty}
\left[\exp\dfrac{\beta}{2}\left[(-l-\frac{1}{2}-u)^{2}-(k+\frac{1}{2}-v)^{2}\right]\cdot\right.\right.\\
&&\left.\left.\cdot A_{l,2n-1-k}\cdot\psi_{-l-\frac{1}{2}-u}\psi^{\ast}_{k+\frac{1}{2}-v}
-\exp\dfrac{\beta}{2}\left[(k+\frac{1}{2}-u)^{2}-(-l-\frac{1}{2}-v)^{2}\right]\right.\right.\\
&&\left.\left.\cdot A_{2n-1-k,l}\cdot\psi^{\ast}_{-l-\frac{1}{2}-v}\psi_{k+\frac{1}{2}-u}\right]\right\},
\end{array}
$$
In particular, $\mathfrak{L}_{-1}$ is
$$
\begin{array}{lllllllll}
\mathfrak{L}_{-1}
&=&\dfrac{1}{2}\sum\limits_{u,v=0}^{\infty}\dfrac{(-1)^{v}}{u!v!}\left\{-2\sqrt{-2}\exp\dfrac{\beta}{2}
\left[(\frac{1}{2}-u)^{2}-(\frac{1}{2}-v)^{2}\right]\cdot\psi_{\frac{1}{2}-u}
\psi^{\ast}_{\frac{1}{2}-v}\right.\\
&&\left.-2\sqrt{-2}\sum\limits_{k=0}^{\infty}\left[\exp\dfrac{\beta}{2}\left[(-k-\frac{1}{2}-u)^{2}
-(\frac{1}{2}-v)^{2}\right]\cdot b^{\{2\}}_{1,k+1}\cdot\psi_{-k-\frac{1}{2}-u}\psi^{\ast}_{\frac{1}{2}-v}\right.\right.\\
&&\left.\left.+\exp\dfrac{\beta}{2}\left[(\frac{1}{2}-u)^{2}-(-k-\frac{1}{2}-v)^{2}\right]\cdot A_{k+1,1}\cdot\psi^{\ast}_{-k-\frac{1}{2}-v}\psi_{\frac{1}{2}-u}\right]\right.\\
&&\left.-2\sqrt{-2}\sum\limits_{k=0}^{\infty}\left[\exp\dfrac{\beta}{2}\left[(-k-\frac{1}{2}-u)^{2}
-(k+\frac{3}{2}-v)^{2}\right]\cdot\psi_{-k-\frac{1}{2}-u}\psi^{\ast}_{k+\frac{3}{2}-v}\right.\right.\\
&&\left.\left.-\exp\dfrac{\beta}{2}\left[(k+\frac{3}{2}-u)^{2}-(-k-\frac{1}{2}-v)^{2}\right]
\cdot\psi^{\ast}_{-k-\frac{1}{2}-v}\psi_{k+\frac{3}{2}-v}\right]\right.\\
&&\left.+\sum\limits_{k=0}^{\infty}(k+\frac{3}{2})\left[\exp\dfrac{\beta}{2}\left[(-k-\frac{5}{2}-u)^{2}
-(k+\frac{1}{2}-v)^{2}\right]\cdot\psi_{-k-\frac{5}{2}-u}\psi^{\ast}_{k+\frac{1}{2}-v}\right.\right.\\
&&\left.\left.+\exp\dfrac{\beta}{2}\left[(k+\frac{1}{2}-u)^{2}-(-k-\frac{5}{2}-v)^{2}\right]
\cdot\psi^{\ast}_{-k-\frac{5}{2}-v}\psi_{k+\frac{1}{2}-u}\right]\right\},
\end{array}
$$

Therefore, we get a Virasoro constraint for the Hurwitz partition function. We express this Virasoro constraint in terms of fermions, in principle, we could get its bosonic version by bosons-fermions correspondence. By solving the constraint $\mathfrak{L}_{-1}$, we can get the Schur polynomial expression of the Hurwitz partition function.


\begin{thebibliography}{99}

\bibitem{A.Alexandrov1} Alexandrov, A.: {Cut-and-Join Operator Represetation for Kontsevich-Witten Tau-Function}. Mod.Phys.Lett.A \textbf{26}, 2193--2199(2011) [hep-th/1009.4887]

\bibitem{A.Alexandrov2} Alexandrov, A.: {From Hurwitz Numbers to Kontsevich-Witten Tau-Function: a Connection by Virasoro Operators}. Lett.Math.Phys. {\bf 104}, 75--87(2014) [hep-th/1111.5349]

\bibitem{A.Alexandrov3} Alexandrov, A., Mironov, A., Morozov, A., Natanzon, S.: {Integrability of Hurwitz Partition Functions,}. J.Phys.A: Math.Theor. {\bf 45}, (2012) [hep-th/1103.4100]

\bibitem{A.Alexandrov4} Alexandrov, A., Mironov, A., Morozov, A.: {M-Theory of Matrix Models}. Theor.Mat.Fiz. \textbf{150}, 179--192(2007) [hep-th/0605171]

\bibitem{A.Alexandrov5} Alexandrov, A., Mironov, A., Morozov, A.: {Instants and Merons in Matrix Models}. Physica.D  \textbf{235}, 126--167(2007) [hep-th/0608228]

\bibitem{A.Alexandrov6} Alexandrov, A., Mironov, A., Morozov, A.: {BGWM as Second Constituent of Complex Matrix Model}. JHEP {\bf 0912}, 053(2009) [hep-th/0906.3305]

\bibitem{A.Alexandrov7} Alexandrov, A., Zabrodin, A.: { Free Fermions and Tau-Functions}.  Journal of Geometry and Physics \textbf{67}, 37--80 (2013) [math-ph/1212.6049]

\bibitem{A.Alexandrov8} Alexandrov, A.: {Enumerative Geometry, Tau-functions and Heisenberg-Virasoro Algebra}. [hep-th/1404.3402].

\bibitem{O.Babelon} Balelon, O., Bernard, D., Talon, M.: {Introduction to Classical Integrable Systems}. Cambridge University Press, 2003

\bibitem{R.Blumenhagen} Blumenhagen, R., Plauschinn, E.: {Introduction to Conformal Field Theory: With Applications to String Theory}. Lect.Notes Phys. {\bf 779}. Berlin Herdelberg: Springer, 2009

\bibitem{G.Borot} Borot, G., Eynard, B., Mulase, M., Safnuk, B.: {A Matrix Model for Simple Hurwitz Numbers and Topological Recursion}. J. Geom. Phys. {\bf 61}, 522--540(2011).[math-ph/0906.1206]

\bibitem{B.Brezin} Br\'ezin, E., Hikami, S.: {Intersection numbers of Riemann surfaces from Gaussian Matrix models}. JHEP {\bf 10}, 096(2007)

\bibitem{S.Chadha} Chadha, S., Mahoux, G., Mehta, M.: {A Method of Integration Over Matrix Variables 2}.  J.Phys.A \textbf{14}, 579(1981).

\bibitem{L.A.Dickey} Dickey, L.: { Soliton Equations and Hamiltonian Systems}. 2nd edition, Advanced Series in Mathematical Physics, Vol.{\bf 26}. World Scientific, 2003

\bibitem{E.Date} Date, E., Kashivara, M., Jinbo, M., Miwa, T.: {Transformation Groups for Soliton Equations}. In: Proc. of RIMS Symposium on Non-Linear Integrable Systems, pp.39--119 Singapore: World Science Publ.Co. 1983

\bibitem{R.Dijkgraaf} Dijkgraaf, R.: {\it Intersection Theory, Integrable Hierarchies and Topological Field Theory}, Cargese Lectures 16--27 July 1991, Publ.in NATO ASI, Cargese 1991

\bibitem{T.Ekedahl} Ekedahl, T., Lando, S., Shapiro, M., Vainshtein, A: {Hurwitz Numbers and Intersections on Moduli Spaces of Curves}. Invent.Math. \textbf{146}, 297--327(2001) [math/0004096]

\bibitem{C.Faber}Faber, C., Shadrin, S., Zvonkine, D.: {Tautological Relations and The r-Spin Witten Conjecture}.  Ann.Sci.Ec.Norm.Super. {\bf 43}(4), 621--658(2010)

\bibitem{P.Di Francesco}  Francesco, P., Ginsparg, P., Justin, J.Z.: {2D Gravity and Random Matrices}. Phys.Rept. {\bf 254}, (1995) [hep-th/9306153]

\bibitem{M.Fukuma}Fukuma, M., Kawai, H., Nakayama, R.: { Infinite Dimensional Grassmannian Structure of Two-Dimensional Quantum Gravity}.  Commun.Math.Phys. {\bf 143}, 371--403(1992)

\bibitem{P.Ginsparg} Ginsparg, P., and Moore, G.: {Lectures on 2D Gravity and 2D String Theory}. [hep-th/9304011]

\bibitem{A.Hurwitz} Hurwitz, A. : {\"{U}ber Riemann¡¯sche Fl\"{a}chen mit gegebene Verzweigungspunk ¡§ ten}. Mathematische Annalen {\bf 39}, 1--66(1891)

\bibitem{C.Itzykson} Itzyskon, C., Zuber, J.: {The Planar Approximation II}. J.Math.Phys.  {\bf 21}, 411(1980)

\bibitem{J.Goeree}Goeree, J.: {W-Constraints in 2D Quantum Gravity}. Nucl.Phys. {\bf B358}(3), 737--757(1991)

\bibitem{V.Kac}Kac, V., Schwarz, A.: {Geometric Interpretation of the Partation Function of 2-D Gravity}. Mod.Phys.Lett.A {\bf 6}, 611(1991)

\bibitem{V.Kac1} Kac, V., Raina, A.: {Bombay Lectures on Highest Weight Reprsentations of Infinite dimensional Lie Alegebra}. Advanced Series in Mathematical Physics Vol.2. Singapore: World Scientific Punblishing, 1987

\bibitem{V.Kac2} Kac, V.: {Vertex algebras for beginners}. Second edition, University Lecture Series 10. American Mathematical Society, Providence, RI, 1998

\bibitem{M.Kazarian} Kazarian, M.: {KP Hierachy for Hodge Integrals}. Adv.Math. \textbf{221}, 1--21(2009) [math.AG/0809.3263]

\bibitem{S.Kharchev} Kharchev, S.: {Kadomtsev-Petviashvili Hierarchy and Generalized Kontsevich Model}. [hep-th/980091]

\bibitem{S.Kharchev2} Kharchev, S., Marshakov, A., Mironov, A., Morozov, A., Zabrodin, A.: {Unification of All String Models with c$<$1}. Phys.Lett. \textbf{B275}, 311--314(1992) [hep-th/9111037]

\bibitem{S.Kharchev1} Kharchev, S., Marshakov, A., Mironov, A., Morozov, A., Zabrodin, A.: {Towards Unified Theory of 2-d Gravity}. Nucl.Phys. \textbf{B380}, 181--240(1992) [hep-th/920103]

\bibitem{M.Kontsevich} Kontsevich, M.: {Intersection Theory on The Moduli Space of Curves and the Matrix Airy Function}. Comm.Math.Phys. \textbf{147}, 1--23(1992).

\bibitem{K.Liu}Liu, K., Vakil, R., Xu, H.: {Formal Pseudodifferential Operators and Witten's $r$-Spin Numbers}. [math.AG:1112.4601]

\bibitem{Liu}liu, S.-Q., Yang, D., Zhang, Y.: {Uniqueness Theorems of W-Constraints for Simple Singularitie}. [math.QA:1305.2593]

\bibitem{X.Liu}Liu, X., Wang, G.: {Connecting the Kontesevich-Witten and Hodge Tau-Function by the $\widehat{GL}$ operators}. [math.ph:1503.05268]

\bibitem{Macdonald} Macdonald, I.: {Symmetric Functions and Hall Polynomials}. 2nd edition. Claredon Press,(1995).

\bibitem{A.Mironov1} Mirozov, A., and Moronov, A.: {Virasoro Constraints for Kontsevich-Hurwitz Partition Function}.  JHEP {\bf 0902}, 024(2009) [hep-th/0807.2843]

\bibitem{A.Mironov2} Mironov, A., Morozov, A., Natanzon, S.: {Integrability Prorerties of Hurwitz Partition Functions. II. Multiplication of the Cut-and-Join Operators and WDVV Equations}. JHEP {\bf 1111}, 097(2011) [hep-th/1108.0885]

\bibitem{A.Mironov3} Mironov, A., Morozov, A., Semenoff, G.: {Unitary matrix integrals in the framework of generalized Kontsevich moldel 1.Brezin-Gross-Witten model}. Int.J.Mod.Phys. \textbf{A11}, 5031--5080(1996) [hep-th/9404005]

\bibitem{T.Miwa} Miwa, T., Jinbo, M., and Date, E.: {Solitons. Differential Equations, Symmetries and Infinite-Dimensional Algebras}. Cambridge Tracts in Mathematics, 135. Cambridge: Cambridge University Press, 2000

\bibitem{T.Miwa1}Miwa, T.: On Hirota's Differential equations.  Proc.Japan.Acad. {\bf 58A}, 9--12(1982)

\bibitem{A.Morozov1} Morozov, A.: {Integrability and Matrix Model}. Phys.Usp. {\bf 37}, 1--55(1994) [hep-th/9303139]

\bibitem{A.Morozov} Morozov, A., Shakirov, S.: {Generation of Matrix Model by W-Operators}. JHEP {\bf 0904}, 064(2009) [hep-th/0902.2627]

\bibitem{A.Okounkov} Okounkov, A., Pandharipande, R.: {Gromov-Witten Theory, Hurwitz Numbers and Matrix Models.1}. [math.AG/0101147]

\bibitem{A.Okounkov1} Okounkov, A., Pandharipande, R.: {The Equivariant Gromov-Witten the of $P^{1}$}. [math.AG/0207233]

\bibitem{S.V.Shadrin}Shadrin, S.: {Geometry of Meromorphisic Funcions and Intersection on Moduli Spaces of Curves}.  Int.Math.Res.Not. {\bf 38}, 2051--2094(2003)

\bibitem{E.Witten} Witten, E.: {Two-Dimensional Gravity and Intersection Theory on Moduli Space}. Survey in Differential Geometry {\bf 1}, 243--310(1991).

\bibitem{E.Witten1}Witten, E.: Algebraic Geometry Associated with Matrix Models of Two-Dimensional Gravity. In Topological Mathods in Mordern Mathematics (Stony Brook, NY 1991), pp.235--269, Publish or Perish, Houston, TX, 1993

\bibitem{Zhou} Zhou, J.: {Explicit Formula for Witten-Kontsevich Tau-Function}. [math.AG/1306.5429]

\bibitem{Zhou1}Zhou, J.: Solution of $W$-Constraints for r-Spin Intersection Numbers. [math-ph:1305.6991]

\end{thebibliography}
\end{document}